\documentclass[aps,prb,superscriptaddress,reprint]{revtex4-2}
\usepackage{graphicx,amsfonts,amssymb,amsmath,hyperref,hypcap,enumerate}
\usepackage{braket}
\usepackage{lipsum} 
\usepackage{amsthm}
\usepackage{multirow}
\usepackage{graphicx}
\usepackage{dcolumn}   %
\usepackage{bm}        %
\usepackage{amssymb}   %
\usepackage{amsmath}
\usepackage{braket}
\usepackage{epstopdf}
\usepackage{color}
\usepackage{subfigure}
\usepackage{diagbox}

\usepackage{multirow}
\usepackage{makecell}
\usepackage{bbold}

\DeclareMathAlphabet{\mathitb}{OT1}{cmr}{bx}{sl}
\usepackage[utf8]{inputenc}

\begin{document}
\def\blue#1{\textcolor{blue}{#1}}
\newcommand{\magenta}[1]{{\textcolor{magenta}{#1}}}

\title{Level statistics of real eigenvalues in non-Hermitian systems}

\author{Zhenyu Xiao}
\affiliation{International Center for Quantum Materials, Peking University, Beijing 100871, China}

\author{Kohei Kawabata}
\affiliation{Department of Physics, Princeton University, Princeton, New Jersey, 08540, USA}

\author{Xunlong Luo}
\affiliation{Science and Technology on Surface Physics and Chemistry Laboratory, Mianyang 621907, China}

\author{Tomi Ohtsuki}
\affiliation{Physics Division, Sophia University, Chiyoda-ku, Tokyo 102-8554, Japan}

\author{Ryuichi Shindou}
\email{rshindou@pku.edu.cn}
\affiliation{International Center for Quantum Materials, Peking University, Beijing 100871, China}

\date{\today}
\begin{abstract}
Symmetries associated with complex conjugation and Hermitian conjugation, 
such as time-reversal symmetry and pseudo-Hermiticity, have great 
impact on eigenvalue spectra of non-Hermitian random matrices. 
Here, we show that time-reversal symmetry and pseudo-Hermiticity lead 
to universal level statistics of non-Hermitian random matrices 
on and around the real axis. From the extensive numerical 
calculations of large random matrices, we obtain the five 
universal level-spacing and level-spacing-ratio distributions 
of real eigenvalues, each of which is unique to the symmetry class.
Furthermore, 
we analyse %
spacings 
of real eigenvalues
in
physical models, such as bosonic many-body systems and free %
fermionic
systems 
with disorder and dissipation. 
We clarify that 
the level spacings %
in ergodic (metallic) phases
are
described by the universal distributions 
of non-Hermitian random matrices in the same symmetry classes,
while
the level spacings in many-body localized and Anderson localized phases %
show the Poisson statistics. 
We also find that %
the number of real eigenvalues %
shows distinct scalings in the ergodic %
and 
localized phases in these symmetry classes. These results serve as 
effective tools for detecting quantum chaos, many-body localization, 
and real-complex transitions in non-Hermitian systems with symmetries. 
\end{abstract}
\maketitle

\section{Introduction}

An understanding of spectral correlations under symmetries is useful in classifying %
phases
of matter~\cite{mehta04}.
In closed quantum systems, the spectral statistics of non-integrable systems typically coincide with those of Hermitian random matrices with symmetries, which serves as an effective tool for detecting quantum chaos~\cite{wigner1951, Bohigas84, haake1991quantum, dalessio2016quantum}.
The spectral statistics also provide a measure of the Anderson transitions~\cite{al89,Shklovskii93,mirlin00,Evers08} and many-body-localization (MBL) transitions~\cite{Oganesyan07, pal10, Serbyn2016, Abanin19}.
When a disordered many-body Hermitian system is in the ergodic phase, 
statistics of spacing between its eigenenergy levels 
are described by the Wigner-Dyson distribution of Hermitian random matrices. 
The Wigner-Dyson distribution is universally 
classified by time-reversal symmetry (TRS). 
Hermitian random matrices without TRS and 
with TRS whose sign is $+1$ and $-1$ 
respectively
belong to 
the Gaussian unitary, orthogonal, and symplectic ensembles, 
each of which exhibits 
the distinct spectral statistics. 

While state-of-the-art quantum experiments %
facilitate probing quantum many-body 
physics including the MBL~\cite{choi16, Abanin19}, 
energy gain and loss naturally exist in these optical 
systems, removing the Hermiticity condition from 
their many-body Hamiltonians. 
Consequently, 
open quantum systems described by non-Hermitian operators have attracted growing interest.
Researchers have studied the non-Hermitian physics of optical phenomena~\cite{Bender98, Makris08, guo2009, Ruter10, Feng11, Zeuner15, pan2018photonic, Li19, Konotop16, Miri19},
topological phases~\cite{Lee16, Xu17, Harari18, Gong18, Yao18, Yao18_2, Kawabata19NC, Kawabata19, Longhi19, Zeng20, Li2020},
and Anderson and many-body  localization~\cite{hatano1996localization,Efetov97,Feinberg99,Hatano16,Xu16,hamazaki2019non,Tzortzakakis20,Wang20,Huang20,Huang20SR,Kawabata20,Luo21,Luo21TM,luo2021unifying}.
These works %
have led
to a remarkable %
advance
in %
spectral properties %
of non-Hermitian operators~\cite{Ginibre65, Grobe88, Grobe89, Lehmann1991, edelman1995, Feinberg97, Sommers98, Kanzieper05, Forrester07, Hamazaki20, Akemann19, Ribeiro19, Sa20, Sa21, Li-Chan-21, GarciaGarcia-22}. 
Still, it remains to be fully explored how symmetries influence the universal spectral properties of non-Hermitian operators.

The level statistics analyses of %
Hermitian systems cannot be directly applied to non-Hermitian %
systems. 
Due to the absence of Hermiticity, the 10-fold Hermitian symmetry classification~\cite{Altland97} is enriched into 38-fold symmetry classification~\cite{BL02,Kawabata19,Zhou19}. 
Eigenvalues of non-Hermitian %
systems
distribute in the two-dimensional (2D) complex plane. 
The statistics of complex level spacings $s_{\alpha}$, defined as the distance with the closest eigenvalues in the complex plane (i.e., $s_{\alpha} \equiv \min_{\beta} |E_{\alpha} - E_{\beta}|$ for all complex eigenvalues $E_{\alpha}, E_{\beta}$ with $\alpha \neq \beta$), were previously studied to capture the spectral correlations of non-Hermitian systems~\cite{Ginibre65,Hamazaki20,Akemann19,Sa20,Sa21}.
Non-Hermitian random matrices without any symmetry show a universal distribution of the spacing of complex eigenvalues, known as the Ginibre distribution~\cite{Ginibre65}. An introduction of the transposition version of TRS, %
$H = {\cal U_T^{\dagger}} H^T {\cal U_T} , \, {\cal U_T^*}  {\cal U_T} = \pm 1 $, 
which is called TRS$^{\dagger}$~\cite{Kawabata19},
changes the distribution into two distinct distributions, depending on the sign of ${\cal U_T^*}  {\cal U_T} = \pm 1$~\cite{Hamazaki20}. 
This is similar to the three-fold Wigner-Dyson distribution for Hermitian random matrices.
Meanwhile, an introduction of TRS,
$H = {\cal U_T^{\dagger}} H^* {\cal U_T}, \, {\cal U_T^*}  {\cal U_T} = \pm 1 $, does not alter the spacing distribution away from the real axis~\cite{Ginibre65}. 
In fact, unlike 
TRS$^\dagger$, %
TRS 
only
relates an eigenvalue with its complex conjugate, so that it has no impact on the correlation between two neighboring eigenvalues away from the real axis. 
This fact makes the role of %
symmetries
in the universality classes of non-Hermitian random matrices elusive.

In this paper, we show that TRS leads to universal level statistics on and around the real axis.
In addition to TRS, we also identify the relevant symmetries that give rise to universal level statistics of real eigenvalues in non-Hermitian random matrices. 
The universal level statistics provide an effective tool for detecting quantum chaos in open quantum systems with the symmetries. 
In the 38-fold symmetry classification of non-Hermitian random matrices, we show that there exist seven symmetry classes in which eigenstates with real eigenvalues preserve all the symmetries of the symmetry class whereas eigenstates away from the real axis break some symmetries. 
They are a class only with pseudo-Hermiticity 
(class A + $\eta$; class AIII), classes with TRS whose sign is either 
$\pm 1$ (classes AI and AII), and classes with both TRS and 
pseudo-Hermiticity (classes AI + $\eta_{\pm}$ and AII + $\eta_{\pm}$);
see Table~\ref{symmetry_class_A}.
In the last classes, TRS commutes or 
anti-commutes with pseudo-Hermiticity. %
The subscript of ${\eta}_{\pm}$ %
denotes the commutation ($+$) or anti-commutation ($-$) 
relation between TRS and pseudo-Hermiticity. Note that 
random matrices with particle-hole symmetry 
($H = - {\cal U_P^{\dagger}} H^T {\cal U_P}$) 
and/or sublattice symmetry ($H = - {\cal U_S^{\dagger}} H {\cal U_S} $) 
do not give rise to the universal level statistics of real
eigenvalues because only states with zero eigenvalue 
respect the symmetries.  

\begin{table*}[bt]
\caption{
Ten-fold symmetry classification based on time-reversal symmetry (TRS), time-reversal symmetry$^{\dagger}$ (TRS$^{\dagger}$), and pseudo-Hermiticity (pH).
TRS and pH are equivalent to particle-hole symmetry$^{\dagger}$ (PHS$^{\dagger}$) and chiral symmetry (CS), respectively.
For the columns of given symmetry, the blank entries mean the absence of the symmetry. 
For TRS and TRS$^{\dagger}$, $\pm 1$ stands for the sign of the symmetry. 
If $H$ belongs to the symmetry class in the first column, ${\rm i} H$ belongs to the  equivalent symmetry class in the second column. 
The column ``soft gap" gives the small $y = \mathrm{Im}(E)$ behavior of the density $\rho_c(x,y)$ of complex eigenvalues if there is a soft gap around the
real axis $y=0$. 
The column ``$\delta(y)$" indicates whether there is a delta function peak on the real axis $y=0$. 
In the presence of the delta function peak, the column ``$\langle r \rangle$" and the column ``$\chi$'' respectively show the mean spacing ratio and spectral compressibility of real eigenvalues (see Eq.~(\ref{r_def}) and Eq.~(\ref{chi_def}) for their definitions) obtained from $4000 \times 4000$ random matrices in the generalized Gaussian ensemble. 
The standard deviation of $\langle r \rangle$ is estimated by the bootstrap method~\cite{press07} and labeled in the parentheses; for example, the standard deviation is 0.0004 for ``0.4194(4)''.
}
\begin{tabular}{cc|ccc|c|ccc}
\hline \hline
\begin{tabular}{c} ~symmetry~ \\ class \end{tabular}  &
\begin{tabular}{c} symmetry \\ ~class (equiv)~ \end{tabular} &
 \begin{tabular}{c} TRS \\ ~~(PHS$^{\dagger}$)~~ \end{tabular} &
 ~~TRS$^{\dagger}$~~
  &\begin{tabular}{c} pH \\ ~~(CS)~~ \end{tabular}& soft gap & 
  ~~$\delta(y)$~~ &    $\langle r \rangle $  &  ~~~~~~$\chi$~~~~~~ \\ \hline
A & A & & & & & & \\
A + $\eta$ & AIII &  & &$\surd $ & $|y|$ &$\surd$  & $0.4194(4)$ & $0.83$  \\ 
AI &D$^{\dagger}$ & +1 &  & & $|y|$ &$\surd$  & $0.4858(3)$  & $0.59$ \\
AII & C$^{\dagger}$ & -1  & &  & $|y|^2$ &  &  &  \\
AI$^{\dagger}$ &AI$^{\dagger}$ & & +1 & & & & & \\
AII$^{\dagger} $ & AII$^{\dagger}$ & & -1 & & & & & \\
AI + $\eta_+$& BDI$^{\dagger}$ & +1 & +1 & $\surd$ & ~~$-|y|\log (|y|)$~~ &$\surd$  & $0.4451(4)$ 
& $0.73$ \\
AI + $\eta_-$& DIII$^{\dagger}$ & +1 & -1 & $\surd$  &$|y|$ &$\surd$ & $0.4943(4)$ & $0.58$ \\
AII + $\eta_+$ &  CII$^{\dagger}$ & -1 & -1 &  $\surd$  &$|y|$ & $\surd$ & $0.3708(7)$ & $1.11$ \\
AII + $\eta_-$ & CI$^{\dagger}$ & -1 & +1 & $\surd$ & $|y|$ &  & & \\
\hline\hline
\end{tabular}
    \label{symmetry_class_A}
\end{table*}

The density of states (DoS) of non-Hermitian random 
matrices and physical Hamiltonians is generally defined 
in the complex plane, $\rho(E \equiv x+{\rm i}y)$. 
Based on analytical and numerical analyses, 
we find that in five symmetry classes out of the seven 
symmetry classes, the DoS in the complex plane has a delta 
function peak on the real axis. They are 
class A + $\eta$, class AI (equivalent to the real 
Ginibre ensemble~\cite{Efetov97,Lehmann1991,edelman1995,Kanzieper05,Forrester07}),
class AI + $\eta_+$, class AI + $\eta_-$, and class AII + $\eta_-$. 
In these symmetry classes, the DoS $\rho( E = x + {\rm i} y)$ is
decomposed into two parts, 
\begin{equation}
    \rho( E = x + {\rm i} y ) = \rho_c(x, y) + \rho_r(x) \delta(y),
\end{equation}
where %
$\rho_c(x,y)$ is the density of complex eigenvalues away from 
the real axis and $\rho_r(x)$ is the density of real eigenvalues.
Since only the states with real eigenvalues 
respect the full symmetries in these symmetry classes, 
$\rho_r(x)$ plays a role similar to the DoS in Hermitian systems. 
We show that the level statistics of real eigenvalues obtained from non-Hermitian random matrices, such as the level-spacing and level-spacing-ratio distributions, are different from those obtained from Hermitian random matrices and belong to the five distinctive universality classes according to the symmetries.
It is also notable that TRS or pseudo-Hermiticity does not 
necessarily lead to %
$\rho_{r}(x) \neq 0$ in the DoS. 
We find that no real eigenvalues appear generally in class AII, which is consistent with the absence of real eigenvalues in the Ginibre symplectic ensemble~\cite{Ginibre65}. We further generalize the absence 
of real eigenvalues to class AII + $\eta_{-}$.

We use random matrix analysis and exact diagonalization %
to identify universal level statistics of real eigenvalues for the five non-Hermitian symmetry classes. 
To demonstrate the universality of the level statistics, 
we apply the analysis to many-body and non-interacting 
physical Hamiltonians with disorder and non-Hermiticity. 
In physical systems that belong to the  
five symmetry classes, a finite density $\rho_r(x)$ of real eigenvalues 
enables comparison %
with those of non-Hermitian random matrices. We introduce non-Hermitian terms into 
interacting spin and hard-core boson models, such that 
many-body Hamiltonians belong to  
classes A + $\eta$, AI, and AI + $\eta_{\pm}$.
By the exact diagonalization, we calculate their 
many-body eigenenergies and their spacing distributions 
on the real axis. %
In these four symmetry classes, the level statistics in 
the ergodic phases %
follow
those of non-Hermitian 
random matrices in the corresponding symmetry classes. 
On the other hand, in class AII + $\eta_{+}$, 
the level statistics %
of a dissipative free fermionic system
deviate
from those of non-Hermitian random 
matrices in class AII + $\eta_{+}$. 
We attribute this 
discrepancy to the %
unconventional level
interaction 
between real eigenvalues, which is unique to non-Hermitian random matrices in class AII + $\eta_+$. 

The %
reality of the spectrum in non-Hermitian Hamiltonians %
was extensively studied~\cite{Bender98}.
References~\cite{edelman1995,Sommers98} showed that the number of real eigenvalues is proportional to the square root of the matrix size for non-Hermitian random matrices in class AI, 
and several previous works~\cite{hatano1996localization,Hamazaki20,Efetov97,Feinberg97,Longhi19,luo2021unifying} found that a nonzero proportion of real eigenvalues %
can appear
in %
non-Hermitian physical systems with TRS. 
However, how the number of real eigenvalues scales with the system size in physical systems,
and its relationship with random matrix theory are still uncovered. %
We find that
the average number $\bar{N}_{\text{real}}$ 
of real eigenvalues show distinctive scalings 
with respect to the dimensions $N$ of Hilbert space %
in the five symmetry classes. 
We clarify %
$\bar{N}_{\text{real}} \propto \sqrt{N}$ in the ergodic 
(metallic) phase and $\bar{N}_{\text{real}} \propto N$ 
in the localized phases. 
Our results show that 
the level statistics analyses %
are powerful tools for detecting quantum chaos and MBL in 
non-Hermitian
systems.

This paper is organized as follows. 
In Sec.\ref{sec_rm}, we begin with %
reviewing  
the symmetry classification of non-Hermitian matrices
and introduce level-spacing and level-spacing-ratio distributions
of real eigenvalues for non-Hermitian random matrices. We 
numerically obtain the real-eigenvalue spacing and 
spacing-ratio distributions from large non-Hermitian random matrices.
Analyzing small random matrices, we clarify the nature of 
effective interactions between two neighboring eigenvalues 
on the real axis and use them to explain the behavior
of large random matrices. We also show the scaling of the 
number of real eigenvalues with 
respect to the dimensions of random matrices. 
In Sec.\ref{sec_mbl}, we use a hard-core boson model and interacting spin models %
to demonstrate the universality of the real-eigenvalue spacing and spacing-ratio distribution functions. 
We argue that the level statistics of real eigenvalues are
useful for detecting different many-body phases in 
interacting disordered systems. We uncover that in the MBL
phase, the number of real eigenvalues shows a non-universal scaling 
with respect to the dimensions of Hilbert space. We provide
an explanation for the non-universal scaling. In Sec.\ref{sec_free}, 
we apply the analysis to non-Hermitian non-interacting fermionic models in two and three dimensions. 
We find that the number of real eigenvalues shows
distinctive universal scaling properties with respect to the matrix dimensions
in the metal and localized phases. Section~\ref{conclusion} is devoted to 
the conclusion and discussion.  

\section{Random matrices}
\label{sec_rm}

\subsection{Non-Hermitian symmetry classes}

The 38-fold symmetry %
class
of non-Hermitian Hamiltonians %
is 
given by 
the
following 
anti-unitary
symmetries~\cite{Kawabata19}, %
\begin{equation}
\begin{aligned}
& \text{time-reversal symmetry (TRS)}:\, \\
	& \mathcal{U_{T_+}} H^* \mathcal{U}^{\dagger}_{\mathcal{T}_+} = H,\quad 
   \mathcal{U_{T_+}} \mathcal{U}^{*}_{\mathcal{T}_+} = \pm1,\\   
   & \text{particle-hole symmetry (PHS)}:\, \\
	& \mathcal{U}_{\mathcal{P}_-}  H^T \mathcal{U}^{\dagger}_{\mathcal{P}_-} = -H,\quad
	\mathcal{U}_{\mathcal{P}_-} \mathcal{U}_{\mathcal{P}_-}^{*} = \pm1,\\
	& \text{time-reversal symmetry$^{\dag}$ (TRS$^{\dag}$)}:\, \\
	& \mathcal{U_{P_+}} H^T \mathcal{U}^{\dagger}_{\mathcal{P}_+} = H,\quad
	\mathcal{U_{P_+}} \mathcal{U}^{*}_{\mathcal{P}_+} = \pm1,\\
	& \text{particle-hole symmetry$^{\dag}$ (PHS$^{\dag}$)}:\,  \\
	& \mathcal{U}_{\mathcal{T}_-}  H^* \mathcal{U}^{\dagger}_{\mathcal{T}_-} = -H,\quad
	\mathcal{U}_{\mathcal{T}_-} \mathcal{U}^{*}_{\mathcal{T}_-} = \pm1,\\
\end{aligned}    
\end{equation}
and unitary symmetries,
\begin{equation}
\begin{aligned}
   & \text{pseudo-Hermiticity (pH)}:\,  \\
	&\mathcal{U}_{\eta} H^{\dagger} \mathcal{U}_{\eta}^{\dagger} = H,\quad
	\mathcal{U}_{\eta}^2 = 1,\\
	& \text{chiral symmetry (CS)}:\,  \\
	&\mathcal{U_C} H^{\dagger} \mathcal{U}_{\cal C}^{\dagger} = -H,\quad    {\cal U_C}^2 = 1,\\
	& \text{sublattice symmetry (SLS)}:\, \\
	& \mathcal{U_S} H \mathcal{U}_{\cal S}^{\dagger} = -H,\quad
    \mathcal{U_S}^2 = 1,
\end{aligned}    
\end{equation}
where $\mathcal{U_{T_{\pm}}}$, $\mathcal{U}_{\mathcal{P}_{\pm}}$, $\mathcal{U}_{\eta}$, $\mathcal{U_C}$, and $\mathcal{U_S}$ are unitary matrices. When $H$ respects TRS (pH), ${\rm i} H$ respects PHS$^{\dagger}$ (CS), and vice versa. 
In this sense, TRS and PHS$^{\dagger}$ are unified, so are pH and CS~\cite{Kawabata19NC}. 
TRS relates an eigenvalue $z$ with its complex conjugate $z^*$. 
If ${\bm v}$ is a right eigenvector of a %
Hamiltonian $H$ 
with TRS for 
an eigenvalue $z$ ($H {\bm v} = z {\bm v}$), ${\cal U}_{\cal T}^{T}\bm v^*$ is %
another
right eigenvector of $H$ with %
the eigenvalue
$z^*$ ($H \!\ {\cal U}_{\cal T}^{T} {\bm v^*} = z^* \!\ {\cal U}_{\cal T}^{T} {\bm v^*}$). 
Likewise, pseudo-Hermiticity (pH) relates an eigenvalue $z$ with its complex conjugate $z^*$, %
PHS$^{\dagger}$ and CS relate an eigenvalue $z$ with $-z^*$, and PHS and SLS relate an eigenvalue $z$ with $-z$. 
On the other hand, TRS$^{\dagger}$ imposes a constraint on each eigenvector.

When a symmetry relates an eigenvalue $z$ with $z^{\prime} \neq z$ and $\lvert z - z^{\prime}\rvert$ is much larger than 
the
mean level-spacing, such symmetry is expected to have no influence on the local eigenvalue correlation around $z$. 
For example, neither TRS nor PHS changes the nearest-spacing distribution of non-Hermitian random matrices %
for general complex eigenvalues~\cite{Hamazaki20}. 
This is similar to Hermitian random matrices with PHS or CS;
for example,
the eigenvalue spacing distribution away from %
zero energy %
in %
class D is the same as that %
in %
class A~\cite{Altland97}. 

The spectral correlation on or around the real axis 
depends on TRS, pH, and their combination (TRS$^{\dagger}$). 
From TRS, pH, and TRS$^{\dagger}$, a ten-fold symmetry 
classification is derived, as shown in Table~\ref{symmetry_class_A}.  
This ten-fold class %
includes seven
symmetry classes that have at least one symmetry associated 
with complex conjugation (TRS) or 
Hermitian conjugation (pH):
a class with pH (class A + $\eta$), classes with TRS whose sign can be 
$\pm 1$ (classes AI and AII), and classes with both %
pH
and TRS, where the %
sign
of TRS is $\pm 1$ and TRS commutes
 with %
 pH
(classes AI + $\eta_{+}$ and AII + $\eta_{+}$)
or
TRS anti-commutes with %
pH
(classes AI + $\eta_{-}$ and AII + $\eta_{-}$). 
According to the 38-fold symmetry classification of non-Hermitian systems~\cite{Kawabata19}, %
these symmetry classes
are equivalent to 
classes AIII, D$^{\dagger}$, C$^{\dagger}$, BDI$^{\dagger}$, DIII$^{\dagger}$, 
CII$^{\dagger}$, and CI$^{\dagger}$. %
(see Table~\ref{symmetry_class_A}).
The ten-fold symmetry class in Table~\ref{symmetry_class_A} is also equivalent to the Hermitian conjugate of the non-Hermitian Altland-Zirnbauer class (i.e., $\mathrm{AZ}^{\dag}$ class) in Ref.~\cite{Kawabata19}.

\subsection{%
Level statistics of real eigenvalues for 
non-Hermitian
random matrices
}
\label{DoS_LS}
\begin{figure*}[bt]
	\centering
	\subfigure[class AI]{
		\begin{minipage}[t]{0.32\linewidth}
			\centering
			\includegraphics[width=1\linewidth]{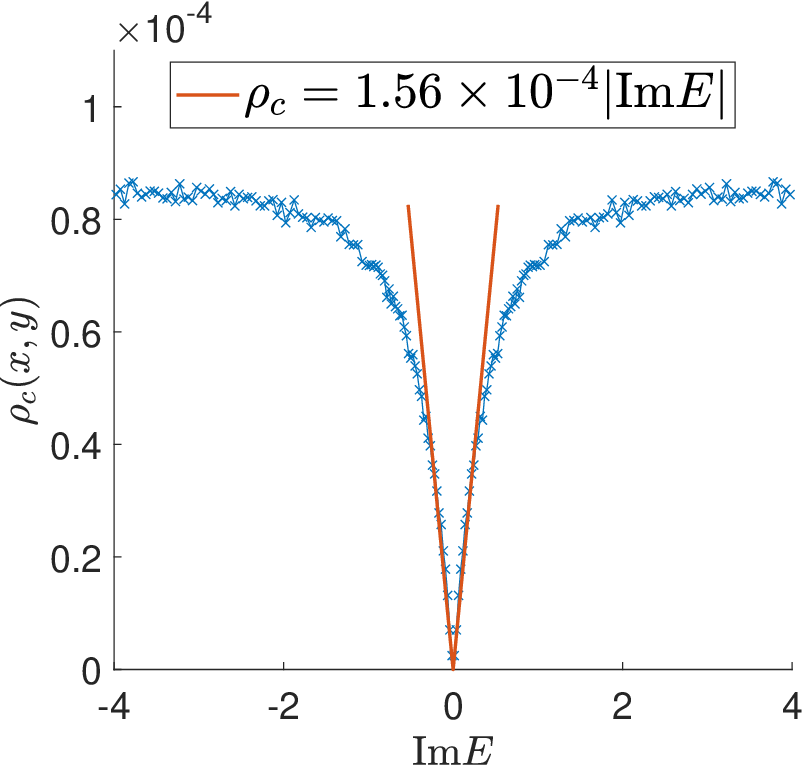}
		\end{minipage}%
	}%
	\subfigure[class AII]{
		\begin{minipage}[t]{0.32\linewidth}
			\centering
			\includegraphics[width=1\linewidth]{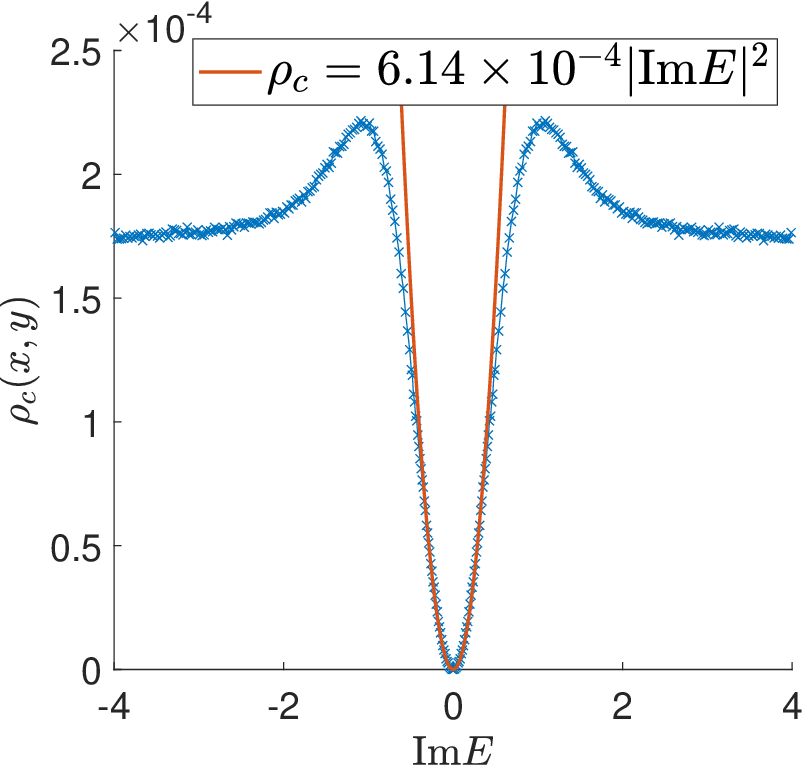}
		\end{minipage}
	}
	\subfigure[class AI + $\eta_+$]{
		\begin{minipage}[t]{0.32\linewidth}
			\centering
			\includegraphics[width=1\linewidth]{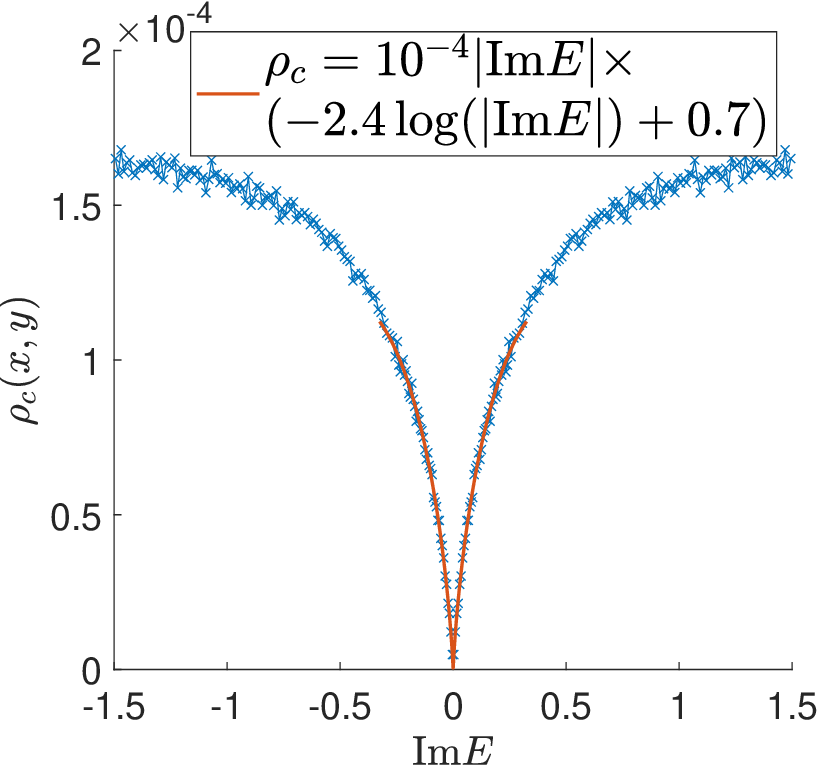}
		\end{minipage}%
	}%
	
	\subfigure[class A + $\eta$]{
		\begin{minipage}[t]{0.25\linewidth}
			\centering
			\includegraphics[width=1\linewidth]{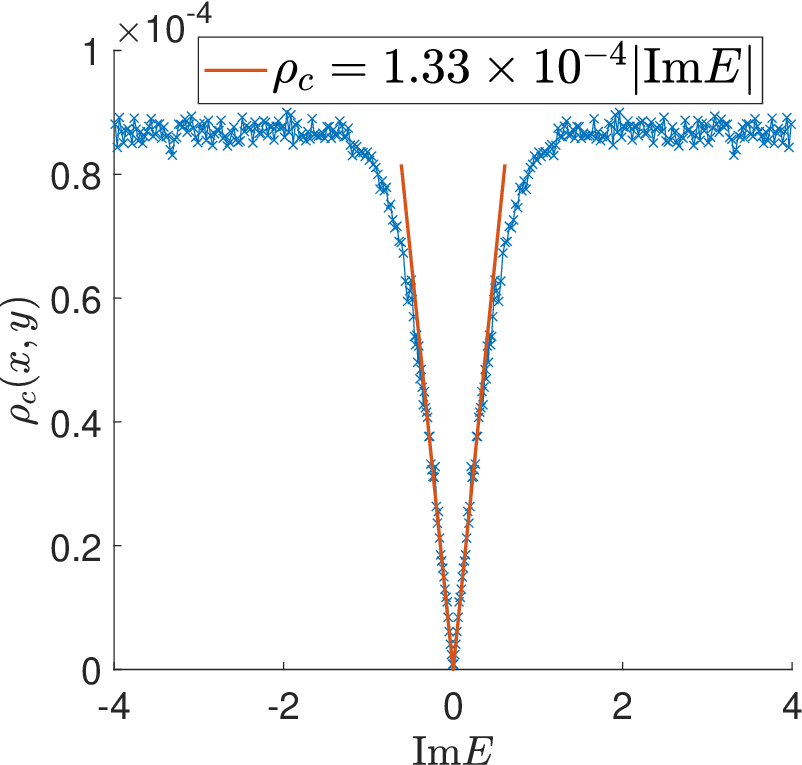}
		\end{minipage}%
	}%
	\subfigure[class AI + $\eta_-$]{
		\begin{minipage}[t]{0.25\linewidth}
			\centering
			\includegraphics[width=1\linewidth]{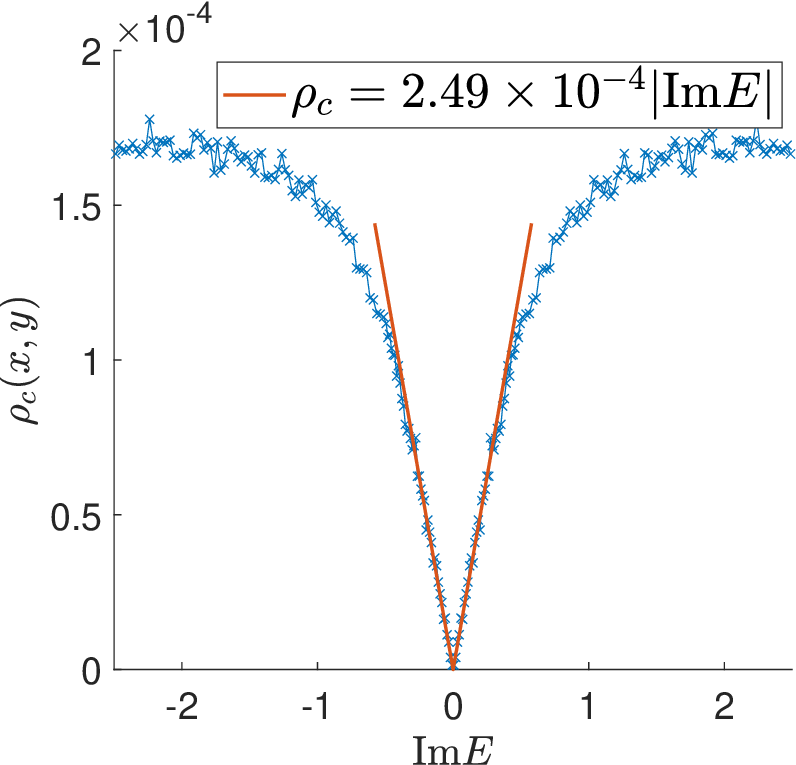}
		\end{minipage}%
	}%
	\subfigure[class AII + $\eta_+$]{
		\begin{minipage}[t]{0.25\linewidth}
			\centering
			\includegraphics[width=1\linewidth]{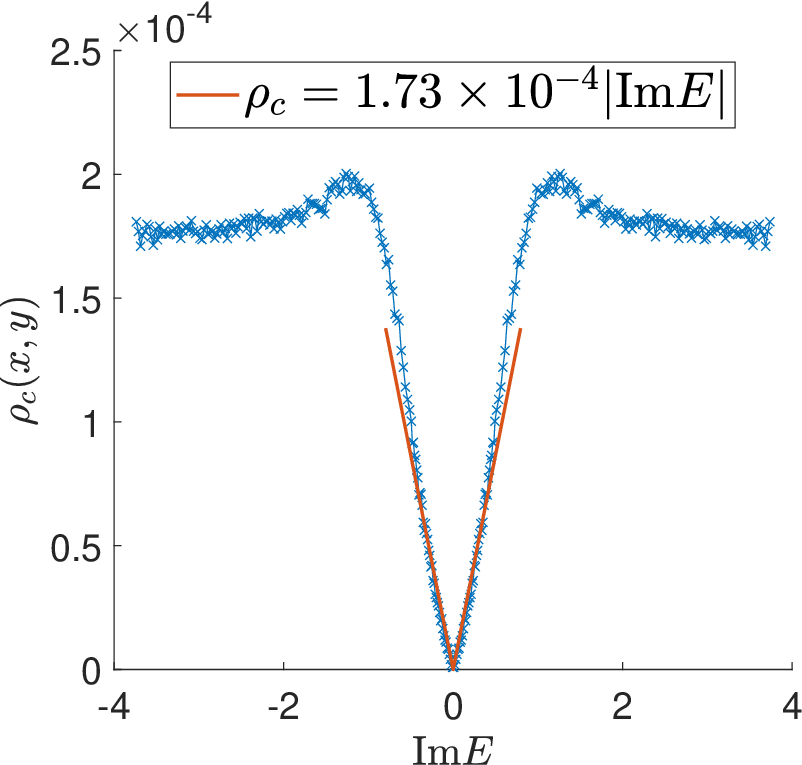}
		\end{minipage}%
	}%
	\subfigure[class AII + $\eta_-$]{
		\begin{minipage}[t]{0.25\linewidth}
			\centering
			\includegraphics[width=1\linewidth]{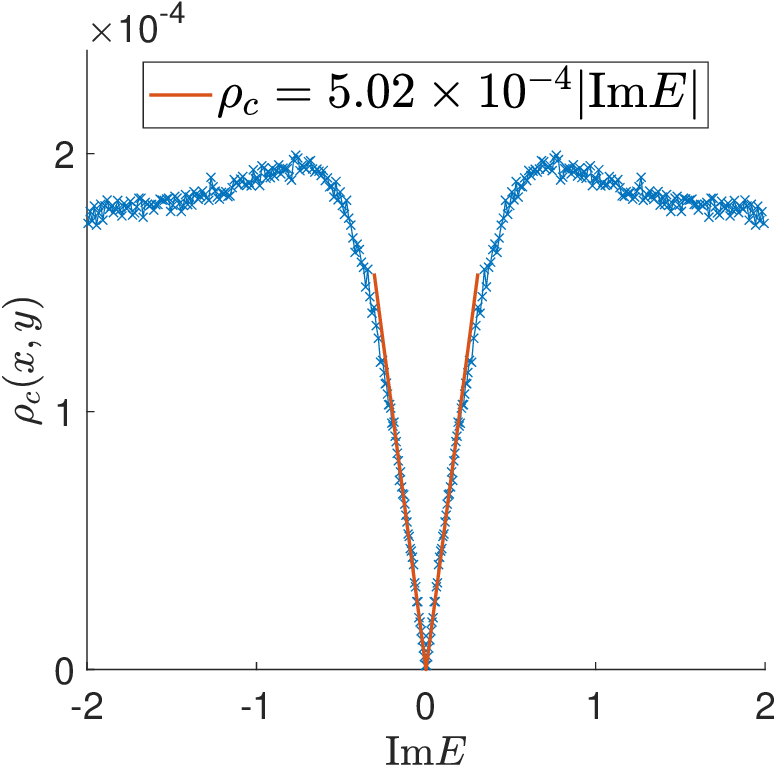}
		\end{minipage}%
	}%

	\caption{Density $\rho_c(x,y)$ of complex eigenvalues
	of non-Hermitian random matrices in the Gaussian ensemble for classes (a) AI, (b) AII, (c) AI + $\eta_+$, (d) A + $\eta$, (e) AI + $\eta_-$, (f) AII + $\eta_+$, and (g) AII + $\eta_-$. 
	Here, $\rho_c(x,y)$ is shown as a  function of $y = \mathrm{Im} \left( E \right)$ for fixed $x = \mathrm{Re} \left( E \right)$ near the real axis of complex energy $E$ (i.e., $y %
	\simeq
	0$). 
	For classes AI, AI + $\eta_{\pm}$, A+ $\eta$, and AII + $\eta_{+}$, the 
	density of states $\rho(E=x+{\rm i}y)\equiv \rho_c(x,y)+\delta(y)\rho_r(x)$ is separated into the 
	density $\rho_c(x,y)$ of complex eigenvalues and the density $\rho_r(x)$ of real eigenvalues. 
	For classes AII 
	and AII +$\eta_{-}$, no real eigenvalues appear, and we have $\rho(x+{\rm i}y) \equiv \rho_{c}(x,y)$. 
	The data of $\rho(x+{\rm i}y)$ %
    are
	obtained from diagonalizations of 5000 samples of $4000 \times 4000$ random matrices in each symmetry class. Note that $\rho_c(x,y)$ is almost independent of $x = \mathrm{Re} \left( E \right)$ when $E$ is away from 
	the boundary of a circle inside which the complex eigenvalues $E$ distribute.
	}
\label{2d_DoS}
\end{figure*}

\begin{figure}[htbp]
		\centering
		\includegraphics[width=1\linewidth]{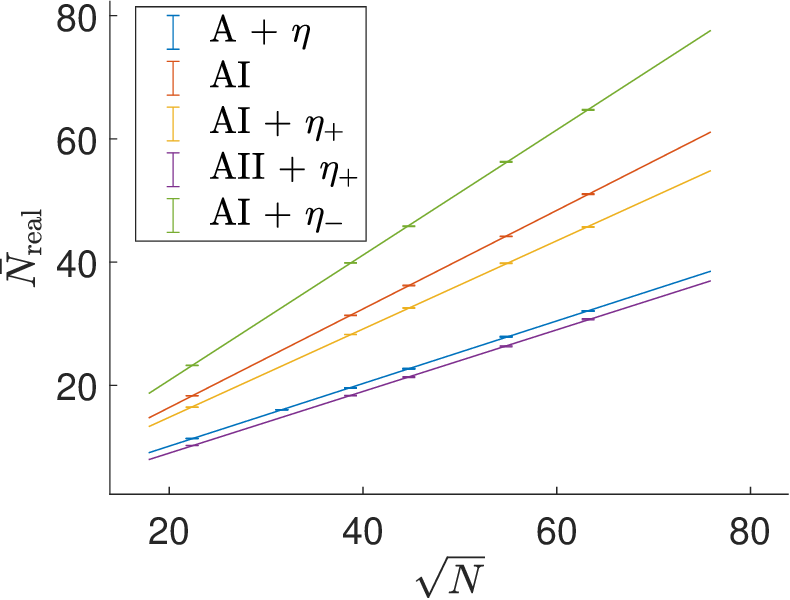}
		\caption{Average number $\bar{N}_{\rm real}$ of real eigenvalues of non-Hermitian random matrices in the Gaussian ensemble %
		as a function of $\sqrt{N}$ for the five symmetry classes. Here, $N$ is %
	    the
		dimensions of random matrices. 
		The error bars in the plot stand %
		for the twice of %
		the
		standard deviation of ${\bar N}_\text{real}$. The standard deviation $\sigma_{N_{\rm real}}$ is estimated by  $N_{\rm sample}\sigma_{N_{\text{real}}}^2 \equiv  \sum_{i=1}^{N_{\text{sample}}} (N_{\text{real}}^{(i)} - {\bar N}_\text{real} )^2 /(N_{\text{sample}}-1)$, where $N_{\text{real}}^{(i)}$ is the number of real eigenvalues in the $i$-th random matrix and $N_{\text{sample}}$ is the number of the random matrices in the ensemble. $N_{\text{sample}}$ is at least $5000$ for each matrix size. The plot clearly demonstrates the square-root scaling ${\bar N}_{\text{real}} \sim \sqrt{N}$ in all the five symmetry classes. %
		}
		\label{N_real}
\end{figure}

\begin{figure*}[bt]
	\centering
	\begin{minipage}[t]{0.199\linewidth}
		\subfigure[class A + $\eta$]{
				\centering
				\includegraphics[width=1\linewidth]{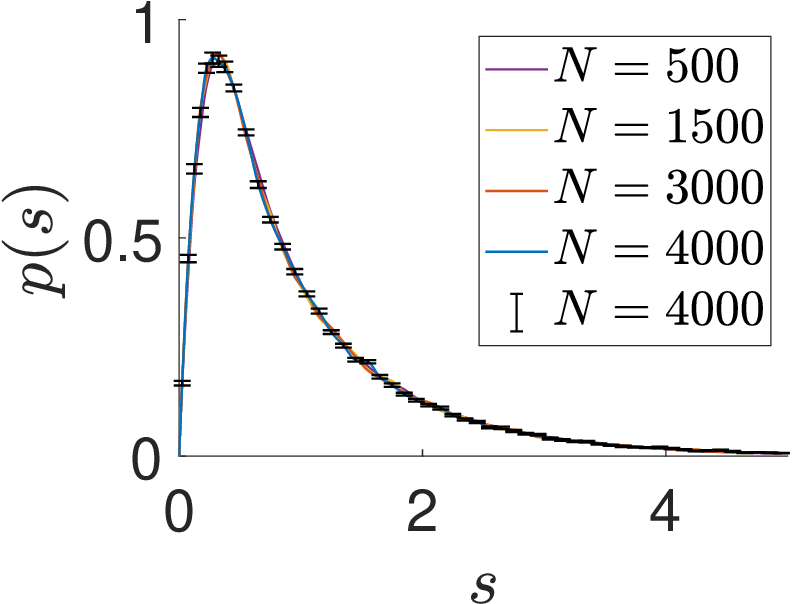}
		}
	\end{minipage}%
	\begin{minipage}[t]{0.199\linewidth}
	\subfigure[class AI]{
			\centering
			\includegraphics[width=1\linewidth]{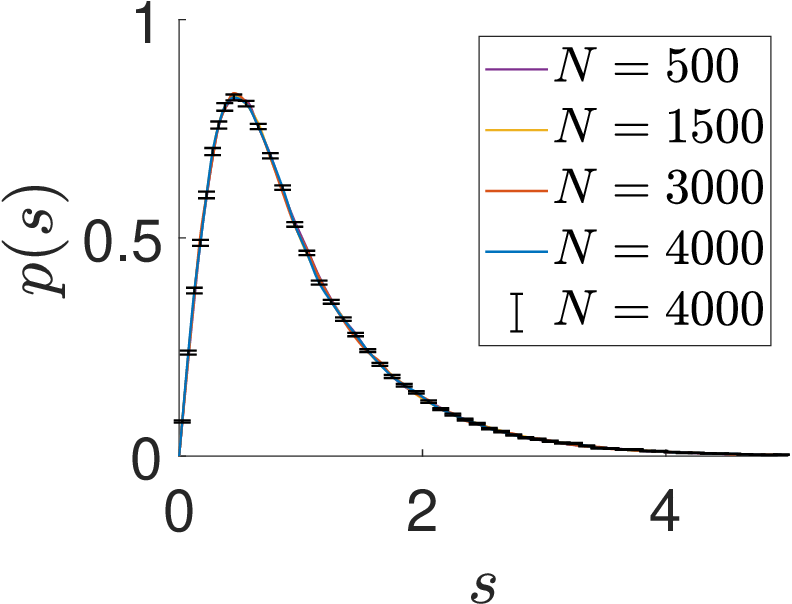}
	}
	\end{minipage}%
    \begin{minipage}[t]{0.199\linewidth}
	\subfigure[class AI + $\eta_+$]{
			\centering
			\includegraphics[width=1\linewidth]{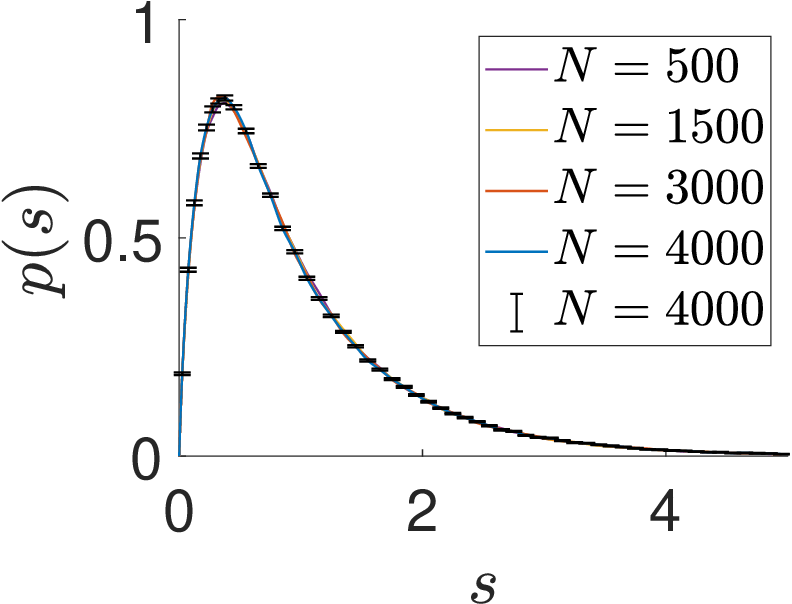}
			}
	\end{minipage}%
    \begin{minipage}[t]{0.199\linewidth}
	\subfigure[class AII + $\eta_+$]{
			\centering
			\includegraphics[width=1\linewidth]{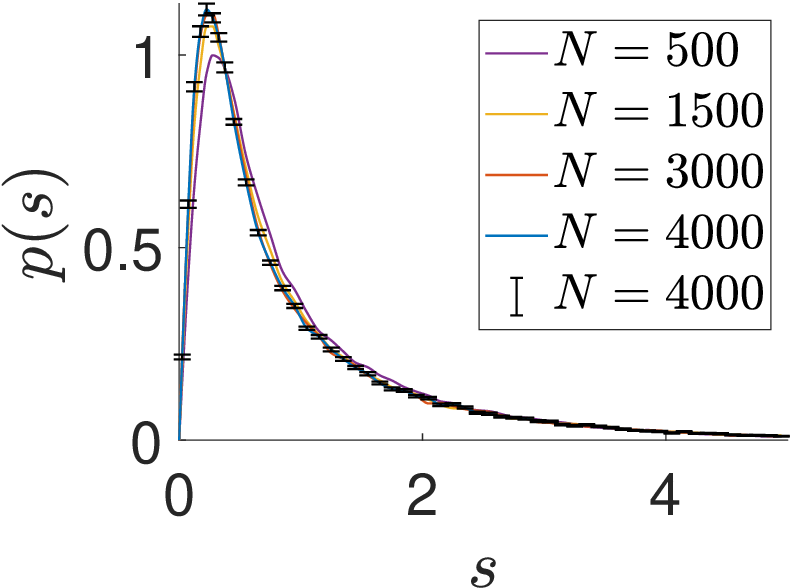}
			\label{ps_CII_GE}
			}
	\end{minipage}%
    \begin{minipage}[t]{0.199\linewidth}
	\subfigure[class AI + $\eta_-$]{
			\centering
			\includegraphics[width=1\linewidth]{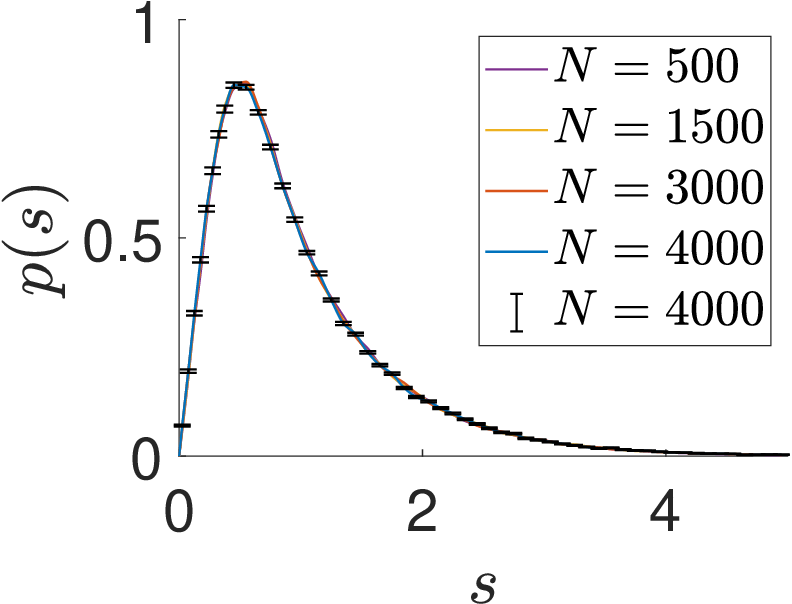}
			}
	\end{minipage}%

	\centering
	\begin{minipage}[t]{0.199\linewidth}
		\subfigure[class A + $\eta$]{
				\centering
				\includegraphics[width=1\linewidth]{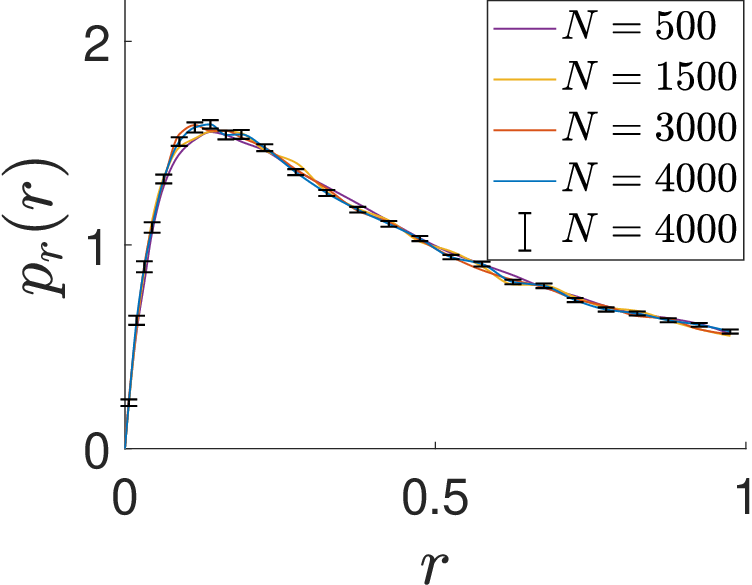}
		}
	\end{minipage}%
	\begin{minipage}[t]{0.199\linewidth}
	\subfigure[class AI]{
			\centering
			\includegraphics[width=1\linewidth]{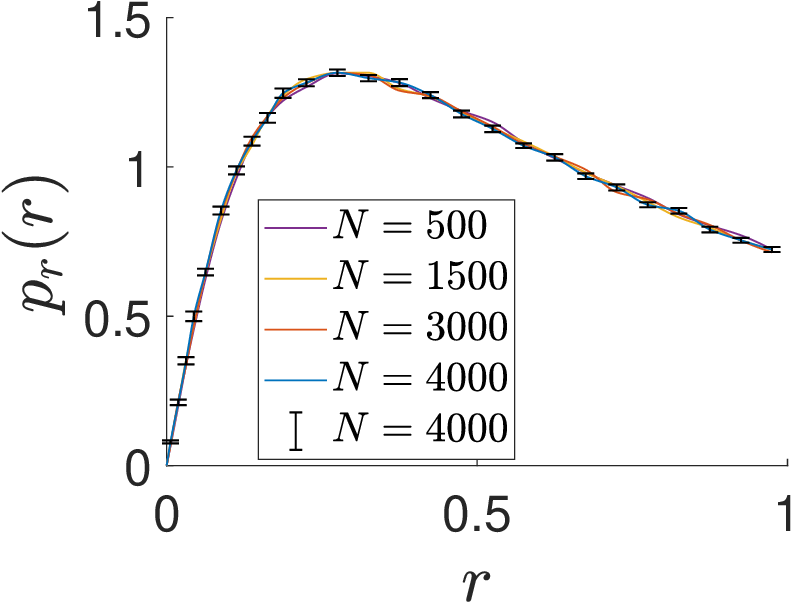}
	}
	\end{minipage}%
    \begin{minipage}[t]{0.199\linewidth}
	\subfigure[class AI + $\eta_+$]{
			\centering
			\includegraphics[width=1\linewidth]{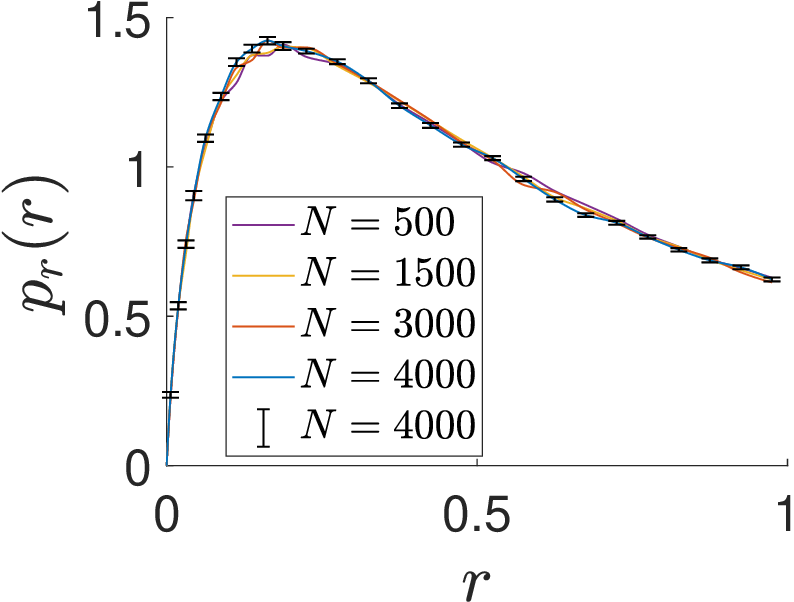}
			}
	\end{minipage}%
    \begin{minipage}[t]{0.199\linewidth}
	\subfigure[class AII + $\eta_+$]{
			\centering
			\includegraphics[width=1\linewidth]{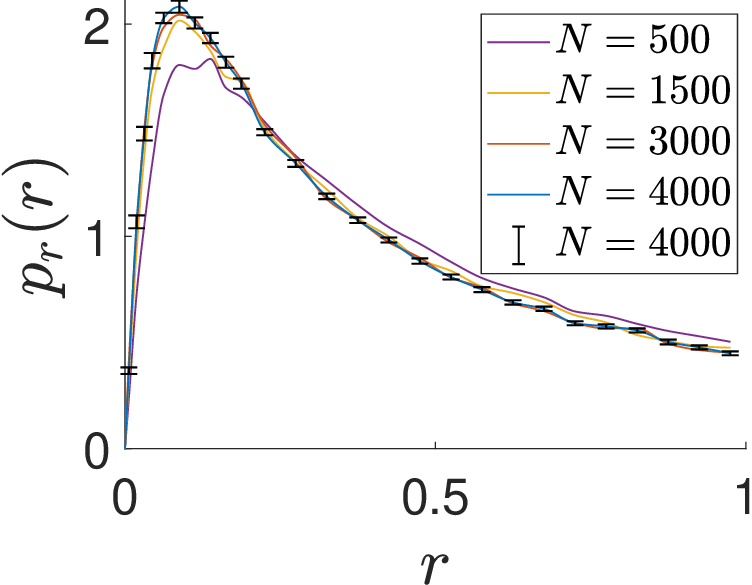}
			\label{pr_CII_GE}
			}
	\end{minipage}%
    \begin{minipage}[t]{0.199\linewidth}
	\subfigure[class AI + $\eta_-$]{
			\centering
			\includegraphics[width=1\linewidth]{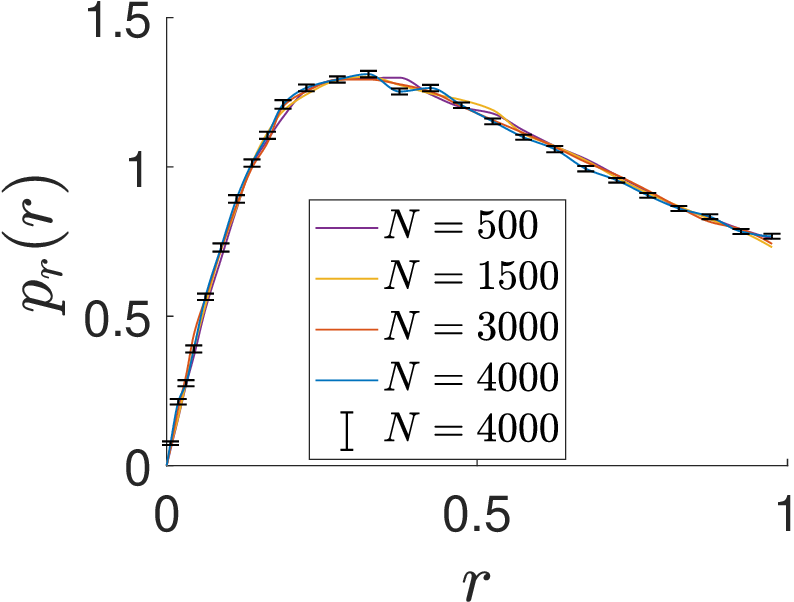}
			}
	\end{minipage}%
	\caption{Level-spacing distributions $p(s)$ of real eigenvalues of $N \times N$ non-Hermitian random matrices in the Gaussian ensemble for (a)~class A + $\eta$, (b)~class AI, (c)~class AI + $\eta_+$, (d)~class AII + $\eta_+$, and (e)~class AI + $\eta_{-}$. 
	Level-spacing-ratio distributions $p_r(r)$ of real eigenvalues of $N \times N$ non-Hermitian random matrices in the Gaussian ensemble for (f)~class A + $\eta$, (g)~class AI, (h)~class AI + $\eta_+$, (i)~class AII + $\eta_+$, and (j)~class AI + $\eta_{-}$. 
	For each $N$ and for each symmetry class, $p(s)$ and $p_r(r)$ are averaged over at least 5000 random matrices in the ensemble. 
	The black points for $p(s)$ and $p_r(r)$ are obtained from $4000\times 4000$ random matrices, where the standard deviation error bars are evaluated by the bootstrap method~\cite{press07}. 
	The error bars for the smaller matrices are smaller than the error bars for $N=4000$ and not shown.}
	\label{ps_and_pr_GE}
\end{figure*}

 \begin{figure*}[bt]
	\centering
	\begin{minipage}[t]{0.68\linewidth}
	\subfigure[comparison among five symmetry classes]{
			\centering
			\includegraphics[width=1\linewidth]{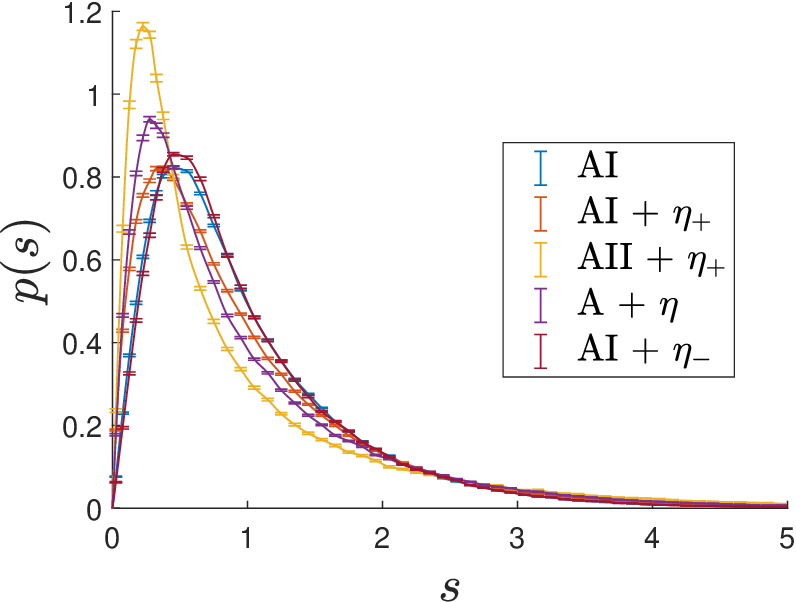}
			}
	\end{minipage}%
	\begin{minipage}[t]{0.3\linewidth}
		\subfigure[class A + $\eta$]{
			\centering
			\includegraphics[width=1\linewidth]{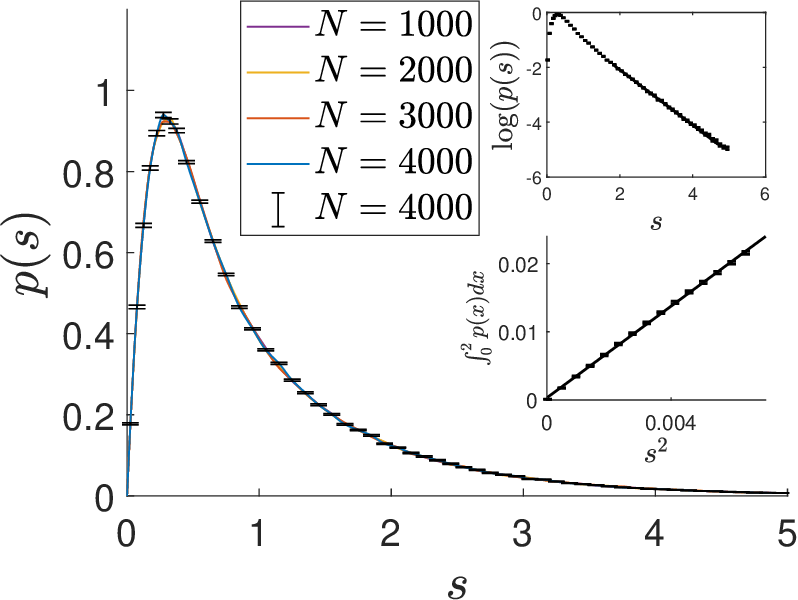}}
		\subfigure[class AI]{
			\centering
			\includegraphics[width=1\linewidth]{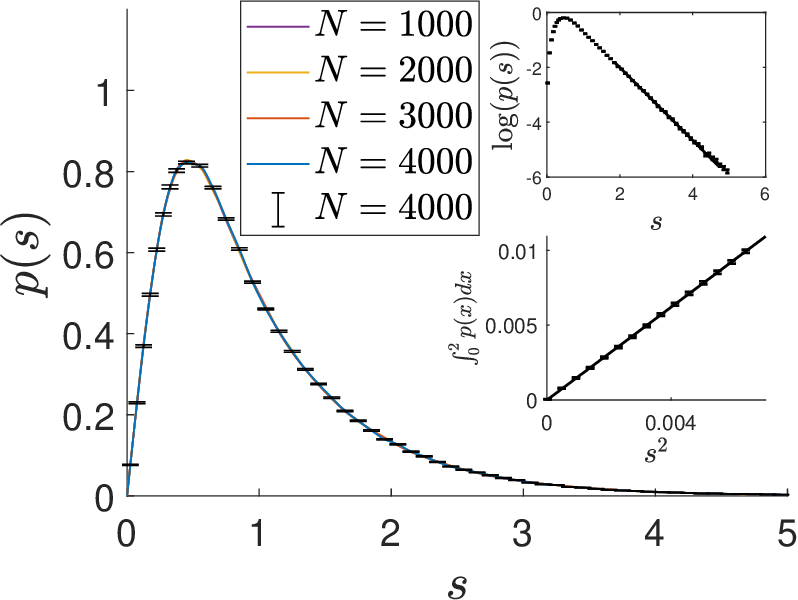}
	}%
	\end{minipage}%
    
    \begin{minipage}[t]{0.33\linewidth}
	\subfigure[class AI + $\eta_+$]{
			\centering
			\includegraphics[width=1\linewidth]{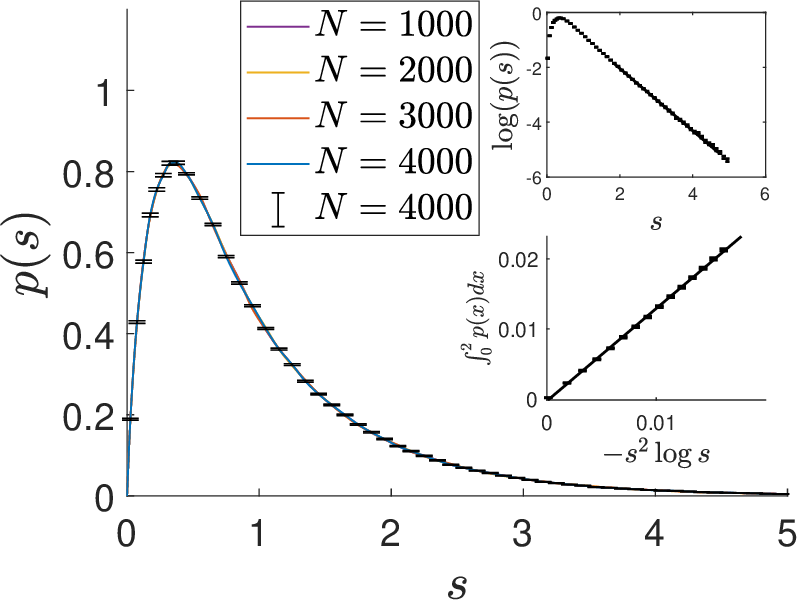}
			}
	\end{minipage}%
    \begin{minipage}[t]{0.33\linewidth}
	\subfigure[class AII + $\eta_+$]{
			\centering
			\includegraphics[width=1\linewidth]{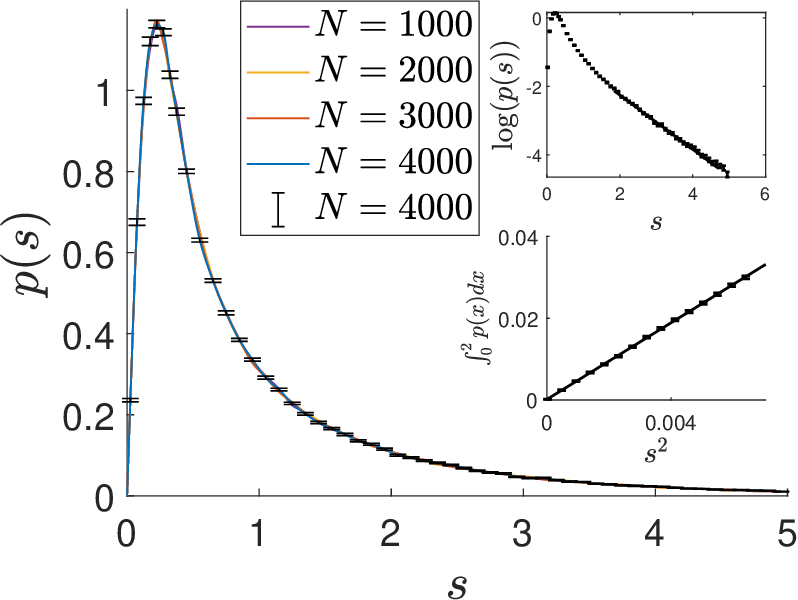}
			}
	\end{minipage}%
    \begin{minipage}[t]{0.33\linewidth}
	\subfigure[class AI + $\eta_-$]{
			\centering
			\includegraphics[width=1\linewidth]{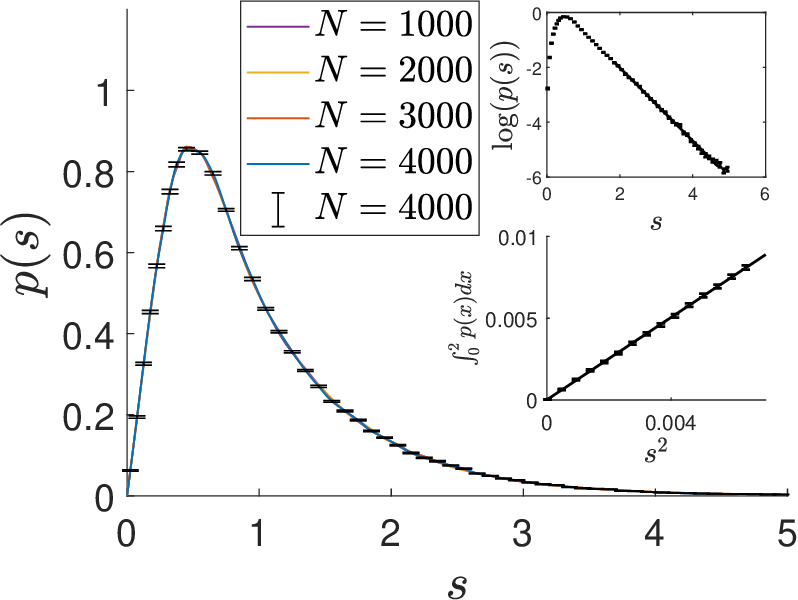}
			}
	\end{minipage}%
	\label{ps_pr_GE}
	\caption{Level-spacing distributions $p(s)$ of real eigenvalues of $N \times N$ non-Hermitian random matrices in the generalized Gaussian ensemble for (a) all the five symmetry classes, (b) class A + $\eta$, (c) class AI, (d) class AI + $\eta_+$, (e) class AII + $\eta_+$, and 
	(f) class AI + $\eta_{-}$. %
	For each $N$ and for each symmetry class, the spacing distribution function is averaged over at least $5000$ random matrices in the ensemble. 
	The black points with the error bars are $p(s)$ obtained from $4000\times 4000$ random matrices, where the standard deviation error bars are evaluated by the bootstrap method~\cite{press07}. 
	The error bars for the smaller matrices are smaller than those for $N=4000$ and not shown.
	In each symmetry class, $p(s)$ of random matrices with different sizes $N$ ($N>1000$) almost overlap with each other.  
	Insets: Asymptotic behaviors of the distribution function for $s \gg 1$ and for $s \ll 1$. For small $s$, the cumulative distribution function $\int_0^s p(s^{\prime}) ds^{\prime}$ is plotted as a function of either $s^2$ or $-\log(s) s^2 $.
    For $s \gg 1$, $\log(p(s))$ is linear in $s$, indicating
 the Poisson-like tail.
    The comparison of $p(s)$ among the five symmetry classes %
    shows that $p(s)$ for classes AI and
    AI + $\eta_-$ are close to each other, and that $p(s)$ for the other three classes %
    are clearly
	distinguished from $p(s)$ for classes AI and AI + $\eta_{-}$.}
	\label{ps_all}
\end{figure*}

\begin{figure*}[bt]
	\centering
	\begin{minipage}[t]{0.68\linewidth}
	\subfigure[comparison among five symmetry classes]{
			\centering
			\includegraphics[width=1\linewidth]{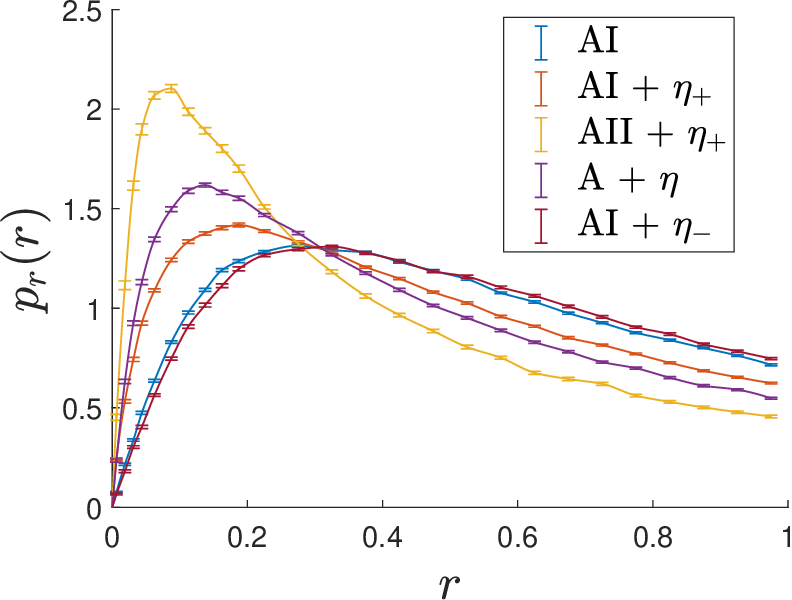}
			}
	\end{minipage}%
	\begin{minipage}[t]{0.3\linewidth}
		\subfigure[class A + $\eta$]{
			\centering
			\includegraphics[width=1\linewidth]{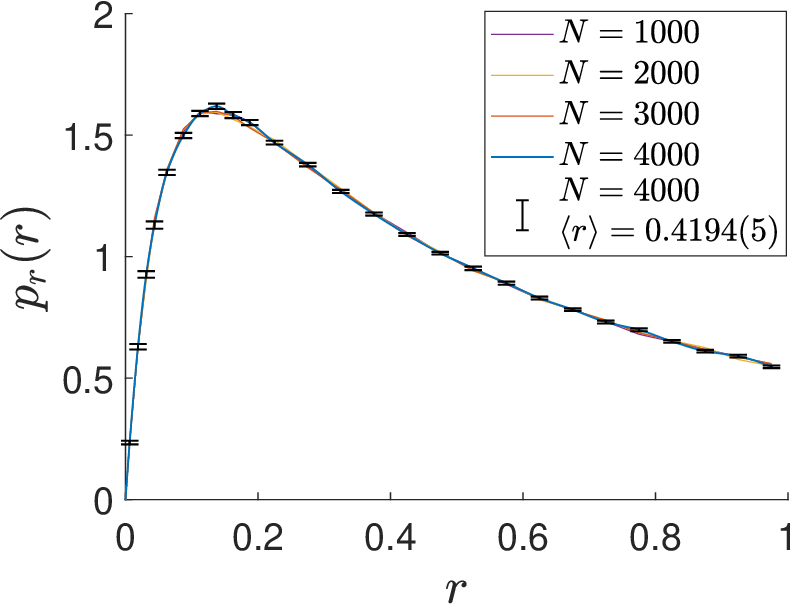}}
		\subfigure[class AI]{
			\centering
			\includegraphics[width=1\linewidth]{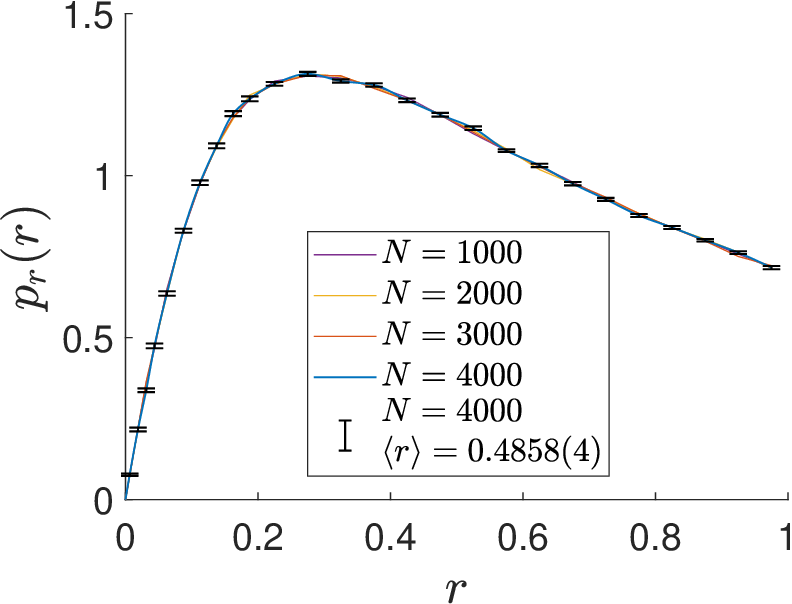}
	}%
	\end{minipage}%
    
    \begin{minipage}[t]{0.33\linewidth}
	\subfigure[class AI + $\eta_+$]{
			\centering
			\includegraphics[width=1\linewidth]{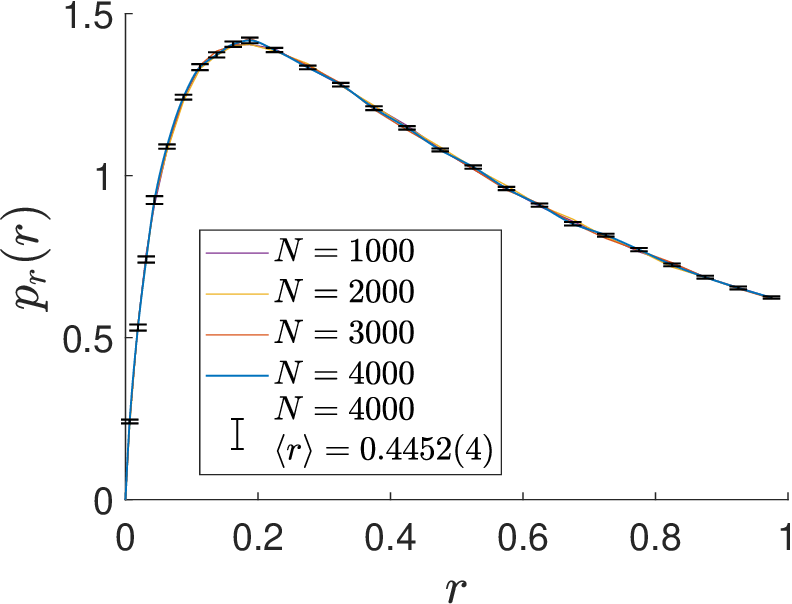}
			}
	\end{minipage}%
    \begin{minipage}[t]{0.33\linewidth}
	\subfigure[class AII + $\eta_+$]{
			\centering
			\includegraphics[width=1\linewidth]{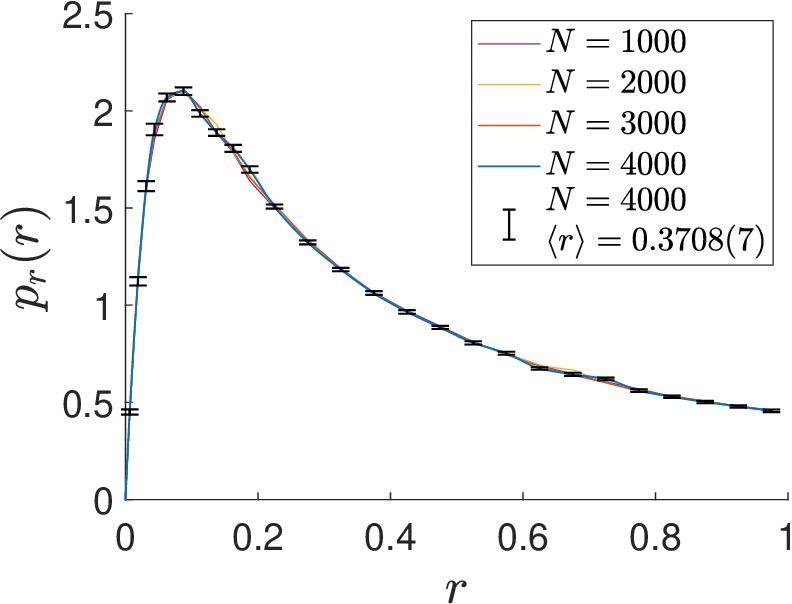}
			\label{pr_ciid}
			}
	\end{minipage}%
    \begin{minipage}[t]{0.33\linewidth}
	\subfigure[class AI + $\eta_-$]{
			\centering
			\includegraphics[width=1\linewidth]{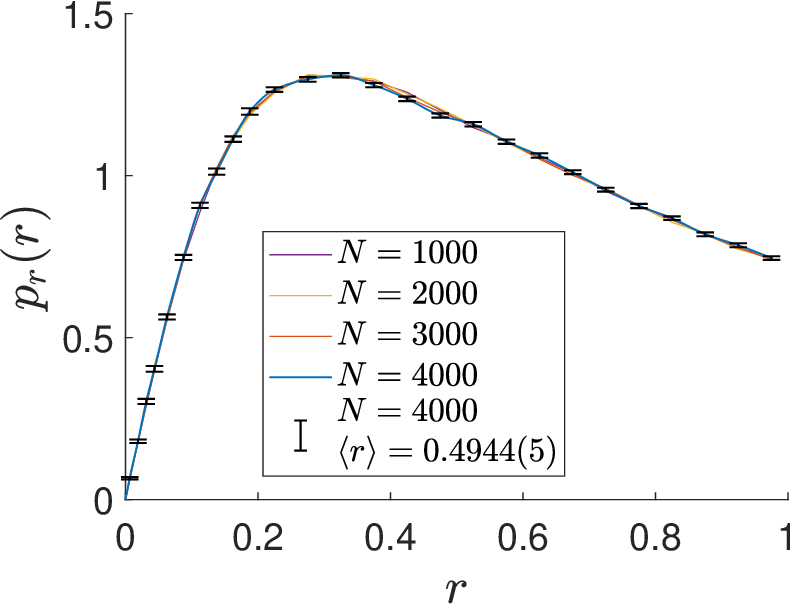}
			}
	\end{minipage}%
	
	\caption{Level-spacing-ratio distributions $p_r(r)$ of real eigenvalues of $N \times N$ 
    non-Hermitian
	random matrices in the generalized Gaussian ensemble 
	for
	(a)~all the five symmetry classes, 
	(b)~class A + $\eta$, 
	(c)~class AI, 
	(d)~class AI + $\eta_+$, 
	(e)~class AII + $\eta_+$, and 
	(f)~class AI + $\eta_{-}$. 
	For each $N$ and for each symmetry class, the level-spacing-ratio distribution function is averaged over at least $5000$ random matrices in the ensemble. 
	The black points with the error bars are $p_r(r)$ obtained from $4000\times 4000$ random matrices, where the standard deviation error bars are evaluated by the bootstrap method~\cite{press07}. 
	The error bars for the smaller matrices are smaller than the error bars for $N=4000$ and not shown.
	In each symmetry class, $p_r(r)$ of random matrices with different sizes $N$ ($N \geq 1000$) almost overlaps with each other.
	The mean value $\langle r \rangle = \int_0^1 p_r(r)dr$ of the spacing ratios and its standard deviation
	are given for $N =4000$ for each symmetry class. 
	The comparison of $p_r(r)$ among the five symmetry classes %
    shows that $p_r(r)$ %
    distinguishes between different symmetry classes. %
	}
	\label{pr_all}
\end{figure*}

We consider
non-Hermitian random
matrices ${\cal H}$ in symmetry 
classes A + $\eta$, AI, AII, AI + $\eta_{\pm}$, and AII + $\eta_{\pm}$ 
in 
the
Gaussian ensemble with the following
probability distribution function $p({\cal H})$: %
\begin{equation}
p({\cal H}) =  C_N^{-1} e^{- \beta \text{Tr}( {\cal H}^{\dagger} {\cal H} )} \, ,
\label{GE}
\end{equation}
where $\beta$ is a positive constant and $C_N$ is a normalized constant.
Without loss of generality, we choose $\beta=1/2$ for the rest of this paper.
Non-Hermitian matrices ${\cal H}$ are required to belong to symmetry classes in Table~\ref{symmetry_class_A}
(see %
Appendix~\ref{sec_be}
for details).
Diagonalizations of large random matrices show that eigenvalues distribute 
almost uniformly in a circle except around the real axis and its 
circumference (not shown here). This distribution is consistent with 
the circular law of the Ginibre ensemble~\cite{Ginibre65}. 
For non-Hermitian random matrices in classes 
A + $\eta$, AI, AI + $\eta_{\pm}$, and AII + $\eta_+$, 
a sub-extensive number of eigenvalues are real, 
and the DoS $\rho(x,y)$ has a delta function peak on the real axis:
\begin{equation}
\rho(E=x+{\rm i}y)=\rho_{c}(x,y)+\rho_r(x)\delta(y).
\end{equation}
In numerical diagonalizations, real eigenvalues and complex eigenvalues are clearly distinguished, although real eigenvalues can artificially have tiny imaginary parts due to machine inaccuracy of a numerical subroutine program.
In fact, 
with proper normalization, the 
apparent
imaginary parts of 
real eigenvalues of Hermitian matrices are less than a certain 
error bound~\footnote{Tiny apparent imaginary parts of %
real eigenvalues stem from the machine 
inaccuracy of a numerical subroutine program. We can estimate 
their upper bound (error bound) by numerical diagonalization of Hermitian 
random matrices based on the same subroutine program. 
By the diagonalization, we observe that when 
eigenvalues of an $N$-by-$N$ Hermitian random matrix 
in the Gaussian ensemble are in a range of 
$[-\Lambda,\Lambda]$ with $\Lambda = {\cal O}(1)$, 
the upper bound of the imaginary parts of the eigenvalues 
is of the order of $10^{-15} \sim 10^{-14}$ for 
$100 \leq N \leq 4000$. 
Based on this observation, 
we set the cut-off $C$ of 
the imaginary part to $C=10^{-13}$. 
To be specific, 
we first normalize an $N$-by-$N$ 
non-Hermitian random matrix in the Gaussian ensemble such that its complex eigenvalues $E$ are distributed within a circle 
of
the order of $1$ (i.e., $|E| \lesssim {\cal O}(1)$). 
Then, we regard any complex eigenvalue of the random matrix whose 
imaginary parts are less than $C=10^{-13}$ as real. The cut-off $C$ is much larger than the error bound, so that it shall 
not miss any 
real eigenvalues of the matrix. 
The probability $P$ of 
complex
eigenvalues being 
mistaken as real depends on $N$ 
and is
negligible 
for $N<10^4$. 
To evaluate this probability $P$, let us suppose that complex eigenvalues $E$ of the non-Hermitian random matrix are distributed uniformly within the circle (i.e., $|E| \lesssim {\cal O}(1)$). 
Then, the probability $p$ that a given %
complex eigenvalue of the random matrix is mistaken 
as real is of the order of $C=10^{-13}$,  
and the probability $P$ that 
complex eigenvalues are   
mistaken as real is estimated as $P=1-(1-p)^{N} \simeq Np$.
This probability is 
negligible for $N<10^4$ owing to $p \sim C = 10^{-13}$. This justifies our setting of 
the cut-off $C$ in this paper, where $N$ is typically less than $10^4$.}.
Meanwhile, to avoid regarding 
real eigenvalues as complex due to the machine 
inaccuracy, we choose a cut-off $C$ larger than the error 
bound. 
The probability that 
complex eigenvalues are 
mistaken as real depends on the dimensions $N$ of the matrix. 
With our choice of the cut-off $C$, this probability is estimated to be 
negligible for $N<10^4$, where $N$ in this paper is 
typically less than $10^4$.

The density $\rho_c(x,y)$ of complex eigenvalues in all the seven symmetry classes 
vanishes toward the real axis and hence has a soft gap around the real axis
(see Fig.~\ref{2d_DoS}). %
The size of the gap is of the same order as a mean level-spacing of eigenvalues in 
the complex plane. When $|y|$ is much smaller than the mean eigenvalue spacing, 
we have $\rho_c(x,y) \sim |y|$ in classes AI, A + $\eta$, AI + $\eta_-$, AII + $\eta_+$, 
and AII + $\eta_-$, while $\rho_c(x,y) \sim |y|^2$ in class AII and 
$\rho_c(x,y) \sim -|y| \log{|y|}$ in class AI + $\eta_+$ (Fig.~\ref{2d_DoS}). 
These small $y$ behaviors are consistent with the small matrix analysis discussed 
in Sec.~\ref{Eff_small_mat}. The logarithmic correction in 
class AI + $\eta_+$ seems to be due to TRS$^{\dag}$ with the sign $+1$~\cite{Grobe89, Hamazaki20}.

The number of real eigenvalues of non-Hermitian real random 
matrices (symmetry class AI) was previously 
 studied~\cite{Efetov97,Lehmann1991,edelman1995,Kanzieper05,Forrester07}. %
However, a systematic study on the other symmetry classes is still 
lacking. In this paper,
we find that the averaged number $\bar{N}_{\rm real}$ of real eigenvalues
of $N \times N$ non-Hermitian random matrices is proportional to the square-root 
of the dimensions of the matrices in all the five symmetry 
classes (see Fig.~\ref{N_real}), 
\begin{equation}
    \bar{N}_{\rm real} \propto 
    \sqrt{N}.
\end{equation}
This sub-extensive number of real eigenvalues enables level 
statistics analyses on the real axis, where the symmetries 
associated with complex conjugation must 
have important effects as in the Hermitian case~\cite{Evers08}. 

Furthermore, we obtain the universal distribution functions of spacings of real eigenvalues.
Let $\lambda_1,\lambda_2,\cdots, \lambda_{N_{\rm real}}$ 
be all the real eigenvalues of %
given ${\cal H}$ 
in the descending order. 
We define a normalized spacing of the real eigenvalues as  
\begin{equation}
    s_i \equiv \frac{\lambda_{i+1} - \lambda_{i}}{ \langle \lambda_{i+1} - \lambda_{i} \rangle}.
\end{equation} 
Here, $\langle ... \rangle$ stands for the average over the 
ensemble and $\langle \lambda_{i+1} - \lambda_{i} \rangle$ 
is evaluated by %
the average density of real eigenvalues %
at $x=(\lambda_i+\lambda_{i+1})/2$,  
\begin{equation}
    \langle \lambda_{i+1} - \lambda_{i}\rangle = \frac{1}{\overline{\rho}_{r}\left(\frac{1}{2}(\lambda_{i+1}+\lambda_i)\right)} \, ,
\end{equation}
where
$\overline{\rho}_r(x)$ is %
the
averaged %
density of real eigenvalues, %
\begin{equation}
    \overline{\rho}_r(x) \equiv %
    \sum_{\lambda_i \in {\cal R}} \langle  \delta(x - \lambda_i) \rangle  %
    \label{rho_r}
\end{equation}
with
the set ${\cal R}$ of real eigenvalues. 
Here, $\overline{\rho}_r(x)$ is estimated by the average over the Gaussian ensemble. 
To exclude a fluctuation due to finite sampling numbers, we follow Refs.~\cite{haake1991quantum,Aguiar05} and replace the delta 
function in Eq.~(\ref{rho_r}) by the
Gaussian distribution, $\exp[-(x-\lambda_i)^2/(2 \sigma^2)]/(\sqrt{2\pi} \sigma)$ with $\sigma = n {\bar s}$.
Here, ${\bar s}$ is the mean level-spacing on the real axis, and $n$ is an ${\cal O}(1)$ constant. 
We verify the validity of this numerical approach by using
different $n$ in the range from $2$ to $5$ and also %
replacing the delta function with the uniform distribution 
\begin{equation}
	\frac{1}{4\sigma} \times 1_{[\lambda_i - 2\sigma,\lambda_i + 2\sigma]}(x) = \begin{cases}
		\frac{1}{4\sigma} & ( x \in [\lambda_i - 2\sigma,\lambda_i + 2\sigma]       ) \, ,\\
		0 &  ( x \notin [\lambda_i -2 \sigma,\lambda_i + 2\sigma]       ) \, . 
    \end{cases}
\end{equation}  
We confirm that $\overline{\rho}_r(x)$ is barely influenced by the approximation scheme, where the maximal difference of $\overline{\rho}_r(x)$ between the different approximation methods is around or smaller than $1\%$. 
Note also that we exclude the real eigenvalues around the edges of the spectrum when studying the distribution of the spacings of real eigenvalues, because $\overline{\rho}_r(x)$ near the edges changes sharply and the estimated $\overline{\rho}_r(x)$ might have larger error bars.

The spacing ratio of real eigenvalues~\cite{Oganesyan07, pal10, Atas13} is also a useful quantity to characterize 
the
level statistics on the real axis. 
It is defined by 
\begin{equation}
	r_i \equiv \min \left( \frac{\lambda_{i+1} - \lambda_i}{\lambda_{i} - \lambda_{i-1}}, \frac{\lambda_{i} - \lambda_{i-1}}{\lambda_{i+1} - \lambda_{i}} \right) \, ,
	\label{r_def}
\end{equation}
satisfying $0 \leq r_i \leq 1$.
Since $r_i$ is a dimensionless quantity and free 
from the normalization, 
it is easier to numerically obtain the distribution 
of $r_i$ than that of $s_i$.

In each of the five symmetry classes (i.e., classes A + $\eta$, AI, AI + $\eta_+$, AI + $\eta_+$, and AI + $\eta_-$), 
we numerically calculate 
the level-spacing distribution $p(s)$ and the level-spacing-ratio distribution $p_r(r)$ of real eigenvalues,  
both of which converge to the characteristic functions (Fig.~\ref{ps_and_pr_GE}). 
Here, $p_r(r)$ and $p(s)$ in class AII + $\eta_+$ converge more slowly than those in the other symmetry classes and do not converge even at the maximal matrix size ($N=4000$; see Figs.~\ref{ps_CII_GE} and \ref{pr_CII_GE}).

To improve the convergence, we also introduce a generalized Gaussian ensemble with the following probability distribution function $p^{\prime}(\cal H)$: 
\begin{equation}
	p^{\prime}({\cal H}) =  C_{N,(\beta_1,\beta_2)}^{ -1} e^{ -{\rm Tr }\,
    [
    \beta_1\,( {\cal H} + {\cal H}^{\dagger} )^{2} -\beta_2\,( {\cal H} - 
    {\cal H}^{\dagger} )^{2} 
    ]} \, ,
	\label{p_gge}
\end{equation}
where $\beta_1$ and $\beta_2$ control the fluctuations of Hermitian and anti-Hermitian parts of $\cal{H}$, respectively. 
For $\beta_1 = \beta_2$, $p^{\prime}({\cal{H}})$ reduces to $p({\cal H})$ in the Gaussian ensemble. 
For $\beta_1 \neq \beta_2$, the eigenvalues $E = x + {\rm i}y$ of $\cal{H}$ distribute almost uniformly in the ellipse in the complex plane,
\begin{equation}
	\frac{x^2}{a^2} + \frac{y^2}{b^2} = 1 \,,
        \label{eq: GGE - ellipse}
\end{equation}
with $a/b = \beta_2 / \beta_1$~\cite{Sommers98}. 
In each symmetry class, random matrices in the generalized Gaussian ensemble 
show the same universal behaviors, such as 
soft gaps around the real axis and the square-root scaling 
of the average number $\bar{N}_{\rm real}$ of real eigenvalues.
We find that in each symmetry class, for a matrix size $N$, 
the average number $\bar{N}^{\prime}_{\rm real}$ of real eigenvalues in 
the generalized Gaussian ensemble with $\beta_1$ and $\beta_2$ 
is approximately scaled by $\bar{N}_{\rm real}$ in the Gaussian ensemble with 
the same matrix size as 
\begin{equation}
	\bar{N}^{\prime}_{\rm real} \simeq \sqrt{\frac{\beta_2}{\beta_1}} \bar{N}_{\rm real} \, ,
\end{equation}
for $\bar{N}_{\rm real}, \bar{N}^{\prime}_{\rm real} \ll N$. 
We also find that 
for $\bar{N}^{\prime}_{\rm real} \simeq \bar{N}_{\rm real}$, 
$p(s)$ and $p_r(r)$ in the two ensembles are close to 
each other 
(see Appendix~\ref{GGE} for details). Thus, $p(s)$ and $p_r(r)$ 
converge much faster in the generalized Gaussian ensemble with $\beta_2>\beta_1$. 
We choose $\beta_2/\beta_1 = 16$ in the following.

In each symmetry class, the error bars of $p(s)$ and $p_r(r)$ of non-Hermitian random matrices in 
the generalized Gaussian ensemble with a larger matrix size ($N>1000$) overlap 
with each other. Both $p(s)$ and $p_r(r)$ converge to the characteristic 
universal functions in the limit of the large matrix size 
(Figs.~\ref{ps_all} and \ref{pr_all}).
$p(s)$ for classes A + $\eta$, AI + $\eta_+$, and AII +  $\eta_+$ in different ensembles 
were previously calculated in Ref.~\cite{marinello19}, %
although the sizes of the matrices are small and $p(s)$ 
in Ref.~\cite{marinello19} does not seem to reach the universal function forms 
shown in Fig.~\ref{ps_all}. $p(s)$ and $p_r(r)$ in the limit of the large matrix size can 
distinguish the different symmetry classes among the five symmetry classes. 
We confirm the universality of $p(s)$ and $p_r(r)$  %
in each symmetry class
by comparisons with $p(s)$ and $p_r(r)$ in the Bernoulli ensemble (see %
Appendix~\ref{sec_be} for details). 
This comparison illustrates that both of the level-spacing and 
level-spacing-ratio distributions of real eigenvalues are 
universal and unique in each symmetry class. 
In Secs.~\ref{sec_mbl} and \ref{sec_free}, we compare 
$p(s)$ and $p_r(r)$ obtained in the random matrix theory with those obtained from physical models.

Note that the level-spacing distributions $p(s)$ in the different 
symmetry classes share the same asymptotic behaviors.
For $s \gg 1$, we have 
\begin{align}
    \log p(s) \propto -s
\end{align}
in all the five symmetry classes. 
On the other hand,
for $s \ll 1$, we have (see the insets of Fig.~\ref{ps_all})
\begin{align}
    p(s) \propto \begin{cases}
    -s \log s & (\text{class AI} + \eta_+), \\
    s & (\text{other four classes}).
    \end{cases}
\end{align}
As a comparison, the level-spacing distributions of 
Hermitian random matrices satisfy $\log p(s) \propto -s^{2}$ for $s \gg 1$ and $p(s) \propto s^{\beta}$
for $s \ll 1$, 
where the Dyson index $\beta = 1,2,4$ characterizes the 
Gaussian orthogonal, unitary, and symplectic ensembles, respectively~\cite{mehta04}. 
The small-$s$ behavior of $p(s)$ and its difference with Hermitian 
random matrices are well understood by analyses of 
effective small matrices (see Sec.~\ref{Eff_small_mat} for details).
In the large $s$ regime, the tail of $p(s)$ in the non-Hermitian case is much heavier than that in the Hermitian case.
This difference in the large-$s$ behavior shows that the correlation between two neighboring real eigenvalues decays more quickly in non-Hermitian random matrices than Hermitian ones. 
When the distance between two neighboring real eigenvalues of a non-Hermitian random matrix is larger, more complex eigenvalues surround them and weaken the correlation between the two real eigenvalues.

For reference, in the Hermitian random matrix theory, 
the level-spacing-ratio distribution $p_r(r)$ is well approximated by~\cite{Atas13}
\begin{equation}
    p_r(r) %
    \simeq
    \frac{1}{C_{\beta}} \frac{(r+r^2)^{\beta}}{(1 + r + r^2)^{(1+\frac{3}{2}\beta)}} \theta(1-r) \, ,  
	\label{pr_hermitian}
\end{equation}
where $\theta(r)$ is the step function, %
$C_{\beta}$ is a normalized constant, %
and
$\beta = 1,2,4$ is the Dyson index.
By contrast, if all real eigenvalues are uncorrelated and follow the Poisson statistics, $p_r(r)$ is given by 
\begin{equation}
	p_r(r) = \frac{2}{(1+r)^2}  \theta(1-r) \, .
	\label{pr_uncorrelated}
\end{equation}
None of these level-spacing-ratio distributions $p_r(r)$ of Hermitian 
matrices can describe any of our universal $p_r(r)$ of non-Hermitian 
random matrices in Fig.~\ref{pr_all},
showing the unique non-Hermitian nature of our universal distribution functions.

Notably, the mean value of the level-spacing ratios 
\begin{equation}
\langle r \rangle = \int_0^1p_r(r)rdr \approx 0.371
\end{equation}
in class AII + $\eta_+$ is smaller than 
\begin{equation}
\langle r \rangle_{\rm Poisson} = -1 + \ln 4 \approx 0.386
\end{equation}
of uncorrelated levels. 
By contrast,
$\langle r \rangle$ of non-Hermitian random 
matrices in the other four symmetry classes are all larger than 
$\langle r \rangle_{\rm Poisson}$ (see Table~\ref{symmetry_class_A} 
and Fig.~\ref{pr_all}). 
In the Hermitian random matrix theory, the mean values of 
level-spacing ratios in the Gaussian orthogonal, unitary, and symplectic ensembles %
are~\cite{Atas13}
\begin{equation}
	\begin{aligned}
		\langle r \rangle_{\rm GOE} \approx 0.531, \,
		\langle r \rangle_{\rm GOE} \approx 0.600 , \,
		\langle r \rangle_{\rm GSE} \approx 0.674, \,
	\end{aligned}
\end{equation}
all of which are larger than $\langle r \rangle_{\rm Poisson} \approx 0.386$.  
In addition, $\langle r \rangle$ increases with the Dyson index $\beta = 1, 2, 4$ that describes the strength of the level repulsion~\cite{mehta04}.
Thus, $\langle r \rangle < \langle r \rangle_{\rm Poisson}$ 
in class AII + $\eta_+$ indicates unusual level interactions 
on the real axis unique to this symmetry class. 

To further clarify the nature of the interactions between real eigenvalues 
in each symmetry class, we also calculate the variance 
$\Sigma_2$ of the number $N_W$ of real eigenvalues in an interval 
on the real axis~\cite{Evers08}, 
\begin{equation}
	\Sigma_2 \equiv \langle N_W^2 \rangle - \langle N_W \rangle^2.
\end{equation}  
We have $\Sigma_2 \propto \log \langle N_W \rangle$ in Hermitian random matrix theory, 
while we have $\Sigma_2 =  \langle N_W \rangle$ for uncorrelated real numbers (i.e., Poisson statistics). 
The spectral compressibility, 
\begin{equation}
	\chi \equiv \lim_{N_W \rightarrow \infty} \frac{d \Sigma_2 %
    }{d\langle N_W \rangle} \, , 
	\label{chi_def}
\end{equation}
quantifies the level interaction between real eigenvalues. 
For Hermitian random matrices, the level repulsion is stronger, leading to
$\chi=0$. 
On the other hand, we have $\chi_{\rm Poisson} = 1$ for the uncorrelated real spectrum.
In the intermediate regime, such as the metal-insulator transition points in Hermitian disordered systems, the 
level repulsion is weaker than the random matrices but stronger than the Poisson statistics, resulting in $0 < \chi <1$~\cite{Evers08}. 
For non-Hermitian random matrices in each of the five symmetry classes,
we find 
\begin{equation}
\Sigma_2 \propto \langle N_W \rangle, 
\end{equation}
meaning that the spectral compressibility $\chi$ 
gives a universal constant unique to each symmetry class 
(see Table~\ref{symmetry_class_A} and Appendix~\ref{sec_nv} for details).

Remarkably, we have $\chi \approx 1.11 >\chi_{\rm Poisson} = 1$ in class AII + $\eta_{+}$, 
again
indicating the unusual level interactions. 
As shown by 
$p(s) \propto s \to 0$ for $s \to 0$,
the interaction is repulsive in the small $s$ 
regime even for class AII + $\eta_{+}$,
although the level repulsion is much smaller than $p(s) \propto s^4$ of Hermitian random matrices in class AII (i.e., Gaussian symplectic ensemble).
Hence, 
from our numerical results of $\langle r \rangle < \langle r \rangle_{\rm Poisson}$ and $\chi>\chi_{\rm Poisson}$,
the attractive interaction should appear %
in 
the finite %
$s$ regime and 
dominate the repulsive interaction in the small
$s$ regime on average.
This is also compatible with the large peak of $p(s)$ and $p_r(r)$ compared with the other symmetry classes (see Figs.~\ref{ps_all} and \ref{pr_all}).
A distinctive feature of class AII + $\eta_{+}$ is the simultaneous presence of TRS and TRS$^{\dag}$ whose signs are $-1$.
While TRS$^{\dag}$ with the sign $-1$ leads to the Kramers degeneracy of generic complex eigenvalues and the consequent strong level repulsion,
TRS with the sign $-1$ leads to the strong level repulsion around the real axis, as shown by the absence of real eigenvalues in class AII.
The combination of TRS and TRS$^{\dag}$ seems to result in the unusual interactions between neighboring levels (Kramers pairs) on the real axis that are repulsive in the small $s$ regime but attractive in the larger $s$ regime.
This is a possible reason for $\langle r \rangle < \langle r \rangle_{\rm Poisson}$ and $\chi>\chi_{\rm Poisson}$.
In Sec.~\ref{Eff_small_mat}, we use effective small matrices to analyze the interactions between neighboring real eigenvalues and find that the degrees of freedom of the attractive interaction in class AII + $\eta_+$ are much larger than those of the other symmetry classes.

Note also that no real eigenvalues generally appear in 
classes AII and AII + $\eta_-$. This should be due to 
TRS with the sign $-1$, which only enforces the Kramers 
degeneracy on the real axis. If a real eigenvalue is 
present in these symmetry classes, a Kramers partner  
with the same real eigenvalue should always appear.
While this Kramers pair is robust against Hermitian perturbations, 
it is sensitive to non-Hermitian perturbations and forms 
a complex-conjugate pair in the complex plane.
Consequently, real eigenvalues are unstable 
in these symmetry classes.
On the contrary, 
in class AII + $\eta_+$, all eigenvalues including 
complex ones exhibit the Kramers degeneracy because 
of the additional presence of TRS$^{\dag}$ with the sign $-1$.
Consequently, the Kramers pair on the real axis is 
robust even against certain degrees of non-Hermitian perturbations, 
which leads to the sub-extensive number of real eigenvalues.
This is different from class AII + $\eta_-$, where 
only TRS$^{\dag}$ with the sign $+1$ is present and no such 
robust Kramers degeneracy is allowed generally.
We also discuss the absence of real eigenvalues in classes 
AII and AII + $\eta_-$ by effective small matrices in Sec.~\ref{Eff_small_mat}.

\subsection{Effective small matrix analysis}
\label{Eff_small_mat}

Interactions between two neighboring eigenvalues can be 
qualitatively understood by effective small 
matrices~\cite{Hamazaki20}. 
When two eigenvalues get close to each other by the change of 
parameters, the interactions between them can be 
studied by nearly degenerate perturbation theory~\cite{haake1991quantum}. 
The %
strength
of the interactions is generally 
determined by symmetry such as TRS and 
pH.
To see  
the influence of symmetry %
in each of the seven symmetry classes, we consider the two adjacent 
eigenvalues that are either both real or complex conjugate to each other. 
Then, we project the variation of a full Hamiltonian onto
a smaller space spanned by eigenvectors that belong to the two 
adjacent eigenvalues. 
The small Hamiltonians thus obtained  
take forms of either $2\times2$ matrix or $4\times4$ matrix, 
depending on the presence of the Kramers degeneracy. 
The symmetry classes of the small matrices are the same as the full Hamiltonians. 

The small matrices in the seven symmetry classes are of the following forms, 
\begin{equation}
\begin{split}
&{\cal H}^{(s)}_{\text{AI}} = \left( \begin{matrix}
a_0 + a_1 & a_2 + a_3\\
a_2 - a_3 & a_0 - a_1 \\
\end{matrix} \right) \, , \\ 
&{\cal H}^{(s)}_{\text{A}+\eta} = \left( \begin{matrix}
a_0 + a_1 & a_3 + {\rm i} a_2\\
-a_3 + {\rm i} a_2 & a_0 - a_1 \\
\end{matrix} \right) \, , \\
&{\cal H}^{(s)}_{\text{AI}+\eta_+} = 
\left( \begin{matrix}
a_0 + a_1 & a_2  \\
-a_2 & a_0 - a_1 \\
\end{matrix} \right) \, , \\ 
&{\cal H}^{(s)}_{\text{AII}} = 
\left( \begin{matrix}
a_0 +  {\rm i} a_1 &  a_3 +  {\rm i} a_2 \\
 -a_3 +  {\rm i} a_2& a_0 -   {\rm i} a_1\\
\end{matrix} \right) \, , \nonumber  
\end{split}
\end{equation}
\begin{equation}
\begin{split}
&{\cal H}^{(s)}_{\text{AI}+\eta_-} = \left( \begin{matrix}
a_0 + a_1 & 0 & a_2 + a_5 & a_4 -a_3  \\
0 & a_0 + a_1 & a_4 + a_3 & a_2 - a_5 \\
a_2 - a_5 & -a_4 + a_3 & a_0 - a_1 & 0 \\
-a_4 - a_3 & a_2 + a_5 & 0 & a_0 - a_1 
\end{matrix} \right) \, , \\ 
&{\cal H}^{(s)}_{\text{AII}+\eta_-} = \left( \begin{matrix}
a_0  &  a_2 + {\rm i}a_1  \\
 -a_2 + {\rm i}a_1 & a_0 \\
\end{matrix} \right) \, , \\
&{\cal H}^{(s)}_{\text{AII}+\eta_+} = \left( \begin{matrix}
a_0+a_1 & 0 & a_2 + {\rm i}a_5 &  a_4 + {\rm i}a_3    \\
0 & a_0 + a_1 & -a_4 +  {\rm i}a_3   & a_2 - {\rm i}a_5 \\
 -a_2 + {\rm i}a_5 &  a_4 + {\rm i}a_3 & a_0-a_1 & 0 \\
-a_4 +  {\rm i} a_3  & -a_2 - {\rm i}a_5 & 0 & a_0-a_1 
\end{matrix} \right) \, , \nonumber 
\end{split}
\end{equation}
where $a_0,a_1,\cdots, a_{m+n}$ are %
real random variables.
The eigenvalues of the small matrices are written in a unified form as 
\begin{equation}
   \begin{aligned}
\lambda & = a_0  \pm  \sqrt{X-Y},  \!\ \!\ \\ 
X &\equiv \begin{cases}
  \sum_{i=1}^{m} a_i^2 &  \!\ (m \neq 0), \\
  0 &  \!\ (m = 0),
\end{cases} 
\!\ \!\  \\
Y &\equiv \sum_{i=m+1}^{m+n} a_i^2,  
\label{eigenvalue}
\end{aligned} 
\end{equation}
for the seven symmetry classes (see Appendix \ref{B} for details).
The two eigenvalues of each of 
these matrices are either both real or complex conjugate to each other. 
For $m=0$ (classes AII 
and AII + $\eta_{-}$), they are always complex conjugate to each other, 
meaning the absence of real eigenvalues.
For $m>0$, 
the probability of two real eigenvalues %
is finite 
and %
equal to the %
probability
for $X \geq Y$. 
This explains the presence and absence of the delta function 
peak on the real axis in the DoS in the seven symmetry classes. 
In Appendix~\ref{B}, we analytically obtain the level statistics of real eigenvalues and the DoS around the real axis for the above effective small matrices in the seven symmetry classes.

The finite probability of the real 
eigenvalues of the random matrices leads 
to the square-root scaling of $\bar{N}_{\rm real}$. According to 
the circular law~\cite{Ginibre65}, the uniform distribution of 
complex eigenvalues within the circle of radius $R$ suggests that 
the number of complex eigenvalues near the real axis within an  
energy window of a mean complex energy spacing $\bar{s}_c$ is scaled by $\sqrt{N}$,
\begin{align}
N \times \frac{2R \bar{s}_c}{\pi R^2} = 2 \sqrt{\frac{N}{\pi}} \propto \sqrt{N}, 
\end{align}
with $\pi R^2/\bar{s}_c^2 = N$. 
The complex eigenvalues near the real axis 
can be regarded as complex-conjugate pairs of eigenvalues each 
of which is described by the small random matrices. 
Due to 
the finite probability of $X>Y$, a complex-conjugate pair of the eigenvalues 
near the real axis is converted into real eigenvalues with a 
finite probability. This gives the square-root scaling, 
$\bar{N}_{\rm real} \sim \sqrt{N}$.  
 
\begin{table}[bt]
\caption{Degrees of freedom of Hermitian and anti-Hermitian parts of traceless effective small random matrices
$H$ for the seven symmetry classes. 
For each class, the traceless parts of the small matrix 
are decomposed into the Hermitian and anti-Hermitian parts, the degrees of which are denoted by $m$ and $n$, respectively. If $H$ belongs to the symmetry class in the first column, ${\rm i} H$ belongs to the equivalent symmetry class in the second column.}
\begin{tabular}{cccc}
\hline \hline
\begin{tabular}[c]{@{}c@{}}symmetry \\ class\end{tabular} &
\begin{tabular}[c]{@{}c@{}}equivalent \\ symmetry class\end{tabular} &
  \begin{tabular}[c]{@{}c@{}}real degree of\\   freedom $m$\end{tabular} &
  \begin{tabular}[c]{@{}c@{}}imaginary degree of\\   freedom $n$ \end{tabular} \\ \hline
AI & D$^{\dagger}$ &2 & 1 \\
A + $\eta$ & AIII & 1 & 2 \\ 
AI + $\eta_+$ & BDI$^{\dagger}$ &  1 & 1 \\
AII & C$^{\dagger}$ &0 & 3 \\
AI + $\eta_-$& DIII$^{\dagger}$ & 3 & 2\\
AII + $\eta_-$ & CI$^{\dagger}$ & 0 & 2 \\
AII + $\eta_+$ &  CII$^{\dagger}$ & 1 & 4 \\
\hline\hline
\end{tabular}
    \label{mn}
\end{table}

For $X < Y$, the two eigenvalues are 
complex conjugate to each other. Thereby, $X$ and $Y$ 
in Eq.~(\ref{eigenvalue}) give an attractive and repulsive 
interaction between the two eigenvalues, respectively. 
The increase of 
$X$ ($Y$) will decrease 
(increase) the distance between the two eigenvalues along the imaginary 
axis. 
In symmetry classes A + $\eta$, AI, AI + $\eta_+$, 
AII + $\eta_+$, and AI + $\eta_-$, we have $m > 0$, and 
both attractive and repulsive interactions are present. 
By contrast, in classes AII and AII + $\eta_-$, we have $m = 0$, and thus 
no attractive interaction is present (see Table \ref{mn}). 

The DoS around the real axis is determined 
by the interaction between the two complex-conjugate eigenvalues. 
When the attractive interaction 
along the imaginary axis
is absent ($m=0$ in classes AII and 
AII + $\eta_{-}$), the two eigenvalues are less likely to appear 
around the real axis than 
the other symmetry classes.
The repulsion between the two 
eigenvalues
becomes larger 
for larger $n$. In fact, our analysis of the small matrices 
in the Gaussian ensemble gives 
\begin{align}
\rho_c(x+ {\rm i}y) \propto \begin{cases}
\left|y \right|^2 & \left( \text{class AII}\right), \\
\left|y \right| & \left( \text{class AII} + \eta_{-}\right).
\end{cases}
\end{align}
In the presence of the attractive interaction ($m>0$ in 
classes A + $\eta$, AI, AI + $\eta_+$, AII + $\eta_+$, and AI + $\eta_-$),
the larger attractive interaction converts the complex-conjugate pair of %
eigenvalues near the real axis %
onto
two real eigenvalues. 
As a result, a sub-extensive number of eigenvalues of the full matrix  
appear on the real axis. %
In fact, our analysis of 
the small matrices in the Gaussian ensemble 
gives 
(see Appendix~\ref{B} for detailed derivations)
\begin{align}
\rho_c(x,y) \propto \begin{cases}
- \left| y \right| \log \left| y \right| & (\text{class AI} + \eta_{+}), \\
\left| y \right| & (\text{other four classes}),
\end{cases}
\end{align}
for small $\left|y \right|$.
These small-$|y|$ %
behaviors
of 
$\rho_c(x,y)$ from the small matrix analyses 
are consistent with the
numerical results of large random 
matrices shown in Fig.~\ref{2d_DoS}.

For $X > Y$, the two 
eigenvalues are real. Thereby, $X$ and $Y$ 
in Eq.~(\ref{eigenvalue}) respectively give a repulsive and attractive interaction 
between the pair of two real eigenvalues. %
The increase of $X$ ($Y$)
increases (decreases) the distance between the two eigenvalues along the real axis.
We analytically calculate 
the 
spacing distribution function of the two real eigenvalues %
for  
the small matrices in the five symmetry classes with $m > 0$ 
(see Appendix~\ref{B} for detailed derivations).  
For small $s$, the real-eigenvalue spacing distribution is obtained as 
\begin{align}
p(s) \propto \begin{cases}
-s\log s & (\text{class AI} + \eta_{+}), \\
s & (\text{other four classes}).
\end{cases}
\end{align}
Note that $p(s)$ from the small matrix analyses 
and that from large random matrices are not exactly the same, while they 
share the same asymptotic %
behavior
for %
small
$s$ for each of the 
five symmetry classes.
Note also that 
in class AII + $\eta_+$, the degrees of freedom of the attractive 
interaction $Y$ (i.e., $n=4$) is much larger than the degrees of freedom $m$ of 
the repulsive interaction $X$ (i.e., $m=1$). 
This is consistent with 
$\langle r\rangle <\langle r\rangle_{\rm Poisson}$ and 
$\chi>1$ in class AII + $\eta_{+}$ for large random matrices (see Sec.~\ref{DoS_LS} for details). %

\section{Dissipative Many-body systems}
\label{sec_mbl}

In the previous section, we study the general behavior of the level statistics of non-Hermitian random matrices in symmetry classes
AI, A + $\eta$, AI + $\eta_{+}$, AI + $\eta_{-}$, and AII + $\eta_{+}$.  
The DoS in these symmetry classes 
shows a delta function peak on the real axis. 
The number of real eigenvalues is scaled by the square-root of the dimensions of the matrices.  

In this section, we study many-body disordered Hamiltonians
that belong to symmetry classes AI, 
A + $\eta$, AI + $\eta_{+}$, and AI + $\eta_{-}$.  
We calculate
the 
level-spacing distributions, level-spacing-ratio distributions, and the 
numbers of real eigenvalues
in the
weak disorder regime (ergodic phase) and 
the strong disorder 
regime (MBL phase). 
In the weak disorder regime, we show that 
the
level-spacing and level-spacing-ratio distributions 
of real eigenvalues
match 
well with
those
of non-Hermitian random matrices 
in the same symmetry classes. 
In addition, we find that 
the %
number 
of real eigenvalues
in 
the weak disorder regime is scaled by the square root of the 
dimensions of the many-body Hamiltonians, which %
is 
also consistent with the random matrix theory. 
In the strong disorder regime, we show that 
the
many-body model 
in class AI
is characterized by non-universal 
scalings of the %
number 
of real eigenvalues
and its standard deviation. 
We also provide
a phenomenological explanation for this 
non-universal %
behavior.

\subsection{Hard-core boson system}
\begin{figure*}[bt]
	\centering
	\subfigure[density $\rho_c(x,y)$ of complex eigenvalues  (ergodic phase)]{
		\begin{minipage}[t]{0.3\linewidth}
			\centering
			\includegraphics[width=1\linewidth]{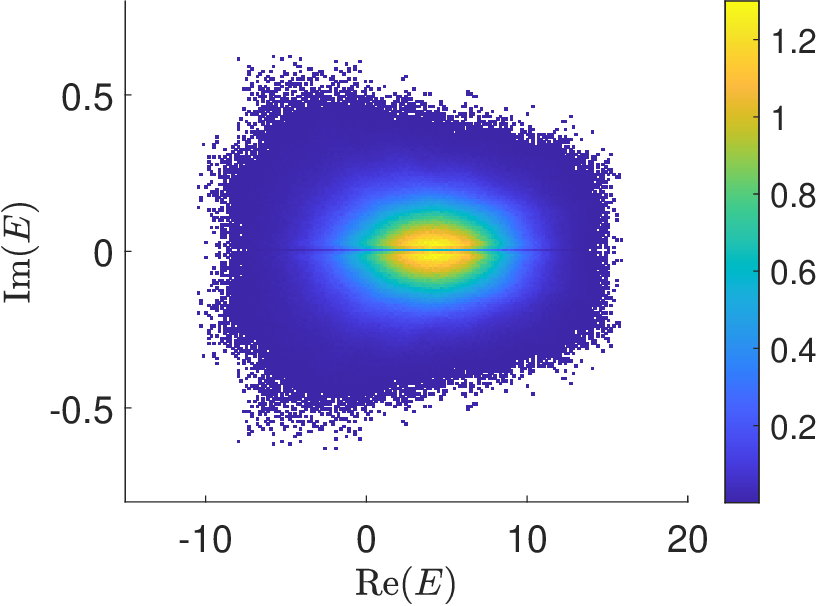}
		\end{minipage}%
		\label{hc_DoS}
	}%
	\subfigure[ $\bar{\rho}_c(y)$ (ergodic phase)]{
		\begin{minipage}[t]{0.3\linewidth}
			\centering
			\includegraphics[width=1\linewidth]{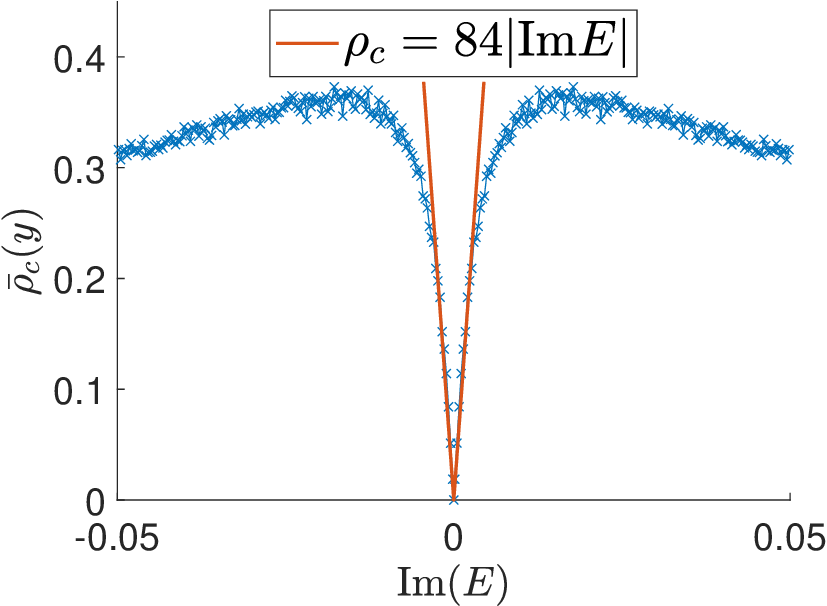}
		\end{minipage}%
		\label{hc_rho}
	}%
	\subfigure[$p(s)$ (ergodic phase) ]{
		\begin{minipage}[t]{0.3\linewidth}
			\centering
			\includegraphics[width=1\linewidth]{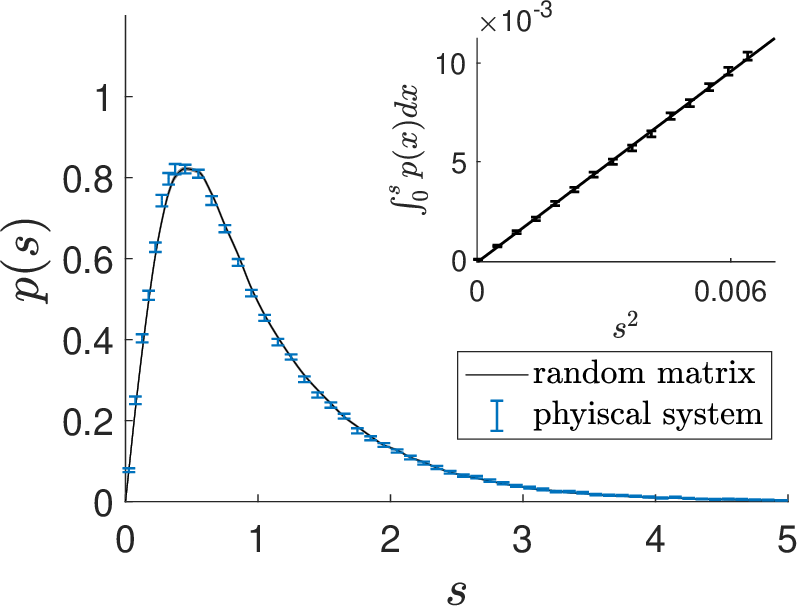}
			\label{hc_ps}
		\end{minipage}%
	}%

	\subfigure[$p(s)$ (MBL phase) ]{
		\begin{minipage}[t]{0.25\linewidth}
			\centering
			\includegraphics[width=1\linewidth]{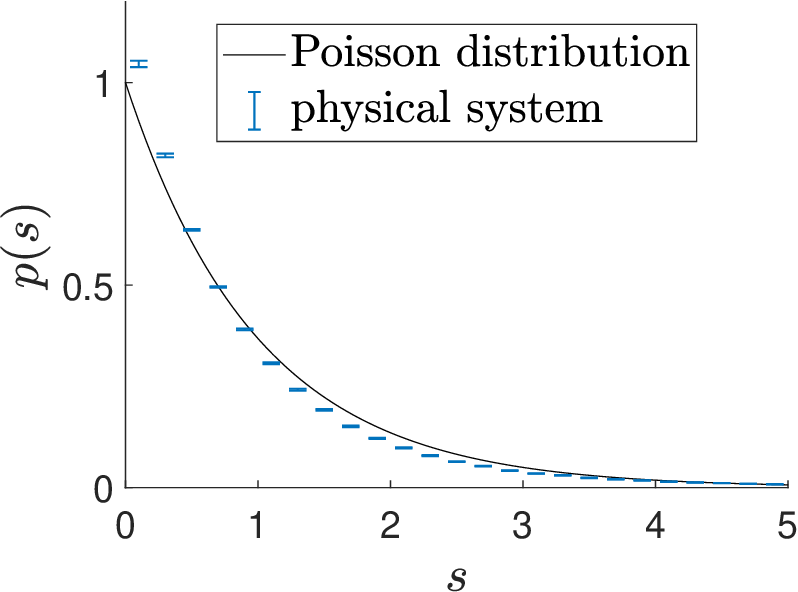}
		\label{hc_ps_loc}
		\end{minipage}%
	}%
	\subfigure[$\langle r \rangle -L_x $ (ergodic phase) ]{
		\begin{minipage}[t]{0.25\linewidth}
			\centering
			\includegraphics[width=1\linewidth]{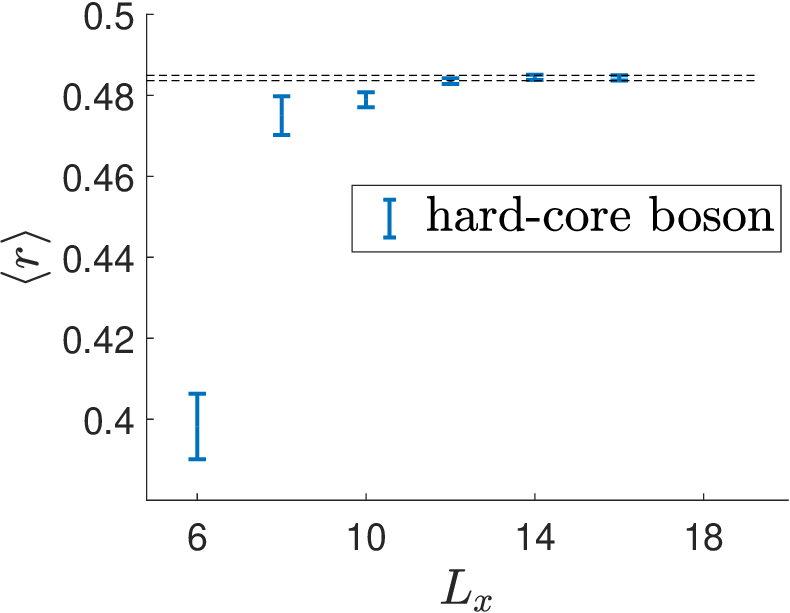}
		\label{hc_meanr_N}
		\end{minipage}%
	}%
	\subfigure[$p_r(r)$ (ergodic phase)]{
		\begin{minipage}[t]{0.25\linewidth}
			\centering
			\includegraphics[width=1\linewidth]{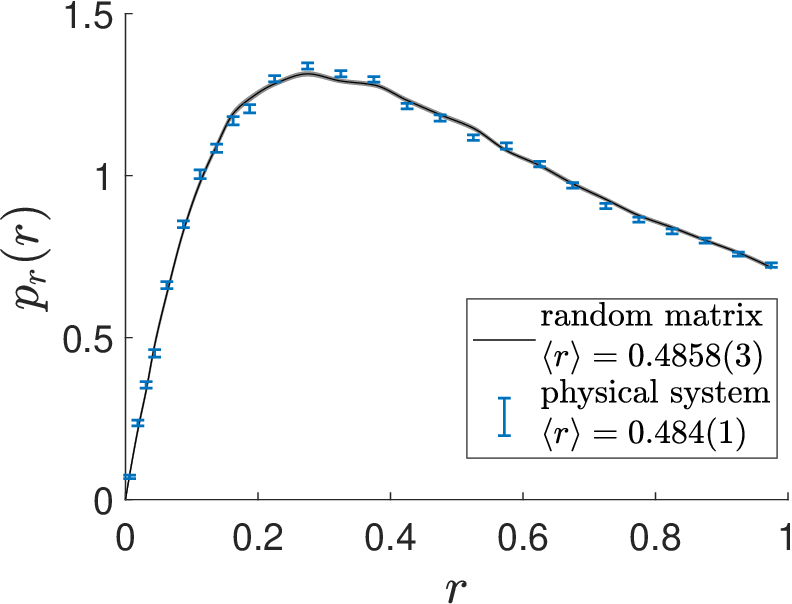}
			\label{hc_pr_erg}
		\end{minipage}%
	}%
	\subfigure[$p_r(r)$ (MBL phase)]{
		\begin{minipage}[t]{0.25\linewidth}
			\centering
			\includegraphics[width=1\linewidth]{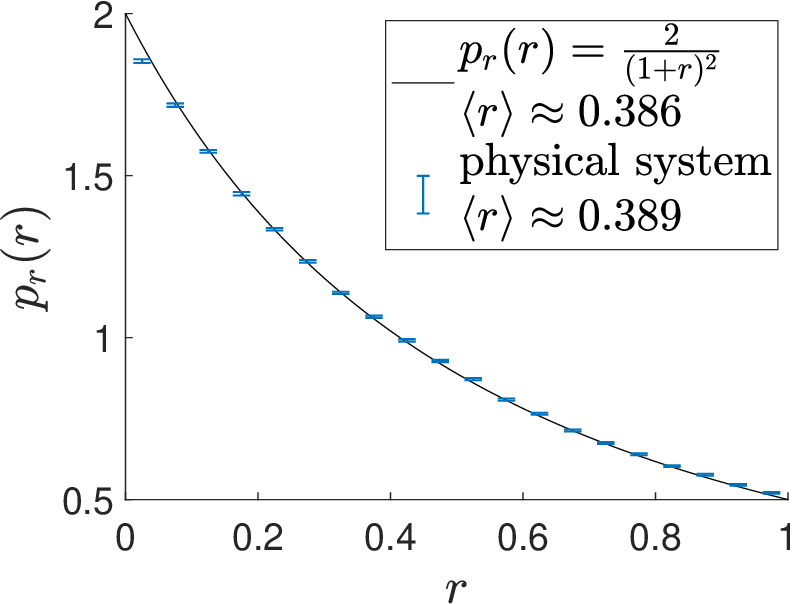}
			\label{hc_pr_loc}
		\end{minipage}%
	}%
	\caption{(a) Heat map of the density $\rho_c(x,y)$ of complex eigenvalues of the hard-core boson model in the
	weak disorder regime ($W=2$, $L_x = 16$). (b)~Integrated density of complex eigenvalues, 
	$\bar{\rho}_c(y) \equiv  \frac{1}{10}\int_{-1}^{9} \rho_c(x, y) dx $. As seen from the heat map, $-1<E<9$
	is well within the ergodic phase for $W = 2$.
	(c)~Level-spacing distribution $p(s)$ 
	of real eigenvalues within the energy window $-1 <E<9$ in the weak disorder regime of the hard-core boson model, and its comparison with $p(s)$ obtained from non-Hermitian random matrices in symmetry class AI (black line). Inset: Asymptotic behavior
	of $\int^s_0 p(s^{\prime}) ds^{\prime}$ for $s \ll 1$. 
	The consistency between $p(s)$ from the hard-core boson model and $p(s)$ from the random matrices justifies that $-1<E<9$ for
	$W=2$ is well within the ergodic phase. 
	(d)~$p(s)$ in the many-body localized phase 
	(all the real energies $E$, $W = 30$), and 
	its comparison with the  Poisson distribution.
	(e)~Mean value of level-spacing ratios $\langle r \rangle = \int_0^1 p_r(r)dr$ as a function of system size $L_x$. 
	The average is taken in the ergodic phase ($W=2$, $-1<E<9$, $L_x=16$).
	(f)~Level-spacing-ratio distribution $p_r(r)$ of real eigenvalues in the ergodic phase %
	($W=2$, $-1<E<9$
	, $L_x=16$) 
	and its comparison to $p_r(r)$ obtained from non-Hermitian random matrices in class AI.  
	(g) $p_r(r)$ in the many-body localized phase and its comparison with $p_r(r)$ of uncorrelated real numbers.
	The mean value $\langle r \rangle $ of each level-spacing-ratio distribution is also shown in the figures. 
	(c)-(g) The error ranges are evaluated by the bootstrap method~\cite{press07}.
    The error ranges of the distributions of random matrices are much smaller than those of the hard-core boson model and are not shown here or in the following figures (i.e., Figs.~\ref{spin_dos_ps}, \ref{bdid_3d_ps_DoS}, and \ref{ciid_2d_ps_DoS}; see also Figs.~\ref{ps_all} and~\ref{pr_all}).
 }%
\label{hard-core_3d_ps_DoS}
\end{figure*}

We consider the following one-dimensional (1D) hard-core boson model with the nonreciprocal hopping~\cite{hatano1996localization,hamazaki2019non}:
\begin{equation}
    {\cal H}_\text{HN} = \sum^{L_x}_{i=1} 
    \Big\{ t \left( e^g c_{i+1}^{\dagger} c_i +  e^{-g} c_{i}^{\dagger} c_{i+1} \right)+
    U n_{i+1}n_i + h_in_i \Big\} \, .\label{HN-H}
\end{equation}
Here, $c_i$ is a boson annihilation operator at site $i$,
$n_i = c_i^{\dagger}c_i$ is its number operator, 
and the periodic boundary conditions are imposed
(i.e., $c_{L_x+1}=c_1$).
Every site is allowed to be occupied by no more than one 
boson under %
the local hard-core boson constraint. 
$g$ controls the degree of %
non-reciprocity %
and non-Hermiticity, and
$h_i$ is %
the
random potential %
at site $i$
that 
distributes uniformly in %
$[-W/2,W/2]$ with $W \geq 0$. 
On the occupation-number basis, ${\cal H}_\text{HN}$ satisfies 
\begin{equation}
{\cal H}_\text{HN} = {\cal H}_\text{HN}^*
\end{equation}
and
belongs to %
class AI. 
This model can be mapped to an interacting spin model with a random magnetic
field
and
realized, for example, in ultra-cold atoms~\cite{hamazaki2019non,Gong18,choi16}.

The Hermitian limit ($g = 0$) of the model %
was
previously studied~\cite{pal10,chertkov21,Serbyn2016,yu16,Abanin19,Kulshreshtha18,Goihl18}, where
the
level-spacing distribution 
obtained
by 
the %
exact diagonalization
is one of the most powerful tools 
for detecting the %
MBL.
To identify 
the
ergodic and MBL phases in the non-Hermitian case ($g \ne 0$), %
Ref.~\cite{hamazaki2019non}
used a scaling of the proportion of the 
number of real eigenvalues, entanglement entropy, and %
level-spacing distribution 
in the complex plane. The proportion of the %
number 
of real eigenvalues
increases as the disorder strength increases. 
Furthermore, 
Ref.~\cite{hamazaki2019non} %
conjectured that complex eigenvalues 
collapse onto the real axis and 
that
the proportion 
of real eigenvalues
becomes approximately one when the system undergoes a 
transition from 
the
ergodic to MBL phases. %
Meanwhile, %
the scaling relationship between 
$\bar{N}_{\text{real}}$ and $N$ in the ergodic phase 
was not clarified.

\begin{figure}
    \centering
	\includegraphics[width=0.95\linewidth]{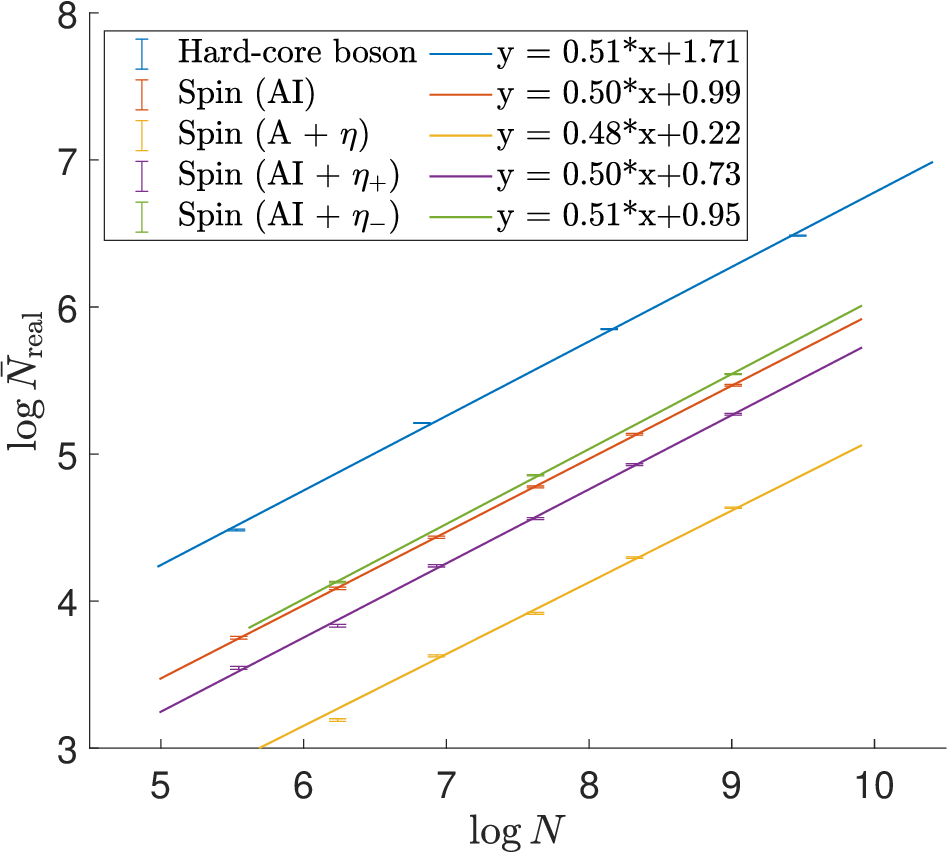}
    \caption{Average number $\bar{N}_{\rm real}$ of all the real eigenvalues as a function of the dimensions $N$ of the Hilbert space %
    for
    the many-body bosonic models in the weak disorder regimes ($W=2$ or  $W_x=W_y=W_z=W_D=1$). The square-root scaling $\bar{N}_{\rm real} \propto \sqrt{N}$ suggests that all the real eigenvalues of the %
    bosonic models in the weak disorder regimes are consistent with the random matrix theory and hence well within the ergodic phase.}
    \label{n_real_mb}
\end{figure}

We study the
weak (strong) disorder regime %
of this model
at the half filling of the boson number 
with %
the
parameters %
$t = 1$, $g=0.1$, $U=2$, 
$W=2$ ($W=30$).
At the half filling, the boson number $M$ is the same as 
the
half of the lattice site number $L_x$, (i.e., $M=L_x/2$).  
At least %
$400$ different disorder realizations of Eq.~(\ref{HN-H}) 
are diagonalized for
each system size (the maximal system size is $L_x = 16$) and 
for each disorder strength. 

In the weak disorder regime (ergodic phase), 
we find that
$\rho(x+{\rm i}y)$ has a delta function peak on the real 
axis, $\rho(x+{\rm i}y)= \rho_c(x,y)+\delta (y) \rho_r(x)$, and
$\rho_c(x,y)$  
shows a soft gap 
$\rho_c(x,y) \propto \lvert y \rvert$
around the real axis $y=0$ (Figs.~\ref{hc_DoS} and \ref{hc_rho}). 
We calculate the level-spacing distribution $p(s)$ and 
level-spacing-ratio distribution $p_r(r)$ of {the} real eigenvalues 
from an energy range around the center of the 
many-body spectrum.
We exclude 
real eigenvalues near the edges of the 
spectrum from the statistics. 
For the system sizes $L_x \geq 12$,  
the error bars of the mean values of the level-spacing ratio with different 
system size $L_x$ already overlap with each other (see Fig.~\ref{hc_meanr_N}). 
This indicates that the level statistics in the ergodic phase reach the convergence for $L_x \ge 12$.

We find that the level-spacing 
distribution and level-spacing-ratio distribution
of real eigenvalues are
well described by those obtained from
non-Hermitian
random matrices in  
class AI 
(Figs.~\ref{hc_ps} and \ref{hc_pr_erg}). 
For reference, we compare the Kolmogorov-Smirnov distances between $p(s)$, $p_r(r)$ of our hard-core boson model in the ergodic  
phase and those from the non-Hermitian random matrices in 
the five symmetry classes in Appendix~\ref{sec_ks}. 
The comparison further confirms the consistency between 
the random matrix theory and the physical Hamiltonian.
We also calculate the number $N_{\rm real}$ of all the real many-body eigenenergies and 
its average $\bar{N}_{\text{real}}$ over the different disorder realizations.
We find that $\bar{N}_{\text{real}}$ 
is scaled by the square root of the dimensions $N$ of the many-body Hamiltonian, 
$\bar{N}_{\rm real} \sim \sqrt{N}$ (Fig.~\ref{n_real_mb}), 
which is also consistent with the random matrix theory.

In the strong disorder regime (MBL phase), 
by contrast, 
almost all the eigenvalues are real, 
and thus we have $\bar{N}_{\rm real} \sim N$. 
While the level-spacing-ratio distribution $p_r(r)$ of real eigenvalues is the same as $p_r(r)$ of uncorrelated real eigenvalues (Eq.~(\ref{pr_uncorrelated})), 
the 
level-spacing distribution $p(s)$ 
of real eigenvalues
is slightly different 
from the
Poisson distribution.
This slight difference is due to the finite-size effect and will vanish if the disorder strength or the system size is increased.

\subsection{Interacting spin system}

\begin{figure*}[p]
	\centering
	\subfigure[density $\rho_c(x,y)$ of complex eigenvalues (class AI)]{
		\begin{minipage}[t]{0.25\linewidth}
			\centering
			\includegraphics[width=1\linewidth]{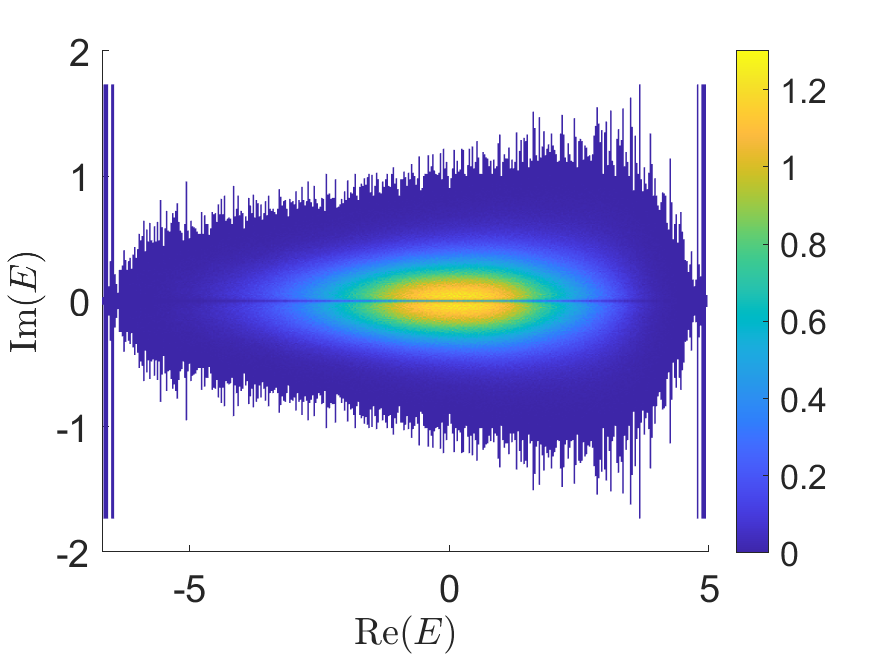}
		\end{minipage}%
	}%
	\subfigure[density $\rho_c(x,y)$ of complex eigenvalues (class A + $\eta$)]{
		\begin{minipage}[t]{0.25\linewidth}
			\centering
			\includegraphics[width=1\linewidth]{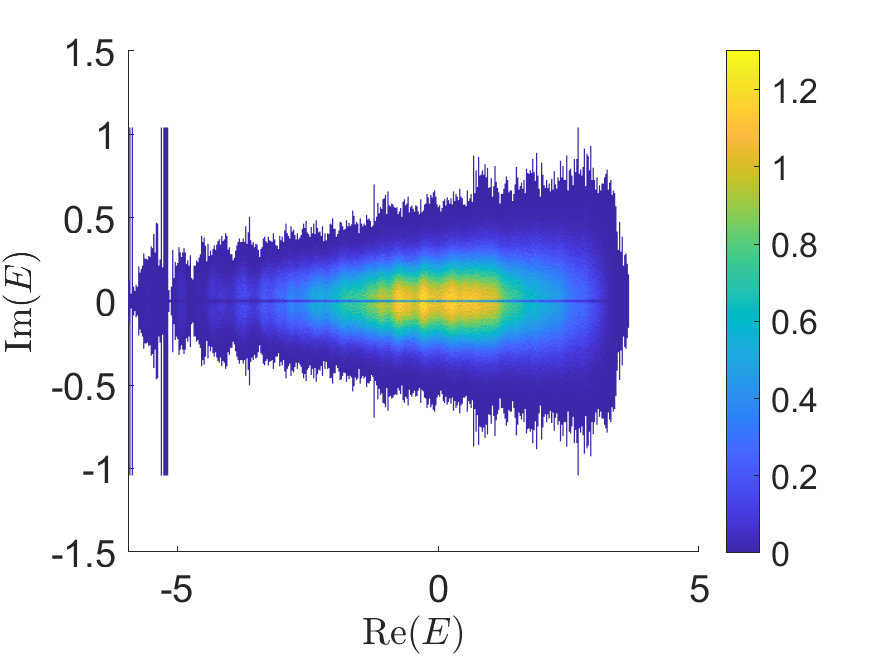}
		\end{minipage}%
 	}%
	\subfigure[density $\rho_c(x,y)$ of complex eigenvalues (class AI + $\eta_+$)]{
		\begin{minipage}[t]{0.25\linewidth}
			\centering
			\includegraphics[width=1\linewidth]{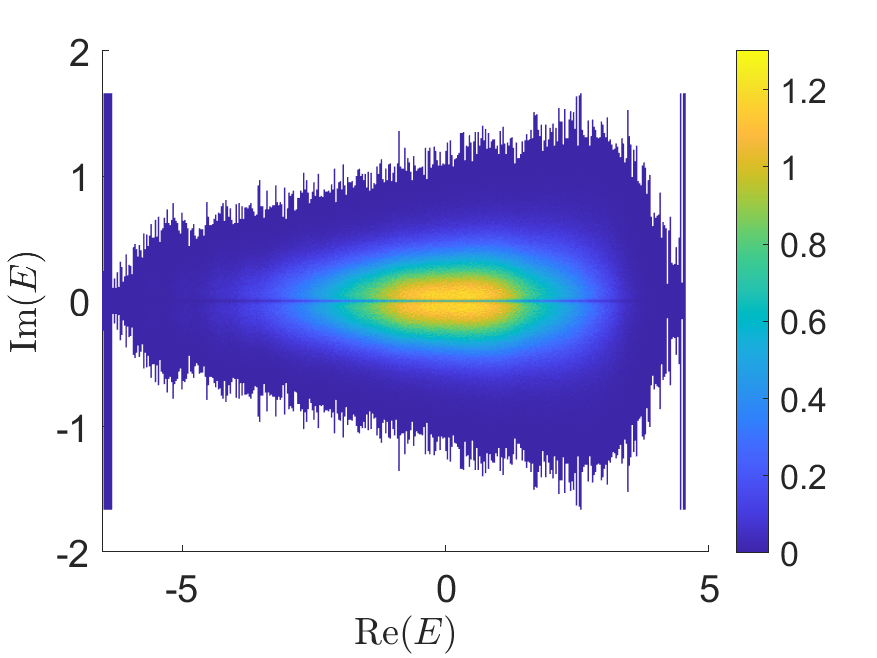}
		\end{minipage}%
	}%
	\subfigure[density $\rho_c(x,y)$ of complex eigenvalues (class AI + $\eta_-$)]{
		\begin{minipage}[t]{0.25\linewidth}
			\centering
			\includegraphics[width=1\linewidth]{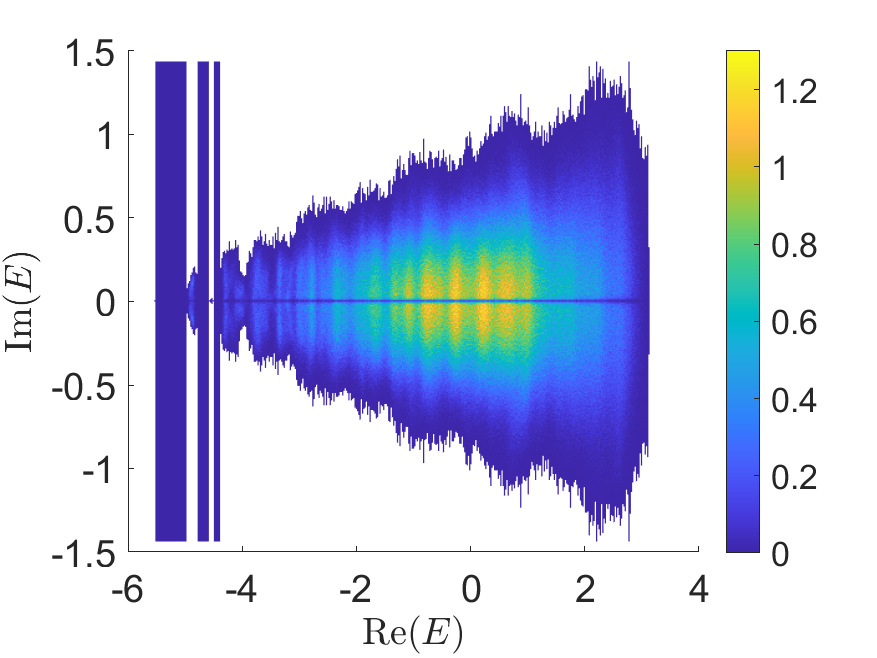}
		\end{minipage}%
	}%
	
	\subfigure[$\bar{\rho}_c(y)$ (class AI)]{
		\begin{minipage}[t]{0.25\linewidth}
			\centering
			\includegraphics[width=1\linewidth]{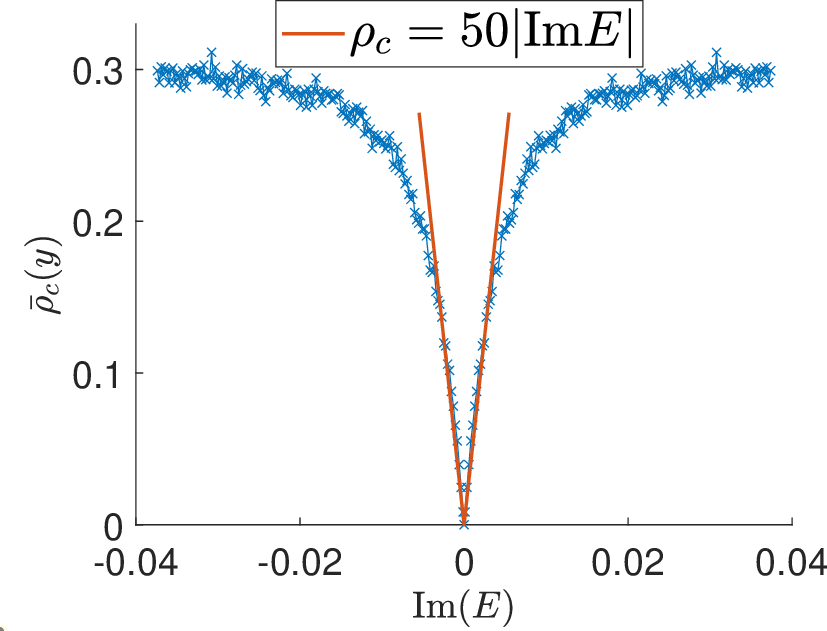}
			\label{spin_rho_AI}
		\end{minipage}%
	}%
	\subfigure[$\bar{\rho}_c(y)$ (class A + $\eta$)]{
		\begin{minipage}[t]{0.25\linewidth}
			\centering
			\includegraphics[width=1\linewidth]{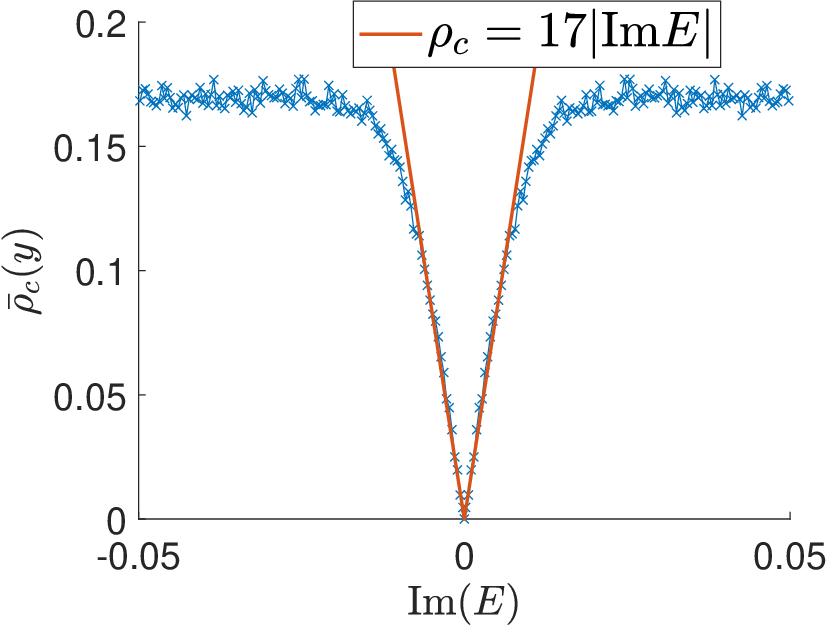}
		\end{minipage}%
 	}%
	\subfigure[$\bar{\rho}_c(y)$ (class AI + $\eta_+$)]{
		\begin{minipage}[t]{0.25\linewidth}
			\centering
			\includegraphics[width=1\linewidth]{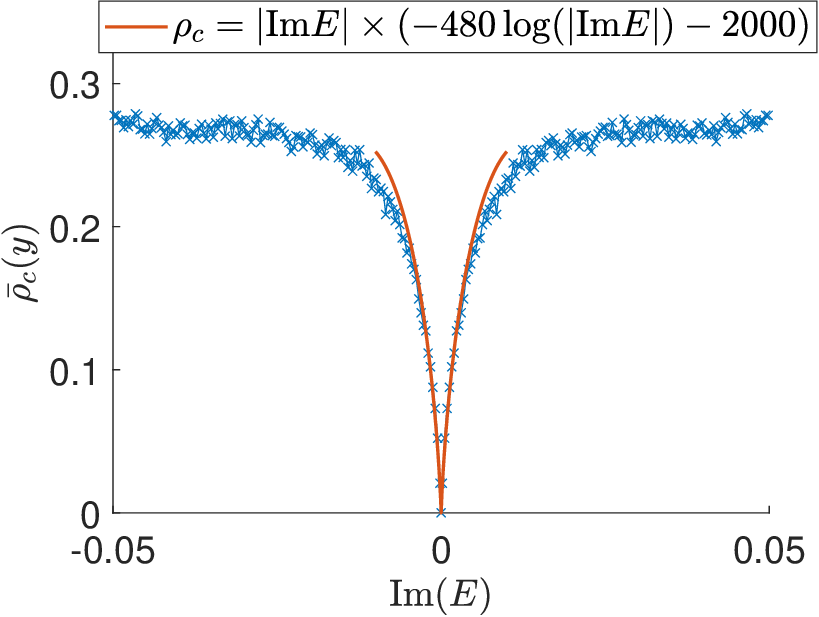}
		\end{minipage}%
	}%
	\subfigure[$\bar{\rho}_c(y)$ (class AI + $\eta_-$)]{
		\begin{minipage}[t]{0.25\linewidth}
			\centering
			\includegraphics[width=1\linewidth]{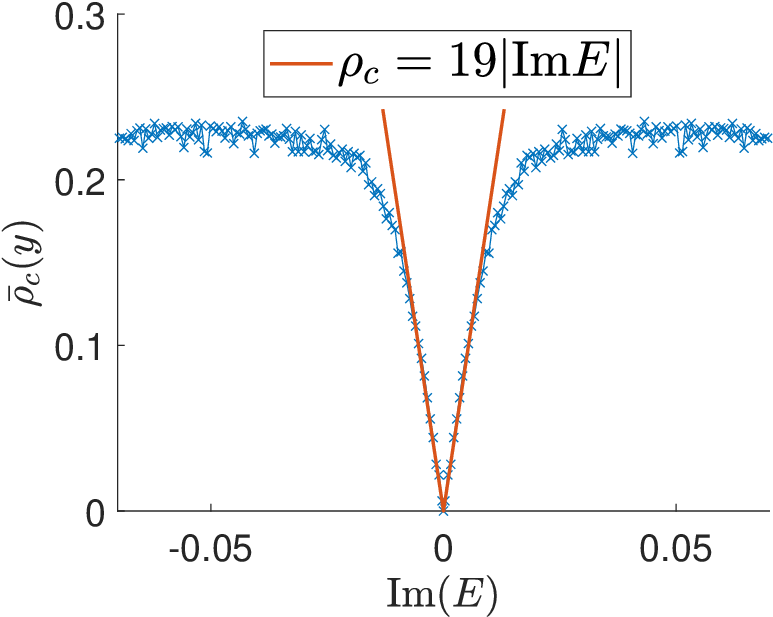}
			\label{spin_rho_DIII}
		\end{minipage}%
	}%
	
	\subfigure[$p(s)$ (class AI)]{
		\begin{minipage}[t]{0.25\linewidth}
			\centering
			\includegraphics[width=1\linewidth]{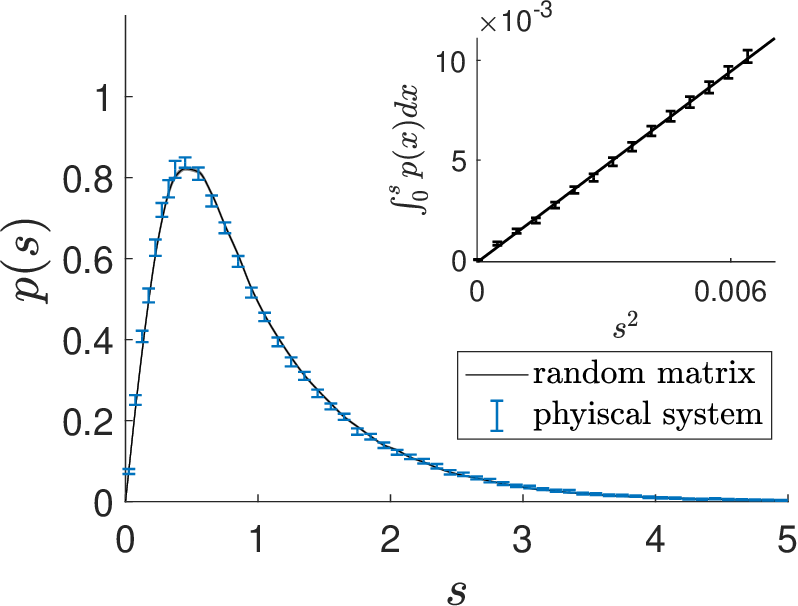}
		\label{spin_ps_AI}
		\end{minipage}%
	}%
	\subfigure[$p(s)$ (class A + $\eta$)]{
		\begin{minipage}[t]{0.25\linewidth}
			\centering
			\includegraphics[width=1\linewidth]{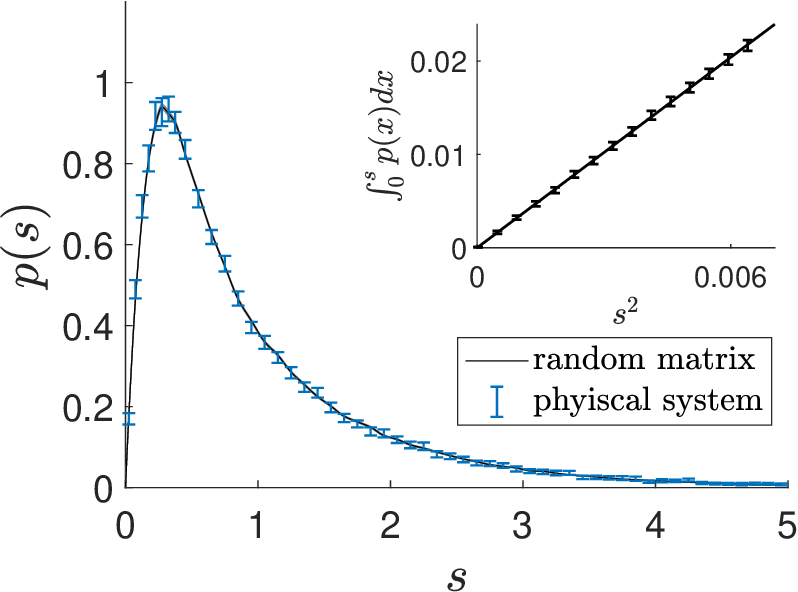}
		\end{minipage}%
	}%
	\subfigure[$p(s)$ (class AI + $\eta_+$)]{
		\begin{minipage}[t]{0.25\linewidth}
			\centering
			\includegraphics[width=1\linewidth]{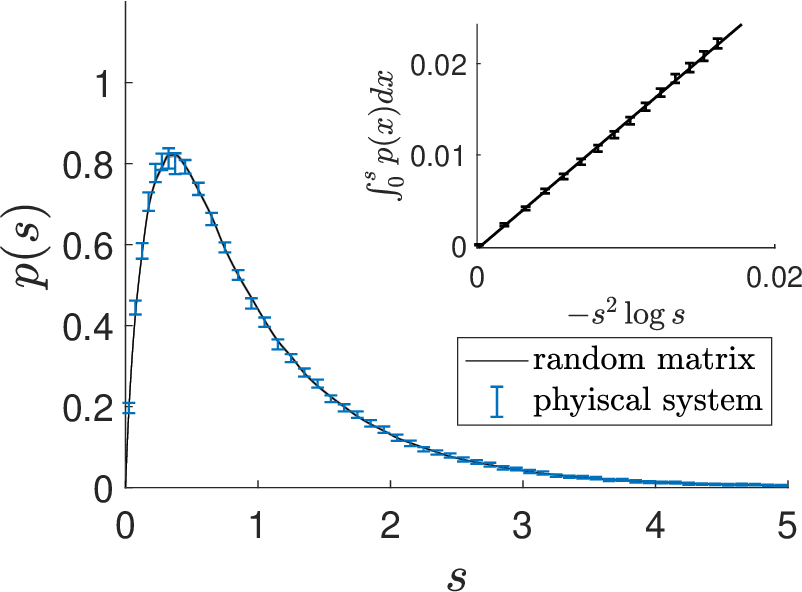}
		\end{minipage}%
	}%
	\subfigure[$p(s)$ (class AI + $\eta_-$)]{
		\begin{minipage}[t]{0.25\linewidth}
			\centering
			\includegraphics[width=1\linewidth]{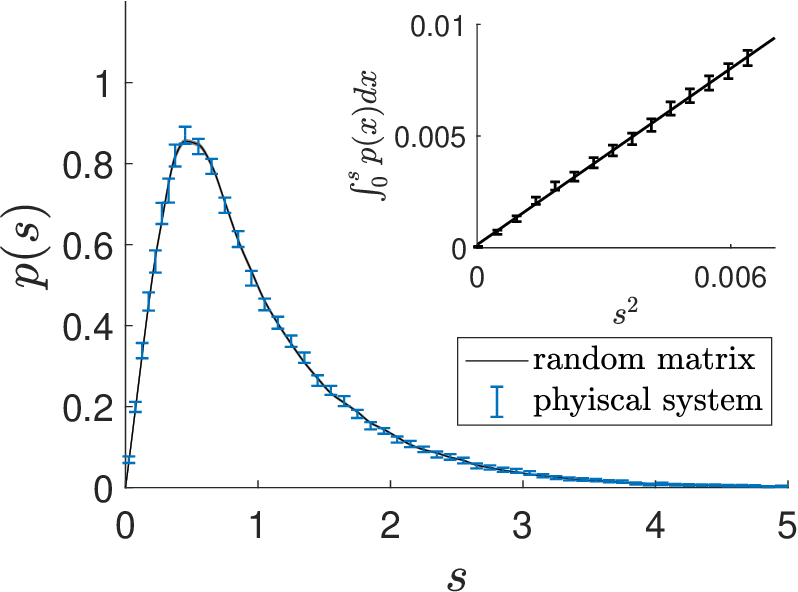}
		\label{spin_ps_DIII}
		\end{minipage}%
	}%

	\subfigure[$p_r(r)$ (class AI)]{
		\begin{minipage}[t]{0.25\linewidth}
			\centering
			\includegraphics[width=1\linewidth]{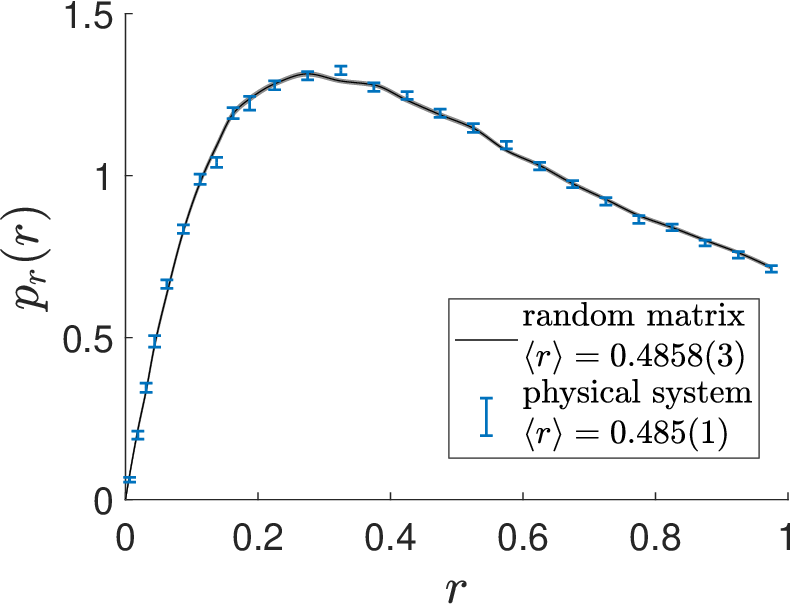}
			\label{spin_pr_AI}
		\end{minipage}%
	}%
	\subfigure[$p_r(r)$ (class A + $\eta$)]{
		\begin{minipage}[t]{0.25\linewidth}
			\centering
			\includegraphics[width=1\linewidth]{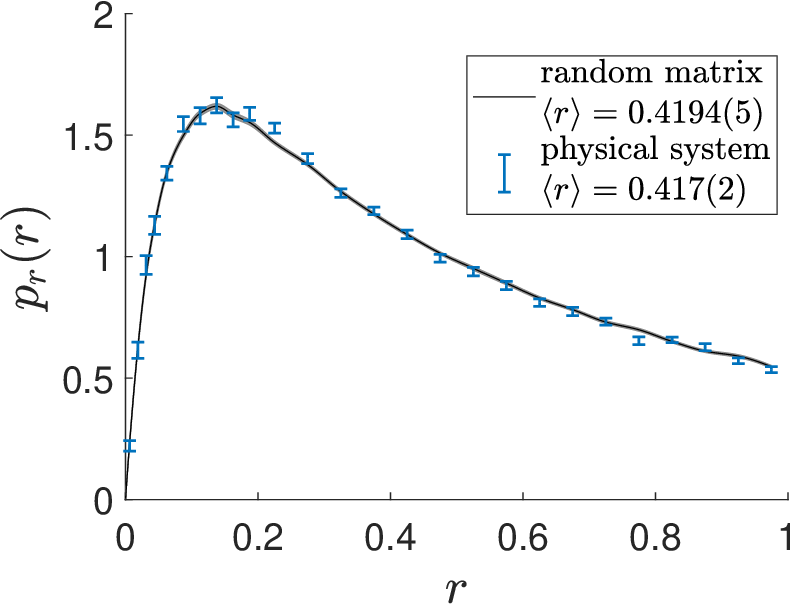}
		\end{minipage}%
		}%
    \subfigure[$p_r(r)$ (class AI + $\eta_+$)]{
		\begin{minipage}[t]{0.25\linewidth}
			\centering
			\includegraphics[width=1\linewidth]{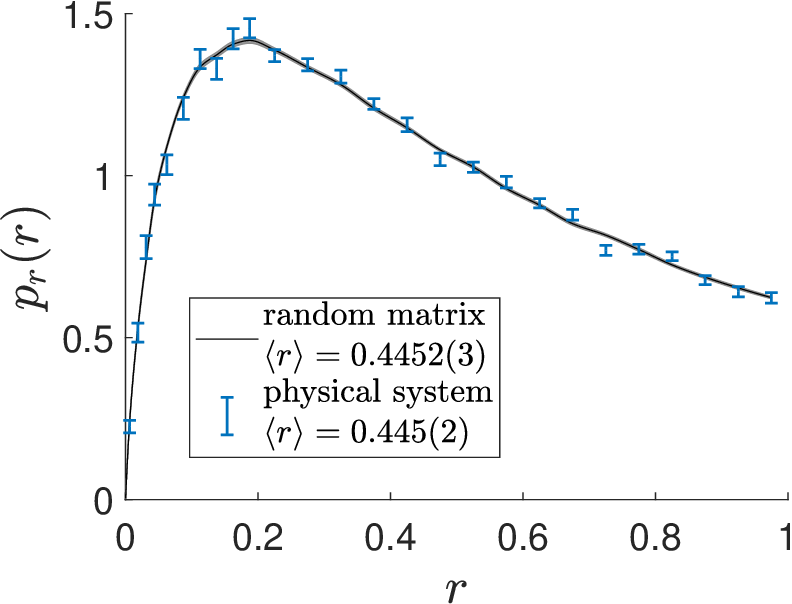}
		\end{minipage}%
	}%
	\subfigure[$p_r(r)$ (class AI + $\eta_-$)]{
		\begin{minipage}[t]{0.25\linewidth}
			\centering
			\includegraphics[width=1\linewidth]{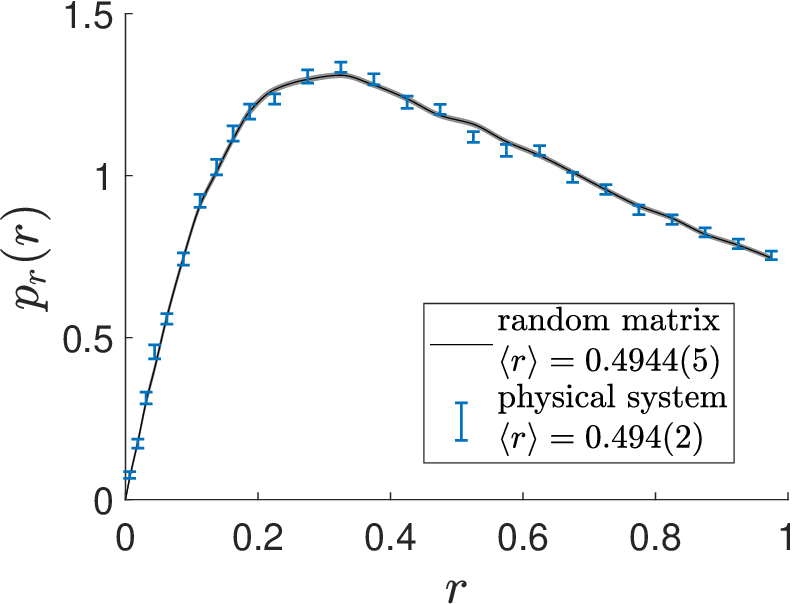}
			\label{spin_pr_DIII}
		\end{minipage}%
	}
	\caption{(a)-(d) Heat maps of the density $\rho_c(x,y)$ of complex eigenvalues of
	the non-Hermitian interacting spin models in 
	the 
	weak disorder regimes ($W_x=W_y=W_z=W_D=1$, $L_x = 13$) %
	for (a)~class AI, (b)~class A + $\eta$, (c)~class AI + $\eta_+$, and (d)~class AI + $\eta_-$. %
	(e)-(h)~Integrated density of complex eigenvalues, 
	$\bar{\rho}_c(y) = \frac{1}{5}\int_{-2.5}^{2.5} \rho_c(x , y) dx$, in the weak disorder regimes. 
	From the heat maps, $-2.5<E<2.5$ for $W_x=W_y=W_z=W_D=1$ is well within 
	the ergodic phase for all the spin models. 
	(i)-(l)~Level-spacing distributions $p(s)$ 
	of real eigenvalues in the weak disorder 
	regimes of the interacting spin models, and their comparison to $p(s)$ 
	from non-Hermitian random matrices in the same symmetry classes (red line). 
	Insets: Asymptotic behavior of $\int^s_0 p(s^{\prime}) ds^{\prime}$ for $s \ll 1$. 
	(m)-(p)~Level-spacing-ratio distributions $p_r(r)$ 
	of real eigenvalues in the weak 
	disorder regimes of the interacting spin models, and their comparison to $p_r(r)$ 
	from non-Hermitian random matrices in the same symmetry classes. 
	The mean value $\langle r \rangle = \int_0^1 p_r(r)dr$ of each level-spacing-ratio distribution is shown in the figures.
	The statistics are
	taken in the weak disorder regime 
  ($W_x=W_y=W_z=W_D=1$ and $-10<E<10$ in all the four spin models). 
	The consistency between $p(s)$ ($p_r(r)$) from the spin models and $p(s)$ ($p_r(r)$) from the random matrices justifies that all eigenstates with real energy $-10 < E < 10$ in the weak disorder regime are in the ergodic phases.
	(i)-(p)~The error ranges are evaluated by the bootstrap method~\cite{press07}.} 
\label{spin_dos_ps}
\end{figure*}

\begin{figure}[bt]
	\centering
	\includegraphics[width=0.9\linewidth]{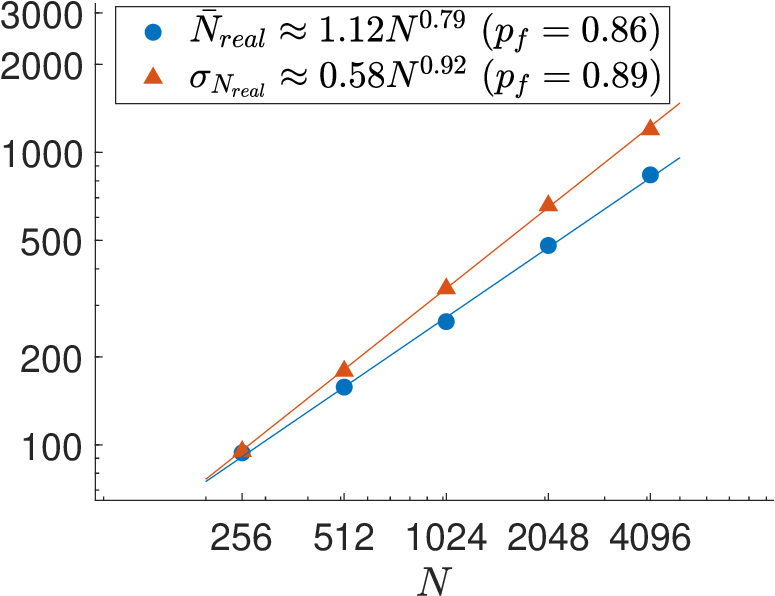}
	\label{n_real_AI_b}
	\caption{Average number $\bar{N}_{\rm real}$ of all the real eigenvalues  
	and its standard deviation $\sigma_{N_{\text{real}}}$ in the non-Hermitian
	interacting spin model ${\cal H}_1$ as functions of the dimensions $N$ 
	of the Hilbert space. 
    The Hermitian and anti-Hermitian disorder strengths are chosen to be on the same order ($W_x = W_z = 15, \,  W_y = 10$). 
	$p_f$ is an estimation of the probability 
	$p$ in %
	Eq.~(\ref{eq_std}) by the linear regression on $\log \bar{N}_{\rm real}$ with $\log N$ or $\log \sigma_{N_{\text{real}}}$ with $\log N$. 
    }
\label{n_real_std}
\end{figure}

\begin{figure}[bt]
	\centering
	\subfigure[$p(s)$ (MBL phase) ($W_x=W_z=20,W_y=2$)]{
		\begin{minipage}[t]{0.5\linewidth}
			\centering
			\includegraphics[width=1\linewidth]{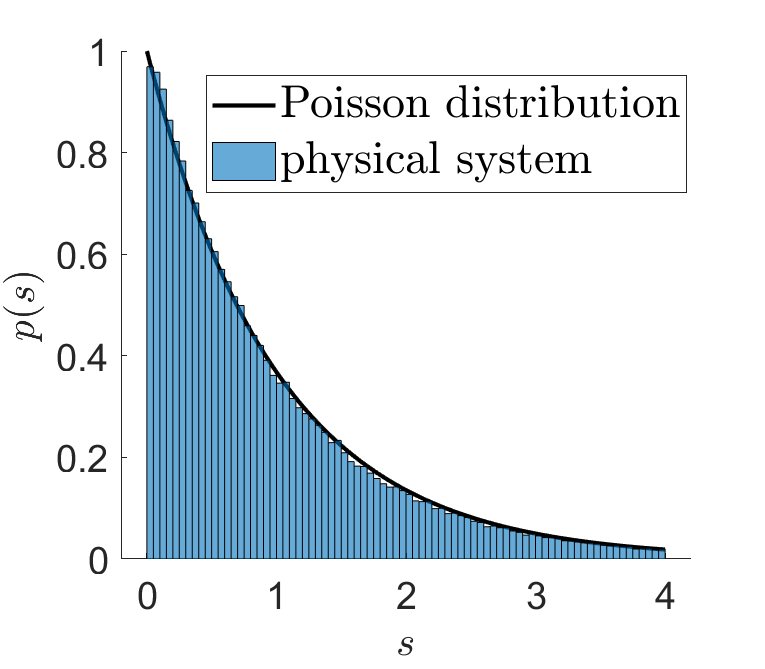}
			\label{ps_spin_AI_loca}
		\end{minipage}%
	}%
	\subfigure[$p(s)$ (MBL phase) ($W_x=W_z=20,W_y=10$) ]{
		\begin{minipage}[t]{0.5\linewidth}
			\centering
			\includegraphics[width=1\linewidth]{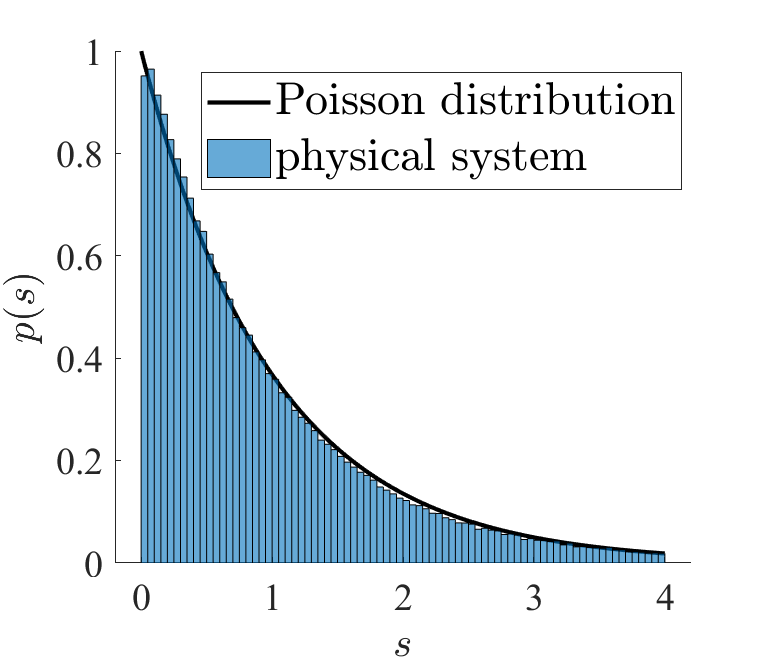}
		\label{ps_spin_AI_locb}
		\end{minipage}%
	}%

	\subfigure[$p_r(r)$ (MBL phase) ($W_x=W_z=20,W_y=2$)]{
		\begin{minipage}[t]{0.5\linewidth}
			\centering
			\includegraphics[width=1\linewidth]{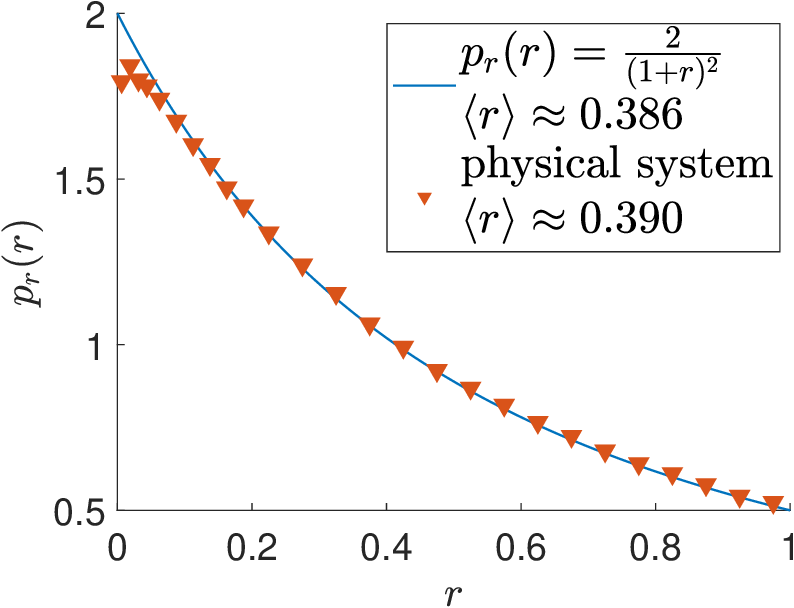}
			\label{pr_spin_AI_locc}
		\end{minipage}%
	}%
	\subfigure[$p_r(r)$ (MBL phase) ($W_x=W_z=20,W_y=10$) ]{
		\begin{minipage}[t]{0.5\linewidth}
			\centering
			\includegraphics[width=1\linewidth]{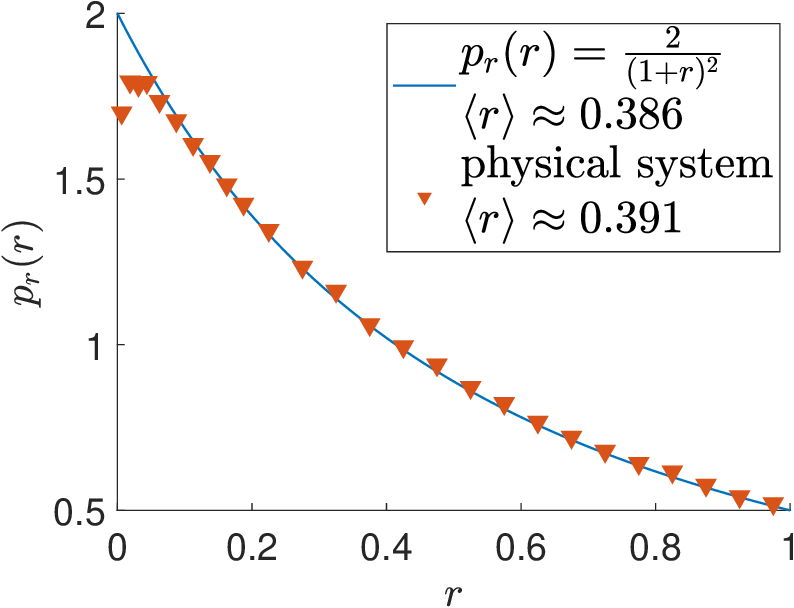}
		\label{pr_spin_AI_locd}
		\end{minipage}%
	}%
	\caption{ Level-spacing distributions $p(s)$ 
	of real eigenvalues in the strong disorder regimes of the non-Hermitian spin model ${\cal H}_1$ for (a) $W_x=W_z=20$, $W_y=2$ and (b) $W_x=W_z=15$, $W_y=10$. Level-spacing-ratio distributions $p_r(r)$ of real eigenvalues in the strong disorder regimes of ${\cal H}_1$ for (c) $W_x=W_z=20$, $W_y=2$ and 
	(d) $W_x=W_z=15$, $W_y=10$. The statistics are
	taken from all the real eigenvalues of ${\cal H}_1$.}
\label{ps_spin_AI_loc}
\end{figure}

We consider the following
1D Heisenberg spin models with random magnetic fields, random energy gain (loss), or random imaginary Dzyaloshinskii–Moriya (DM) interaction: %
\begin{equation}
\begin{aligned}
  {\cal H}_{\text{I}} &= \sum^{L_x}_{i=1} 
  J {\bm S}_i \cdot {\bm S}_{i+1}  \, ,\\
  {\cal H}_{1} &= {\cal H}_{\text{I}}  + \sum_i\Big\{ h_{x}^{(i)} S_x^{(i)} + {\rm i}h_{y}^{(i)} S_y^{(i)} + h_{z}^{(i)} S_z^{(i)} \Big\}  \, ,\\
   {\cal H}_{2} &= {\cal H}_{\text{I}}  + \sum_i \Big\{ 
   {\rm i} h_{x}^{(i)} S_x^{(i)} + {\rm i}h_{y}^{(i)} S_y^{(i)} 
   + h_{z}^{(i)} S_z^{(i)} \Big\} \, ,\\
     {\cal H}_{3} &= {\cal H}_{\text{I}}  + \sum_i \Big\{ 
     {\rm i}h_{y}^{(i)} S_y^{(i)} + h_{z}^{(i)} S_z^{(i)} \Big\}, 
\end{aligned}
\end{equation}
\begin{align}
    {\cal H}_4 & =  {\cal H}_{\text{I}}  + \sum^{L_x}_{i=1} \Big\{ {\rm i}D_{x}^{(i)} \left( S_y^{(i)} S_z^{(i+1)} - S_z^{(i)} S_y^{(i+1)} \right)  \nonumber \\
     & \ \  + {\rm i}D_{z}^{(i)} \left( S_y^{(i)} S_x^{(i+1)} - S_x^{(i)} S_y^{(i+1)} \right) \Big\} . \nonumber 
\end{align}
Here, ${\bm S}_i\equiv (S_x^{(i)},S_y^{(i)},S_z^{(i)}) \equiv \frac{1}{2}(\sigma_x^{(i)},  \sigma_y^{(i)},\sigma_z^{(i)} )$ is the spin-1/2
operators
at site $i$ with the periodic boundary conditions (i.e., ${\bm S}_{L_x+1}={\bm S}_{1}$).
The lattice site number %
$L_x$ %
for ${\cal H}_4$ is chosen to be an odd integer 
to respect TRS whose sign is $-1$ 
(see also Eq.~(\ref{eq: AI+eta- symmetry}) below).
$h_{x}^{(i)}$, $h^{(i)}_y$, and $h^{(i)}_z$ are real random numbers %
that 
describe random magnetic fields or random energy gain and loss. 
$h^{(i)}_{\mu}$ ($\mu=x,y,z$)
distributes independently %
and uniformly in 
$[-{W_{\mu}}/2,W_{\mu}/2]$ with $W_{\mu} \geq 0$.  $D_{x}^{(i)}$ and $D_{z}^{(i)}$ are real random numbers 
(random imaginary DM interactions) 
that distribute independently %
and uniformly 
$[-{W_{D}}/2,W_{D}/2]$ with $W_{D} \geq 0$. These non-Hermitian terms can be realized, 
for example, in continuously-measured %
cold atomic systems~\cite{lee2014heralded,daley2014quantum}.

The random spin model ${\cal H}_1$ satisfies 
\begin{equation}
    {\cal H}_1={\cal H}_1^*
\end{equation}
and
thus belongs
to %
symmetry 
class AI.  ${\cal H}_2$ satisfies
\begin{align}
{\cal H}_2 =\left(\prod_i^{L_x} \sigma_z^{(i)}\right) {\cal H}_2^{\dagger} \left(\prod_i^{L_x} \sigma_z^{(i)}\right) %
\end{align}
and belongs
to %
symmetry class A + $\eta$. 
${\cal H}_3$ satisfies 
\begin{align}
{\cal H}_3 &= {\cal H}_3^*, \\
{\cal H}_3 &= \left(\prod_i^{L_x} \sigma_z^{(i)}\right) {\cal H}_3^{\dagger} \left(\prod_i^{L_x} \sigma_z^{(i)}\right)
\end{align}
with 
\begin{align}
\left(\prod_i^{L_x} \sigma_z^{(i)}\right) = \left(\prod_i^{L_x} \sigma_z^{(i)}\right)^* ,
\end{align}
and thus
belongs to %
symmetry 
class AI + $\eta_+$. 
${\cal H}_4$ satisfies %
\begin{align}
{\cal H}_4 &= {\cal H}^*_4, \\
{\cal H}_4 &= \left(\prod_i^{L_x} \sigma_y^{(i)}\right) {\cal H}_4^{\dagger} \left(\prod_i^{L_x} \sigma_y^{(i)}\right). 
\end{align}
When $L_x$ is an odd integer, the unitary matrix satisfies 
\begin{align}
\left(\prod_i^{L_x} \sigma_y^{(i)}\right) = -\left(\prod_i^{L_x} \sigma_y^{(i)}\right)^*,  
    \label{eq: AI+eta- symmetry}
\end{align}
and hence the TRS operator and pH operator anti-commute with each other. Thus, ${\cal H}_4$ belongs to class AI + $\eta_-$ for odd $L_x$.

We study the weak disorder regimes of the four 1D Heisenberg models 
with the parameters $J = 1$, $W_x = W_y=W_z =W_{D}=1$ and with the
different system sizes (the maximal size is $L_x = 13$). 
From the DoS 
$\rho(E=x+{\rm i}y) = \rho_c(x,y) + \delta(y)\rho_r(x)$, 
the soft gap of $\rho_c(x, y) $ around the real axis $y=0$
is
universally observed with the same asymptotic %
behavior
for
small $y$   
(Fig.~\ref{spin_rho_AI}-\ref{spin_rho_DIII}). 
For different $L_x$ ($L_x$ 
of ${\cal H}_4$ changes only for odd numbers), the  
number $\bar{N}_\mathrm{real}$ of real eigenvalues
shows the square-root scaling with respect to the dimensions $N$ of 
the many-body Hamiltonians,
$\bar{N}_{\text{real}} \propto \sqrt{N}$
(Fig.~\ref{n_real_mb}),
which is consistent with the random matrix theory.
Both level-spacing statistics and level-spacing-ratio statistics
of real eigenvalues for each spin model show  
the same distributions as in the random matrices in the same 
symmetry class (Figs.~\ref{spin_ps_AI}-\ref{spin_ps_DIII} 
and \ref{spin_pr_AI}-\ref{spin_pr_DIII}). 
These results indicate that the square-root scaling 
of $\bar{N}_{\rm real}$ universally holds true in 
the ergodic phases of interacting disordered systems.
It should be noted that 
many-body eigenstates in the ergodic phases are 
extended in many-body Hilbert space, while 
single-particle eigenstates in the metal phase, which are studied in Sec.~\ref{sec_free},
are extended in the spatial coordinate space. 

Moreover, we study the 
strong disorder regimes of ${\cal H}_1$ in 
class AI 
with 
the
parameters  
$J = 1$, 
$W_x = W_z= 20$, $W_y=2$ or $J=1$, $15 \leq W_x = W_z \leq 40$, $W_y=10$. 
We find that
${\cal H}_2$ and ${\cal H}_3$ in 
the
strong disorder regimes %
show %
the level statistics 
of real eigenvalues
similar to those of
${\cal H}_1$ in the strong disorder regimes 
(not shown). 
$W_x$ and $W_z$ in ${\cal H}_1$ describe Hermitian local 
disorder while $W_y$ describes anti-Hermitian local disorder. 
When 
the Hermitian disorder %
dominates over the %
anti-Hermitian disorder
($W_x=W_z = 20$, $W_y =2$), 
almost all %
the eigenvalues are real, where we have $\bar{N}_{\text{real}} \propto %
N$. 
The level-spacings and level-spacing ratios
of real eigenvalues satisfy the Poisson distribution (Fig.~\ref{ps_spin_AI_loca}) and the distribution in Eq.~(\ref{pr_uncorrelated}) (Fig.~\ref{pr_spin_AI_locc}).

\begin{figure}[bt]
    \centering
    \includegraphics[width=0.9\linewidth]{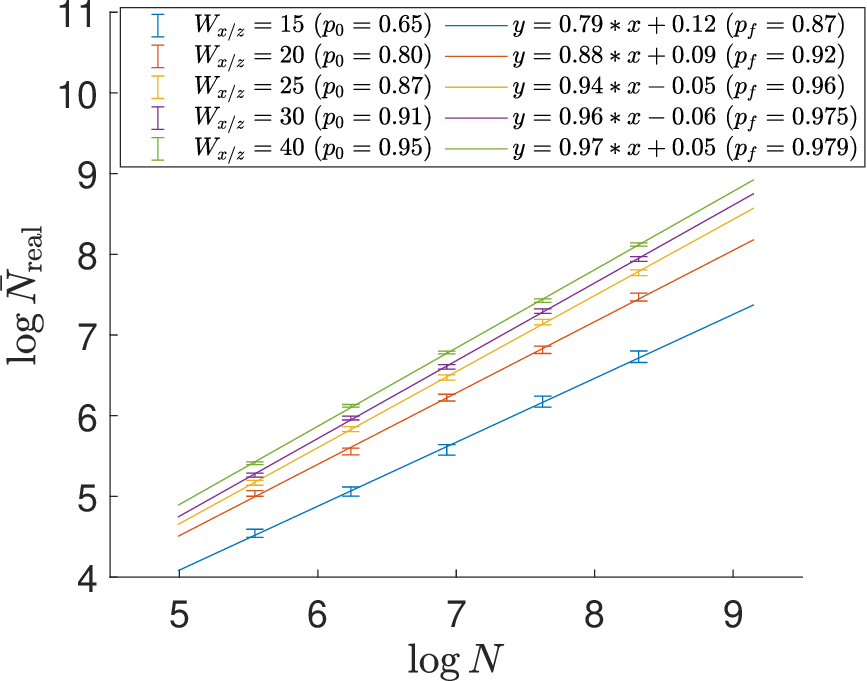}
    \caption{Average number $\bar{N}_{\rm real}$ of real eigenvalues as a function of the dimensions $N$ of the Hilbert space of the interacting spin model ${\cal H}_1$. 
    The Hermitian disorder strength and anti-Hermitian disorder strength are of the same order: 
    the anti-Hermitian disorder strength is fixed to $W_y = 10$, while the Hermitian disorder strength is
    changed from $W_{x/z} =15$ to $W_{x/z} =40$. 
    $p_0$ is the probability that an eigenvalue of the local disordered Hamiltonian $h_{x}^{(i)} \sigma_x^{(i)} + {\rm i}h_{y}^{(i)} \sigma_y^{(i)} + h_{z}^{(i)} \sigma_z^{(i)}$ becomes real.
	$p_f$ is an estimation of the probability 
	$p$ in Eq.~(\ref{eq_std}) by the linear regression on $\log \bar{N}_{\rm real}$ with $\log N$.
	}
    \label{n_real_spin_AI}
\end{figure}

When the anti-Hermitian disorder is of the same 
order as the Hermitian disorders ($W_y = 10$, $10 \leq W_x = W_z \leq 40$), the 
number $N_\text{real}$
of real eigenvalues fluctuates largely from sample to sample.  
The standard deviation of $N_\text{real}$,
$\sigma^2_{N_\text{real}} \equiv \langle N_\text{real}^2  \rangle - \langle N_\text{real} \rangle^2$, 
grows exponentially with the system size %
$L_x$, %
and 
$\sigma_{N_{\rm real}}$ is much 
larger than 
$\bar{N}_\text{real} \equiv \langle N_\text{real} \rangle$ 
(Fig.~\ref{n_real_std}).  
Notably, we find that the scalings of $\bar{N}_{\text{real}}$ and $\sigma_{N_{\rm real}}$ with respect 
to the dimensions $N$ of the Hamiltonian are characterized by non-universal 
powers, such as ${\bar N}_{\text{real}} \sim N^{\alpha}$ (see Fig.~\ref{n_real_spin_AI}).
The powers of both 
$\bar{N}_{\rm real}$ and $\sigma_{N_{\rm real}}$ increase when the Hermitian disorders 
become larger.

The non-universal powers $\alpha$ in the scalings of $\bar{N}_{\text{real}}$ and $\sigma_{N_{\rm real}}$ 
can be explained with a hypothesis that the MBL phase in the 
non-Hermitian case exhibits an emergent integrability as in the Hermitian 
case~\cite{Abanin19,Serbyn2016}. Suppose that the many-body non-Hermitian Hamiltonian 
in the MBL phase can be effectively expanded in terms of $L_x$ 
mutually-commuting %
bit 
operators $\tau^{(i)}_z$ ($i=1,2,\cdots, L_x$) as 
\begin{equation}
\begin{aligned}
    {\cal H}_1^{\text{MBL}} &= \sum^{L_x}_{i=1} \tau_z^{(i)} + 
    \sum_{i,j} J_{ij} \tau_z^{(i)} \tau_z^{(j)} \\
    &+ \sum_{ijk} K_{ijk} \tau_z^{(i)}\tau_z^{(j)}\tau_z^{(k)} + \cdots \, .
\end{aligned}
\end{equation}
Here, we have $[\tau^{(i)}_z,\tau^{(j)}_z]=[{\cal H}^{\rm MBL}_1,\tau^{(i)}_z]=0$ for 
all $i$ and $j$. 
The bit operator $\tau^{(i)}_z$ is %
a two-by-two
non-Hermitian matrix.  
${\cal H}_1^{\text{MBL}}$ %
respects TRS and belongs to class AI.
Given that the coefficients such as $J_{ij}$ and $K_{ijk}$ are 
real numbers, the bit operator must 
also
be real $(\tau^{(i)}_z)^*=\tau^{(i)}_z$: %
\begin{align}
    \tau^{(i)}_z = \left(\begin{array}{cc}
    a^{(i)}_0 + a^{(i)}_1 & a^{(i)}_2 + a^{(i)}_3 \\
    a^{(i)}_2 - a^{(i)}_3 & a^{(i)}_0 - a^{(i)}_1 \\
    \end{array}\right) %
\end{align}
with real numbers $a^{(i)}_{\alpha}$ ($\alpha=0,1,2,3$). The real numbers of 
different $i$ and different components are 
almost independent of one another in the strong disorder regime.
When ${\cal H}_1$ is dominated by the on-site random terms 
($W_x,W_y,W_z \gg J$), the bit operator 
is given by the random magnetic fields at each lattice site, 
$a^{(i)}_0=0$, $a^{(i)}_1=h^{(i)}_z/2$, $a^{(i)}_2=h^{(i)}_x/2$, and 
$a^{(i)}_3=h^{(i)}_y/2$. For $(a^{(i)}_1)^2+(a^{(i)}_2)^2>(a^{(i)}_3)^2$, 
the bit operator $\tau^{(i)}_z$ has real eigenvalues, 
\begin{align}
    \lambda^{(i)}_{\pm 1} &= a_0  \pm \sqrt{(a^{(i)}_1)^2+(a^{(i)}_2)^2-(a^{(i)}_3)^2}. \nonumber 
\end{align}
For $(a^{(i)}_1)^2+(a^{(i)}_2)^2<(a^{(i)}_3)^2$, 
the bit operator has complex eigenvalues, 
\begin{align}
    \lambda^{(i)}_{\pm {\rm i}} &= a_0 \pm {\rm i} \sqrt{(a^{(i)}_3)^2-(a^{(i)}_1)^2-(a^{(i)}_2)^2}. \nonumber 
\end{align}
From them, a many-body eigenvalue of 
${\cal H}^{\rm MBL}_1$ is given 
by 
\begin{align}
 E(\{\beta_{j}\}) = \sum^{L_x}_{j=1} 
 \lambda^{(j)}_{\beta_{j}} + \sum_{i,j} 
 J_{ij} \lambda^{(i)}_{\beta_{i}} \lambda^{(j)}_{\beta_{j}} + \cdots 
 \label{E_mbl}
\end{align}
with $\beta_j=\pm 1,\pm {\rm i}$ for $j=1,2,\cdots,L_x$.

\begin{figure}[bt]
    \centering
    \includegraphics[width=0.9\linewidth]{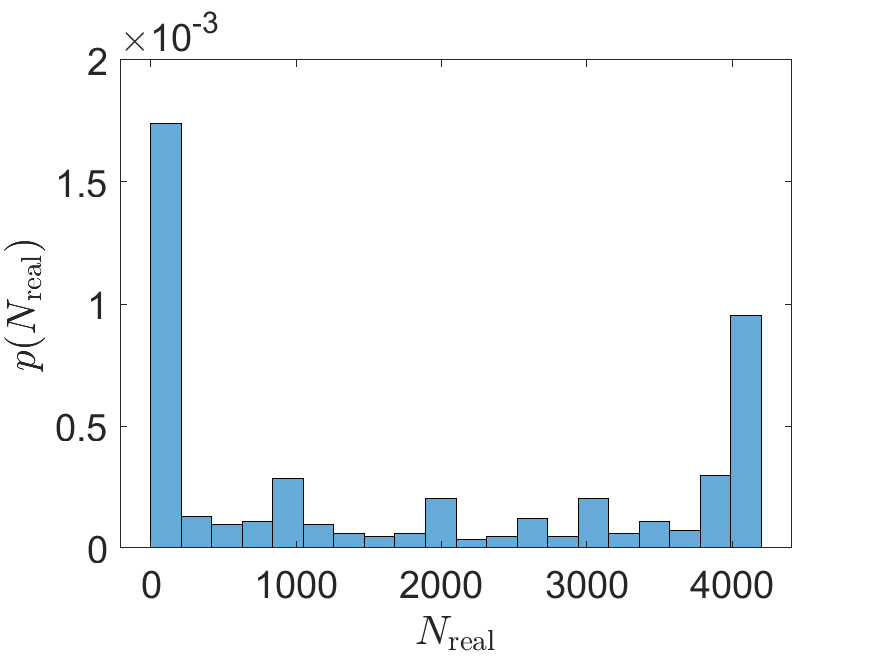}
    \caption{Probability distribution of the number $N_{\rm real}$ of real eigenvalues of the interacting spin model 
    ${\cal H}_1$ in the strong disorder regime ($W_x = W_z = 20, \, W_y = 10$). The distribution is obtained from the diagonalization of $400$ samples with the system size 
    $L_x = 12$. 
    The dimensions of the Hilbert space are $N = 2^{L_x} = 4096$. 
    The two peaks, $N_{\rm real} = 0$ and $N_{\rm real} = N$, 
    appear in the distribution.}
    \label{N_real_dis}
\end{figure}

Let $p$ be a probability of the bit operator of $i$ having real 
eigenvalues. The probability is independent of $i$ 
and two bit operators at different $i$ and $j$ 
are uncorrelated with each other.
Thus, a probability of a given 
many-body eigenvalue being real-valued 
equals a probability of all $\lambda_{\beta_i}^{(i)}$ being real, which is $p^{L_x}$. 
The average and standard deviation of the %
number 
of real eigenvalues
are estimated as %
 \begin{equation}
 \begin{aligned}
  \bar{N}_{\text{real}}  & = \left( 2p \right)^{L_x}, \\
  \sigma_{ N_{\text{real}} } &= \sqrt{\langle N_{\text{real}}^2  \rangle -  {\bar N}_{\text{real}}^{2}} =  2^{L_x} \sqrt{p^{L_x}-p^{2L_x}} \\
  & \simeq (2\sqrt{p})^{L_x} %
 \quad %
 \end{aligned}
 \end{equation}
 for $L_x \gg 1$ and $p < 1$.
Here, $\langle ... \rangle$ means the average over different disorder realizations. 
These evaluations lead to the scalings,  
\begin{equation}
\begin{aligned}
         \bar{N}_{\rm real}& \sim N^{\alpha} , \, \alpha = 1 + \log_2 p < 1, \\ 
         \sigma_{N_{\text{real}}} & \sim N^{\beta} , \, \beta = 1 + \frac{1}{2}\log_2 p < 1 \, .
\end{aligned}
\label{eq_std}
\end{equation}
Note that in the MBL phase with $p<1$, 
$\sigma_{ N_{\text{real}}}$ is much 
larger than $\bar{N}_{\text{real}}$ 
for large $N$, being consistent with the 
numerical observation (Fig.~\ref{n_real_std}). 
When the on-site disorders in 
${\cal H}_1$ become %
much more dominant than
the Heisenberg interaction $J$,
the probability $p$ in the scaling forms can be determined  
only
by $W_x$, $W_y$, and $W_z$ 
(see $p_0$ in the caption of Fig.~\ref{n_real_spin_AI}).
The numerical 
data with the finite disorder strengths are 
approximately well fitted by Eq.~(\ref{eq_std}) with similar 
values of $p$ (Fig.~\ref{n_real_spin_AI}). 
In the strong disorder regime, the probability distribution of $N_{\rm real}$ shows two peaks around $N_{\rm real} = 0$ (i.e., none of eigenvalues are real) and $N_{\rm real} = N$ (i.e., all eigenvalues are real) (Fig.~\ref{N_real_dis}), which %
are
also consistent with 
the
phenomenological explanation 
in
Eq.~(\ref{E_mbl}).

The %
level-spacing statistics 
of real eigenvalues
show %
the Poisson distribution 
in 
the
strong disorder regime. 
To illustrate this %
with 
$\sigma_{N_{\rm real}} \gg \bar{N}_{\rm real}$, we unfold the %
level-spacing of many-body eigenvalues %
by the density $\rho_c(x,y)$ of complex eigenvalues in %
each
sample 
of different disorder realizations,
\begin{align}
s_i &\equiv  (\lambda_{i+1} - \lambda_i) \!\ \bar{\rho}^{(k)}_r\Big(\frac{\lambda_{i+1}+\lambda_i}{2}\Big) 
\nonumber 
\end{align} 
with 
\begin{align}
\bar{\rho}^{(k)}_{r}(x) &\equiv \frac{N^{(k)}_{\rm real}}{\bar{N}_{\rm real}}
\sum_{%
\lambda_i
\in {\cal R}} \langle \delta\big(x-%
\lambda_i
\big) \rangle.  
\end{align}
Here, $\lambda_i$ ($i=1,2,\cdots, 2^L_x$) stands 
for the many-body eigenvalues in the descending order
and $N^{(k)}_{\rm real}$ is %
the number of the 
real eigenvalues in the $k$-th sample. The real-eigenvalue 
spacing thus normalized shows the Poisson distribution in the 
strong disorder regime with $\sigma_{N_{\rm real}} \gg N_{\rm real}$ (Fig.~\ref{ps_spin_AI_locb}). 
The spacing-ratio statistics over samples with very different 
numbers of real eigenvalues inevitably increase statistical errors 
(Fig.\ref{pr_spin_AI_locd}).

\section{Dissipative free fermions}
\label{sec_free}

\begin{figure*}[tb]
	\centering
	\subfigure[density $\rho_c(x,y)$ of complex eigenvalues ($W_1 = 3, W_2 = 1$)]{
		\begin{minipage}[t]{0.3\linewidth}
			\centering
			\includegraphics[width=1\linewidth]{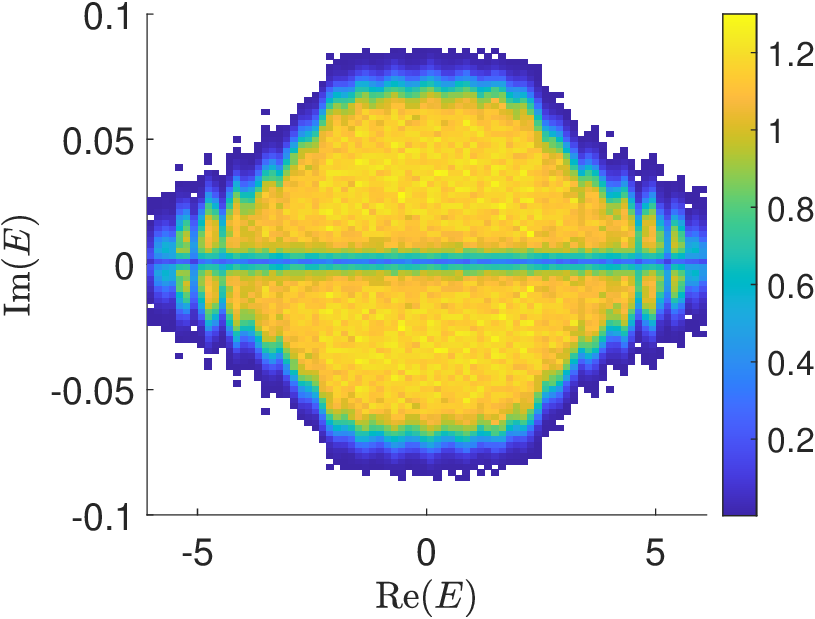}
		\end{minipage}%
		\label{bdid_3d_DoS}
	}%
      \subfigure[$\bar{\rho}_c(y)$   ($W_1 = 3, W_2 = 1$)]{
		\begin{minipage}[t]{0.3\linewidth}
			\centering
			\includegraphics[width=1\linewidth]{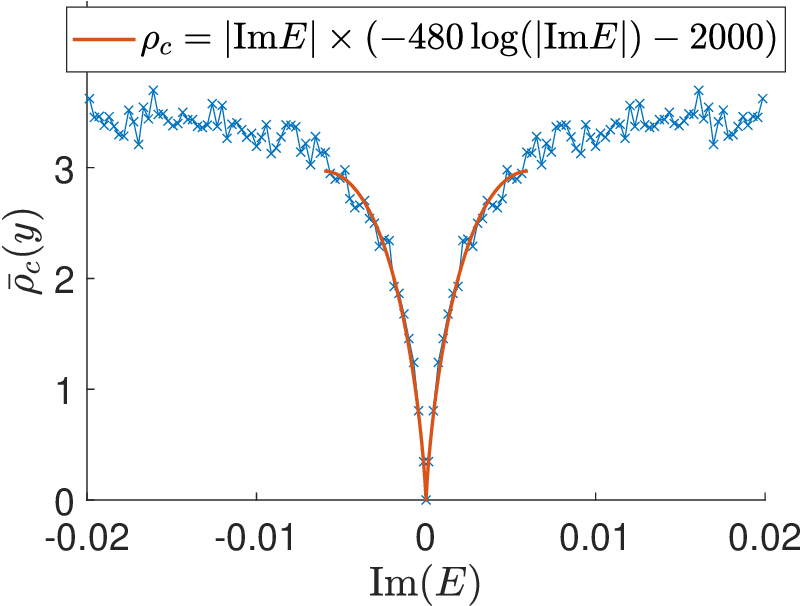}
		\label{bdid_3d_rho_imE}
		\end{minipage}%
	}%
	\subfigure[$p(s)$ ($W_1 = 3, W_2 = 1$, metal phase)  ]{
		\begin{minipage}[t]{0.3\linewidth}
			\centering
			\includegraphics[width=1\linewidth]{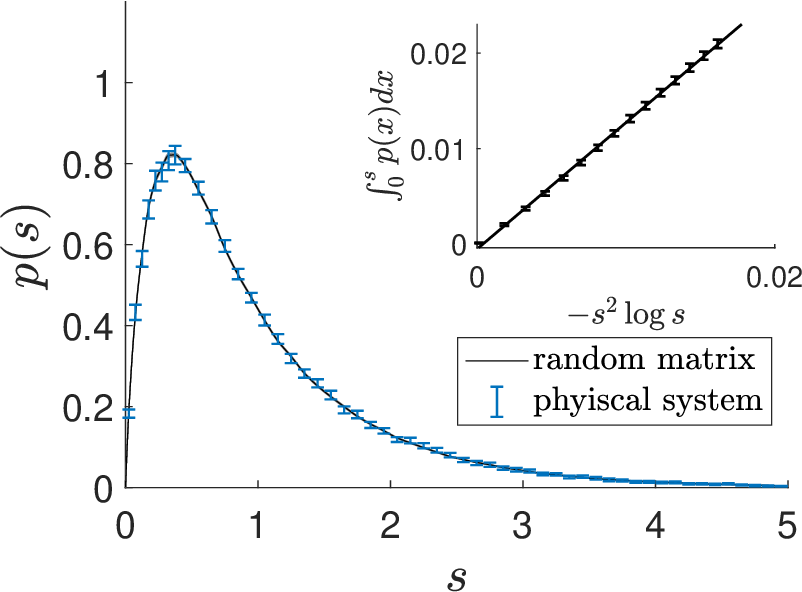}
		\label{bdid_3d_ps}
		\end{minipage}%
	}

	\subfigure[$p(s)$ ($W_1 = 60, W_2 = 60$, localized phase)]{
		\begin{minipage}[t]{0.24\linewidth}
			\centering
			\includegraphics[width=1\linewidth]{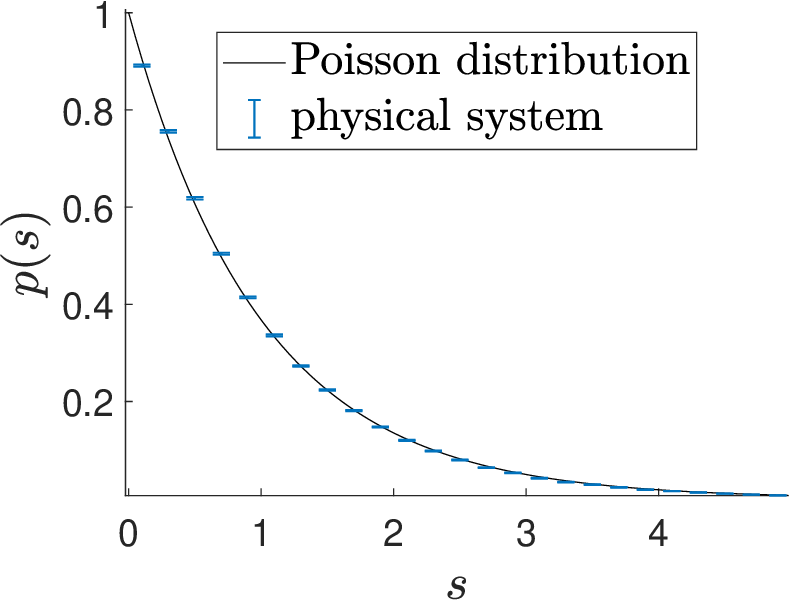}
		\label{bdid_3d_ps_loc}
		\end{minipage}%
	
	}
	\subfigure[$\langle r \rangle-N$  ($W_1 = 3, W_2 = 1$, metal phase)]{
		\begin{minipage}[t]{0.24\linewidth}
			\centering
			\includegraphics[width=1\linewidth]{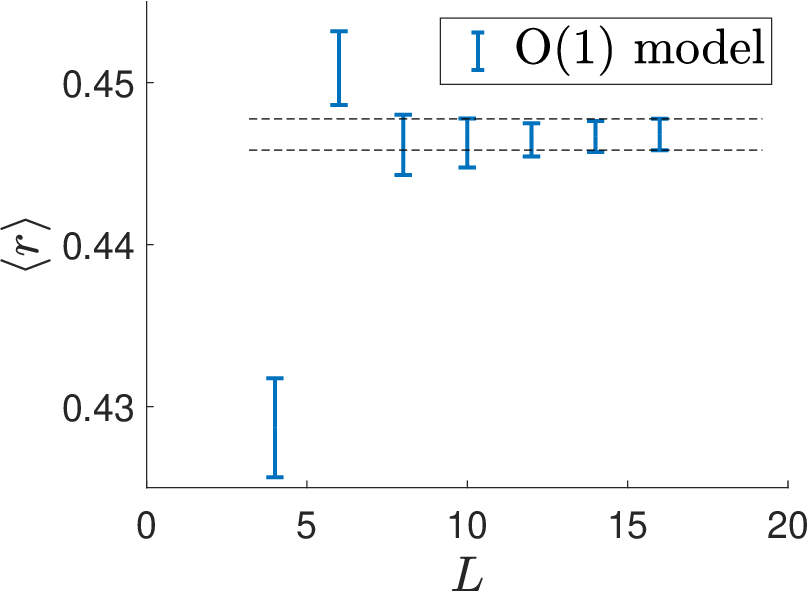}
		\label{bdid_meanr_L}
		\end{minipage}%
	
	}
	\subfigure[ $p_r(r)$ ($W_1 = 3, W_2 = 1$, metal phase) ]{
		\begin{minipage}[t]{0.24\linewidth}
			\centering
			\includegraphics[width=1\linewidth]{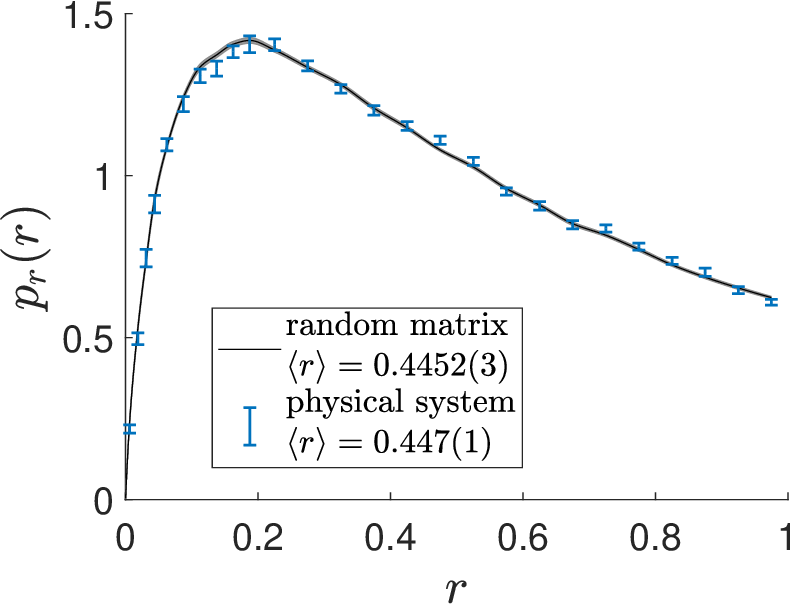}
			\label{pr_free_3d}
		\end{minipage}%
	}%
	\subfigure[$p_r(r)$ ($W_1 = 60, W_2 = 60$, localized phase)]{
		\begin{minipage}[t]{0.24\linewidth}
			\centering
			\includegraphics[width=1\linewidth]{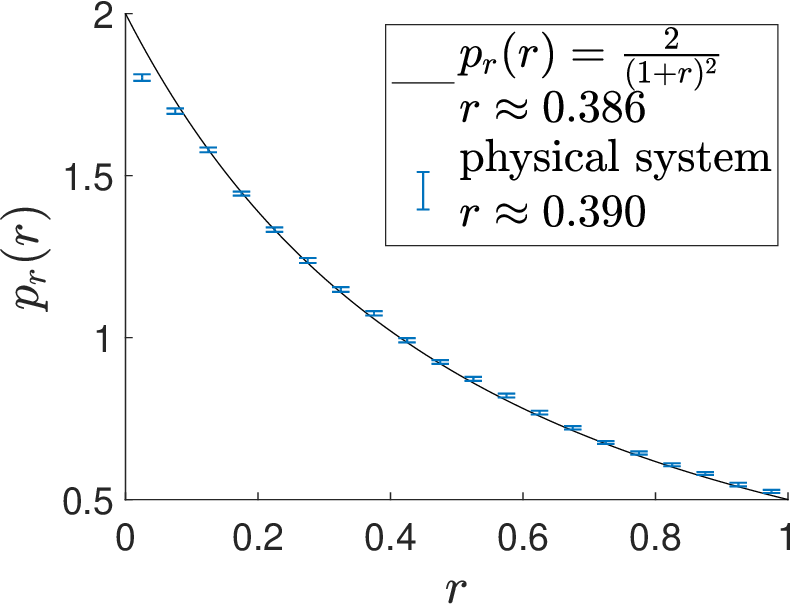}
			\label{pr_free_3d_loc}
		\end{minipage}%
	
	}
	\caption{(a) Heat map of the density $\rho_c(x,y)$ of complex eigenvalues in 3D class AI + $\eta_{+}$ for the
	weak disorder regime ($W_1=3$, $W_2=1$, $16\times 16 \times 16$ sites). (b) Integrated density of complex eigenvalues,
	$\bar{\rho}_c(y) \equiv  \frac{1}{2 x_1}\int_{-x_1}^{x_1} \rho_c(x,y) dx$, around the real axis $y = \mathrm{Im} (E) = 0$ with $x_1 =2$. 
	(c)~Level-spacing distribution $p(s)$ 
	of real eigenvalues in the metal phase 
	($\lvert E \rvert < 4$, $W_1=3$, and $W_2=1$), and its 
	comparison to $p(s)$ from non-Hermitian random matrices 
	in the same symmetry class (black line). Inset: Asymptotic 
	behavior of $\int^s_0 p(s^{\prime}) ds^{\prime}$ for 
	$s \ll 1$ in the metal phase. 
    (d)~$p(s)$ in the
	localized phase of the 3D model (all the real 
	energy $E$, $W_1 = W_2 = 60$) and its comparison to the 
	Poisson distribution.
	(e)~Mean value of level-spacing ratios $\langle r \rangle = \int_0^1 p_r(r)dr$ as a function of system size $L$.
	(f)~Level-spacing-ratio distribution $p_r(r)$ of real eigenvalues in the metal phase ($\lvert E \rvert < 4$, $W_1=3$, $W_2=1$) and its comparison to $p_r(r)$ from non-Hermitian
	random matrices in the same symmetry class. 
	(g)~$p_r(r)$ of real eigenvalues in the localized phase (all real energy $E$, $W_1 = W_2 = 60$) and its comparison to $p_r(r)$ of uncorrelated real numbers. 
	The mean value $\langle r \rangle$ of each level-spacing-ratio distribution is shown in the figures.  
    (c)-(g)~The error ranges are evaluated by the bootstrap method~\cite{press07}.
	}
\label{bdid_3d_ps_DoS}
\end{figure*}

In the previous section, we demonstrate the universal 
level statistics
of real eigenvalues in the ergodic phases of
the bosonic many-body Hamiltonians.
In this section, we study
non-interacting
fermionic
Hamiltonians with disorder and non-Hermiticity that belong to symmetry classes AI + $\eta_{+}$ and AII + $\eta_{+}$.
We calculate the DoS, the %
level-spacing 
statistics, and the %
number 
of real eigenvalues
in the weak and strong disorder regimes. %
In both %
regimes, the DoS has a delta function peak 
on the real axis, $\rho(E\equiv x+{\rm i}y)=\rho_c(x,y)+
\delta(y) \rho_r(x)$. 
In the metal phases, 
we demonstrate that 
the number of real eigenvalue is scaled by the square root of the dimensions of the Hamiltonians, being consistent with the random matrix theory. 
We also find that the real-eigenvalue spacings and spacing ratios for class AI + $\eta_+$ show the 
same distributions as those of the random matrices in the same symmetry class while we find discrepancies %
for class AII + $\eta_+$. 
We discuss possible reasons for these discrepancies. 
In the localized phases, 
by contrast, the level-spacings of real eigenvalues show the Poisson 
distribution, and the number of real eigenvalues is linearly  
scaled by the dimensions of the Hamiltonians. 

\subsection{3D class AI + $\eta_{+}$} %

\begin{figure}[bt]
    \centering
	\includegraphics[width=0.9\linewidth]{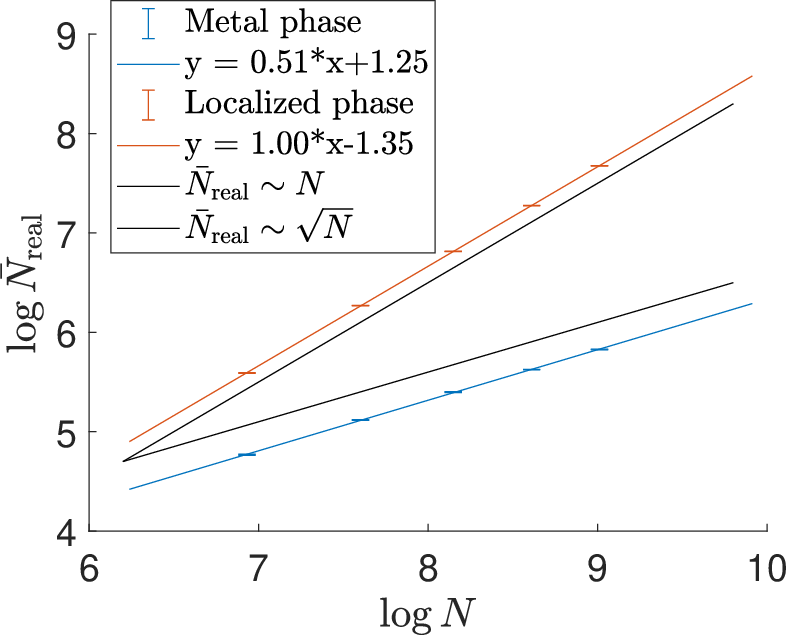}
    \caption{Average number 
    $\bar{N}_{\rm real}$
    of real eigenvalues as a function of the dimensions $N$
    of the non-Hermitian disordered Hamiltonian in 3D class AI + $\eta_{+}$.
    The lines with the different colors are for the different disorder strengths. 
    The blue line is for the metal phase in the weak disorder regime ($|E| \leq 4$, $W_1=3$, and $W_2=1$). The red line is for the
    localized phase in the strong disorder regime (all the real energy $E$, $W_1=W_2=60$). For reference, the black lines are $\bar{N}_{\rm real} \propto N $ and $\bar{N}_{\rm real} \propto \sqrt{N}$, respectively.}
    \label{n_real_bdi_3d}
\end{figure}

We study a %
non-Hermitian 
extension of 
the
Anderson
model 
on 
the
three-dimensional
(3D) %
cubic lattice:
\begin{align}
    {\cal H}_{\text{3D}} &=  \sum_{i} \left( c_{i}^{\dagger} ( \varepsilon_i \sigma_0 + \varepsilon^{\prime}_i \sigma_z) c_{i} + 
    {\rm i} \omega_{i} c_{i}^{\dagger}\sigma_y c_{i} \right) + t \sum_{\langle i, j\rangle} c_{i}^{\dagger} \sigma_0 c_{j}.\label{H_bdid}
\end{align}
Here, $\varepsilon_{i}$ and $\varepsilon_{i}^{\prime}$ describe 
the
Hermitian %
disordered potentials that
distribute independently %
and uniformly in %
$[-W_1/2,W_1/2]$, and 
$\omega_i$ describes %
the anti-Hermitian disordered potential that
distributes 
uniformly in $[-W_2/2,W_2/2]$. 
The non-Hermitian Hamiltonian ${\cal H}_{\text{3D}}$ satisfies %
TRS %
\begin{equation}
{\cal H}_{\text{3D}} = {\cal H}_{\text{3D}}^*,
\end{equation}
and %
pH
\begin{equation}
    {\cal H}_{\text{3D}} = \sigma_z {\cal H}_{\text{3D}}^{\dagger} \sigma_z,
\end{equation} 
where the TRS operator and the 
pH operator commute with each other. 
Thus, this 
model belongs to %
symmetry 
class AI + $\eta_+$.  

We investigate the weak (strong) disorder regime
with 
the
parameters %
$t = 1$, $W_1=3$, $W_2=1$ 
($W_1 = 60$, $W_2 = 60$) and with the periodic boundary conditions. 
We diagonalize ${\cal H}_{\rm 3D}$ with 
$240$ different disorder realizations 
with different system sizes 
(the maximal system size is $16 \times 16 \times 16$). 
We find that
eigenstates %
with
real energy $E$ undergo 
the Anderson transition in the weak disorder regime. 
An energy region near $E=0$ ($\left| E \right| < 4$) is in 
the
metal and localized phases 
in the weak and strong disorder regimes, respectively. 
We calculate
the DoS, the %
level-spacing 
distribution, and the number  
of real eigenvalues in the weak and strong disorder regimes. 
In the weak disorder regime, 
$\rho_c(x,y)$ shows a soft gap %
$\rho_c(x,y) \propto - \lvert y \rvert \log \lvert y \rvert$
around the real axis $y=0$ 
(Figs.~\ref{bdid_3d_DoS} and \ref{bdid_3d_rho_imE}), sharing the 
same scaling as in the random matrix theory for 
symmetry class AI + $\eta_{+}$. 
The mean value of the level-spacing ratios converges 
for the system size $L \geq 8$ (Fig.~\ref{bdid_meanr_L}).
The level-spacing distribution 
$p(s)$ and level-spacing-ratio distribution $p_r(r)$
of real eigenvalues
respectively
match %
well  
with $p(s)$ and $p_r(r)$ from non-Hermitian random matrices %
in class AI + $\eta_{+}$ 
(Figs.~\ref{bdid_3d_ps} and \ref{pr_free_3d}). 
The 
number 
of real eigenvalues
is 
scaled by the square root of the dimensions of the Hamiltonian, 
being consistent with the scaling from
the random matrix theory. 
In the strong disorder regime, 
on the other hand,
the 
level-spacing and level-spacing-ratio statistics
of real eigenvalues
are consistent with those of uncorrelated real eigenvalues
(Figs.~\ref{bdid_3d_ps_loc} and \ref{pr_free_3d_loc}),
and
the soft gap of $\rho_{c}(x,y)$ 
near the real axis disappears (not shown here). 

In the strong disorder regime, $N_\text{real}$ is scaled linearly in $N$ 
(Fig.~\ref{n_real_bdi_3d}). 
This linear
scaling $\bar{N}_{\rm real} \propto N$
in %
the
localized phase is %
explained as follows. A disordered Hamiltonian in 
the localized phase can be viewed as 
$(L/\xi)^d$
almost independent blocks. 
Here, 
$\xi$ is a 
localization length in the localized phase, 
$L$ is the linear dimensions of the system size,
and $d$ is the spatial dimensions.
Each block 
can be regarded as an independent random matrix, 
and 
dimensions of each block 
are
of 
the order of $\xi^d$. 
Thus, ${N}_{\text{real}}$ 
is 
evaluated as 
\begin{equation}
    {N}_{\text{real}} \sim (L/\xi)^d \sqrt{\xi^d} \propto N,
\end{equation}
which is consistent with the numerical results.

\subsection{2D class AII + $\eta_{+}$} %
\begin{figure}
    \centering
    \includegraphics[width=1\linewidth]{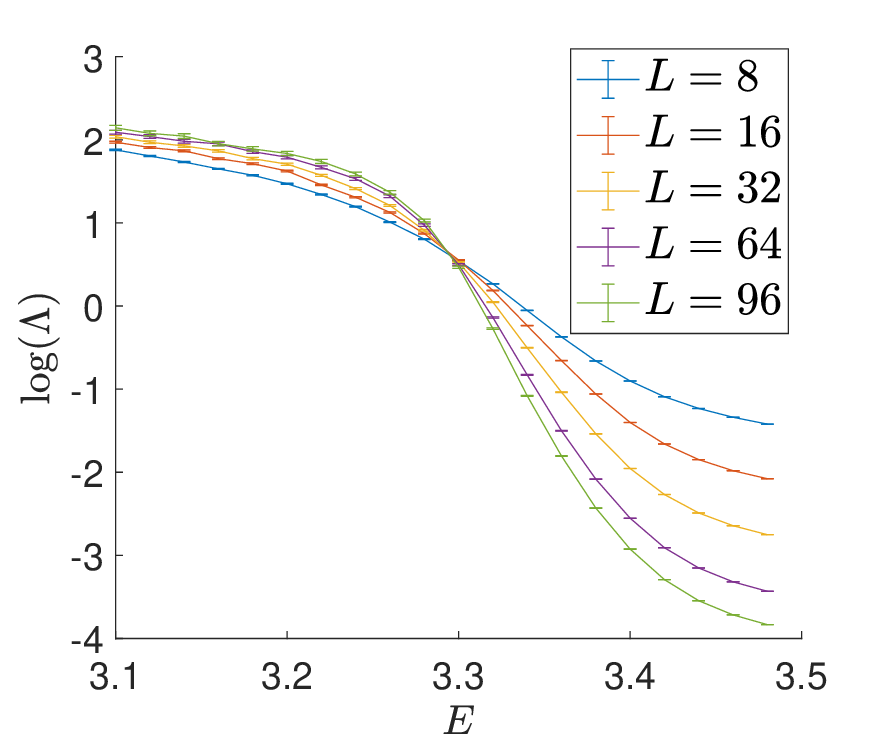}
    \caption{Normalized localization length 
    $\Lambda  = \xi_x/L$
    as a function of real energy $E$ in the 2D non-Hermitian SU(2) model in 
    the 
    weak disorder regime ($W = 0.4$, $W^{\prime}=0$). The localization length $\xi_x$ along 
    the 
    $x$ direction is calculated in the quasi-1D geometry ($L_x \times L$ with $L_x \gg L$).  
    Eigenstates 
    with
    energy $E$ and eigenstates %
    with
    energy $-E$ share the same localization length. 
    For 
    $\lvert E \rvert < E_c$ ($\lvert E \rvert > E_c$) with $E_c \approx 3.3$, $\Lambda$ increases (decreases) as $L$ increases, and thus the system is in 
    the
    metal (localized) phase.
    }
    \label{ciid_2d_len}
\end{figure}

\begin{figure}
    \centering
    \includegraphics[width=1\linewidth]{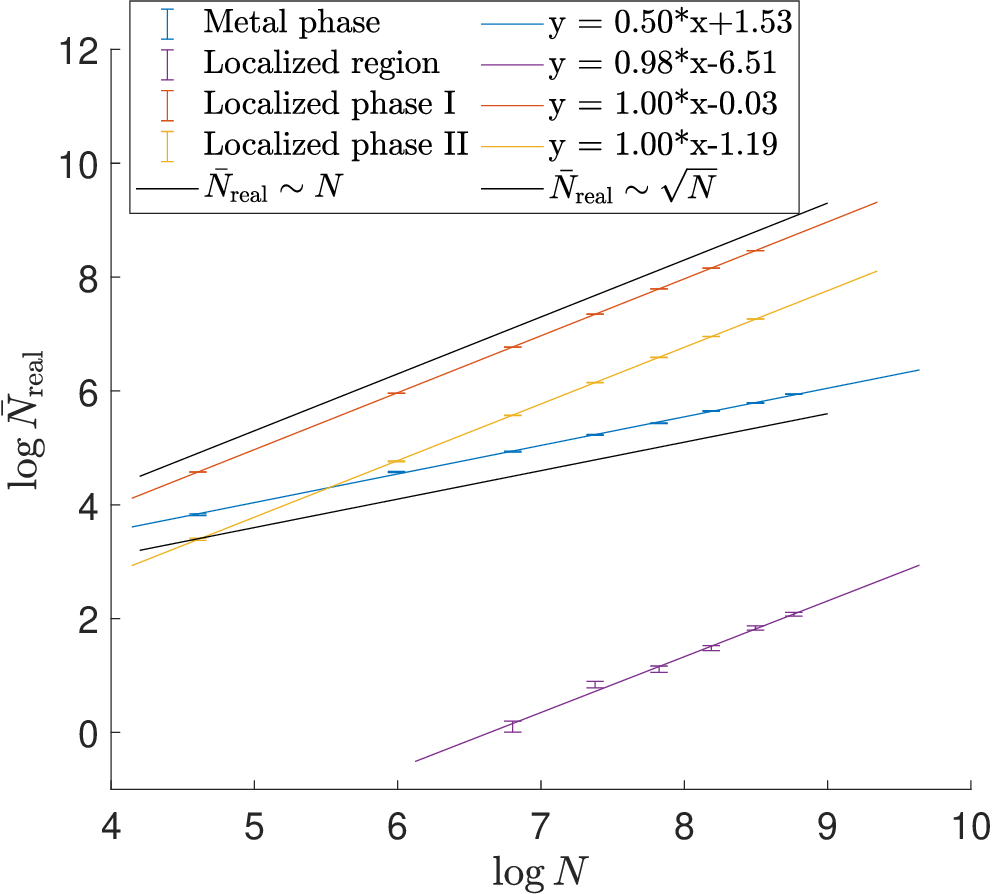}
    \caption{Average number $\bar{N}_{\rm real}$ of real 
    eigenvalues as a function of the dimensions $N$ of the 2D 
    non-Hermitian SU(2) model. The different colored lines are for the
    different disorder strength and energy regions. The blue
    line is for the metal phase ($\left| E \right| \leq 3$ with $W = 0.4$, $W^{\prime}=0$), the purple line is for the
    localized phase ($\left| E \right| \geq 3.5$ with $W = 0.4$, $W^{\prime}=0$), 
    the red line is for the localized phase I (all the real energies $E$ 
    with $W=80$, $W^{\prime}=0$), and the
    yellow line is for the localized phase II (all the real energies $E$ 
    with $W=40$, $W^{\prime}=40$). For reference,  
    the black lines are $\bar{N}_{\rm real} \propto
    N $ and $\bar{N}_{\rm real} \propto
    \sqrt{N}$, respectively.}
    \label{n_real_ciid_2d}
\end{figure}

\begin{figure*}[bt]
	\centering
	\subfigure[density $\rho_c(x,y)$ of complex eigenvalues ($W = 0.4$)]{
	\begin{minipage}[t]{0.23\linewidth}

			\centering
			\includegraphics[width=1\linewidth]{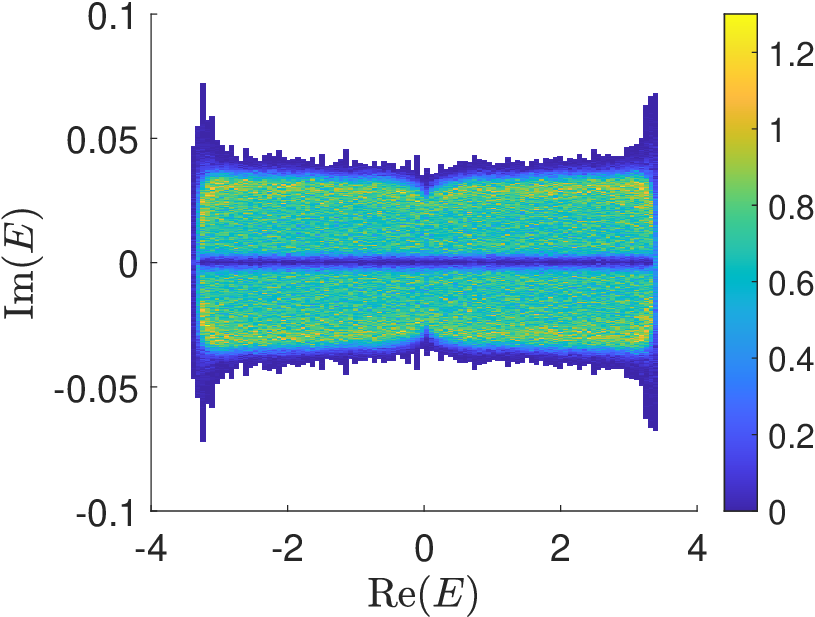}
            \label{ciid_2d_DoS}
	\end{minipage}%
	}%
	\subfigure[$\bar{\rho}_c(y)$ ($W=0.4$)]{
		\begin{minipage}[t]{0.23\linewidth}
			\centering
			\includegraphics[width=1\linewidth]{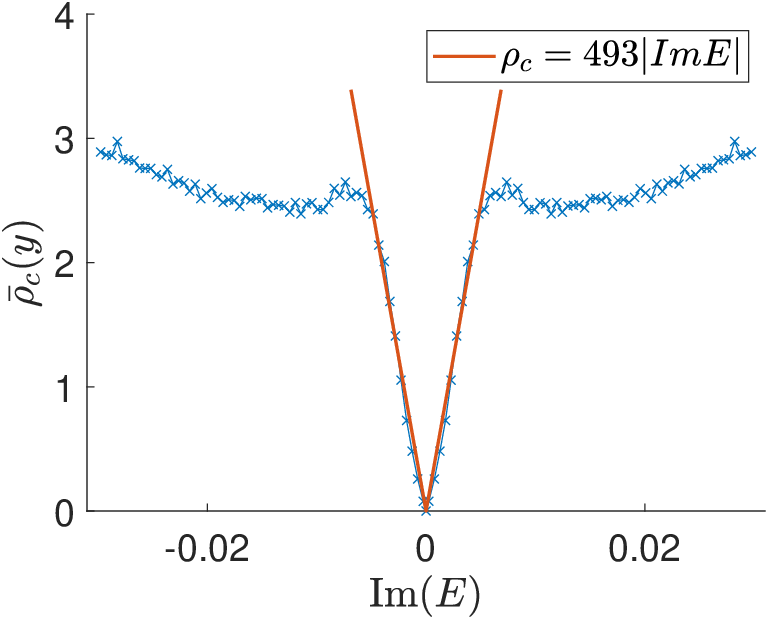}
            \label{ciid_2d_DoS_ImE}
		\end{minipage}%
	}%
	\subfigure[$\bar{\rho}_c(y)$ ($W^{\prime} = W =40$)]{
	\begin{minipage}[t]{0.23\linewidth}
		\centering
		\includegraphics[width=1\linewidth]{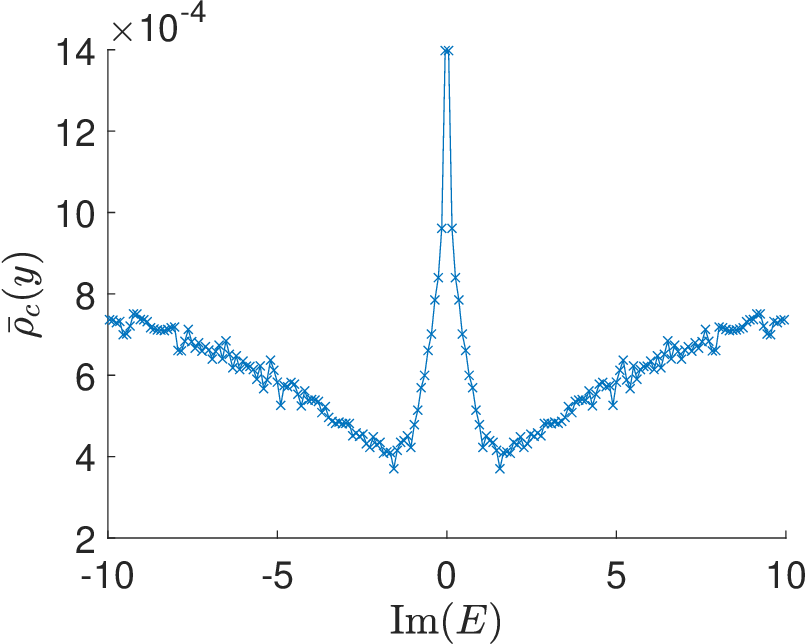}
	\end{minipage}%
	\label{ciid_2d_DoS_ImE_loc}
}
\subfigure[$\langle r \rangle - L (W = 0.4)$ ]{
\begin{minipage}[t]{0.23\linewidth}
			\centering
			\includegraphics[width=1\linewidth]{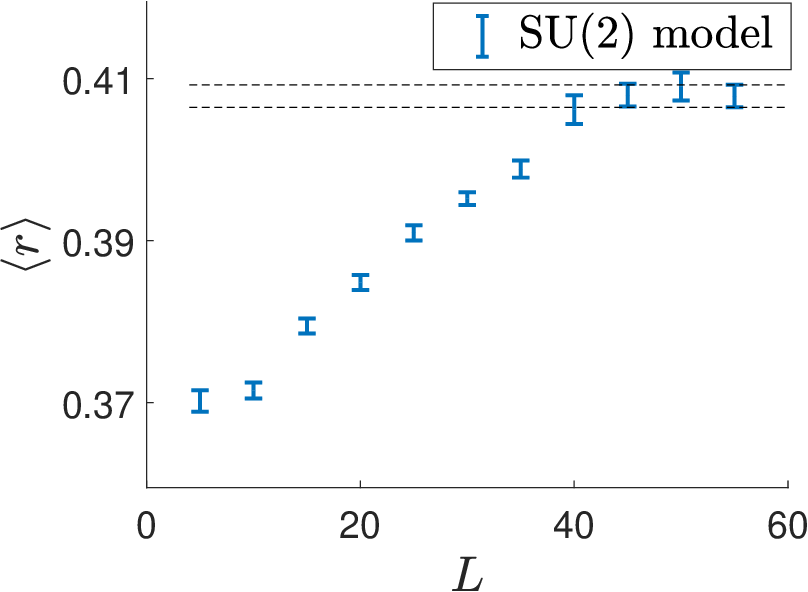}
		\end{minipage}
		\label{ciid_2d_meanr_L}
}

\subfigure[$p(s)$ ($W=0.4$, metal phase)]{
\begin{minipage}[t]{0.23\linewidth}
			\centering
			\includegraphics[width=1\linewidth]{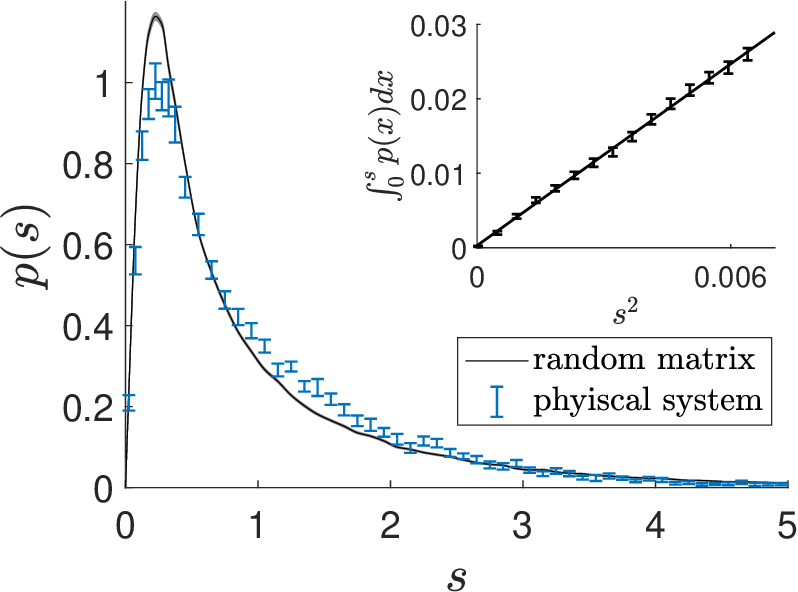}
		\end{minipage}
		 \label{ciid_2d_ps}
}
\subfigure[$p(s)$ ($W=80$, localized phase)]{
	\begin{minipage}[t]{0.23\linewidth}
		
			\centering
			\includegraphics[width=1\linewidth]{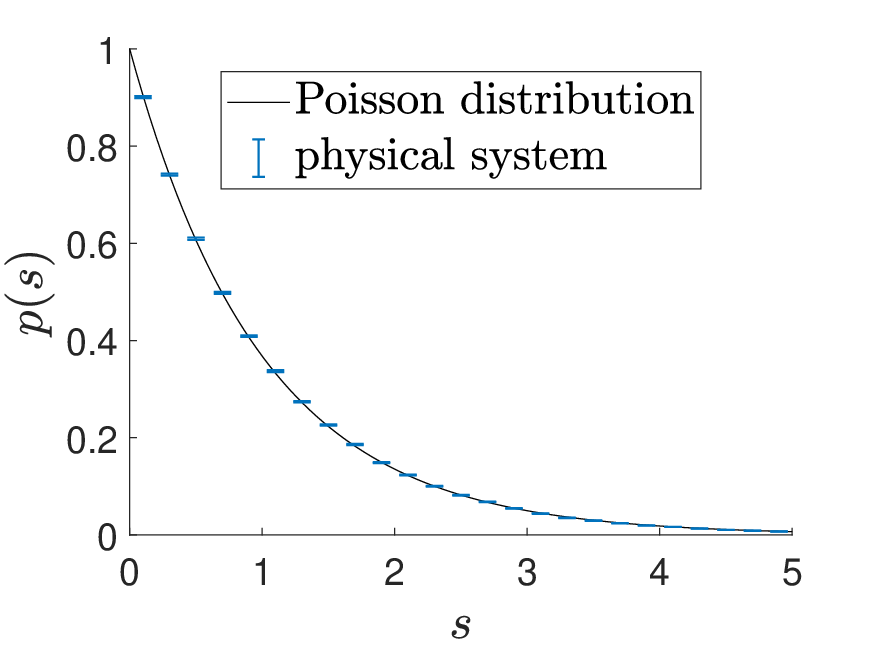}
						\label{ciid_2d_ps_l}
	\end{minipage}
	}
\subfigure[$p_r(r)$ 
($W = 0.4$, metal phase)]{
	\begin{minipage}[t]{0.23\linewidth}
		\centering
		\includegraphics[width=1\linewidth]{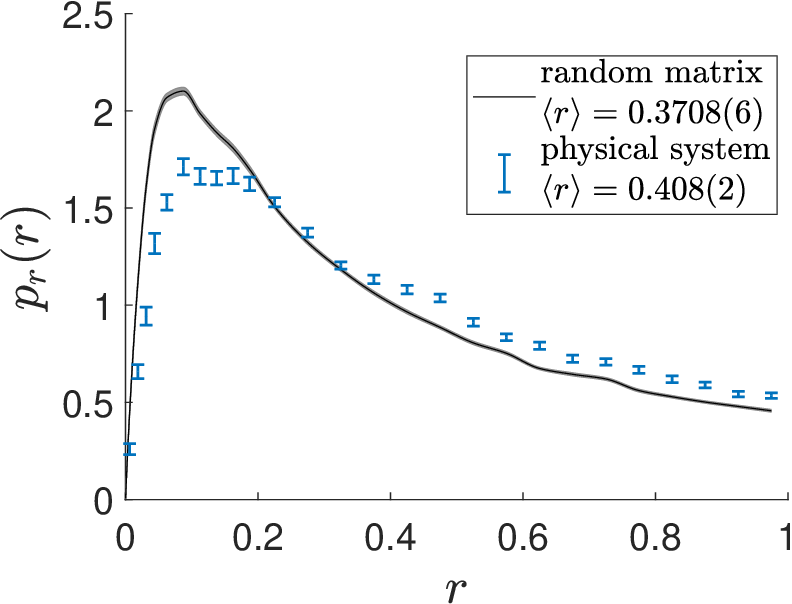}
	\end{minipage}%
		\label{ciid_2d_pr}	
}%
	\subfigure[$p_r(r)$ 
	($W=80$, localized phase)]{	
\begin{minipage}[t]{0.23\linewidth}
		\centering
		\includegraphics[width=1\linewidth]{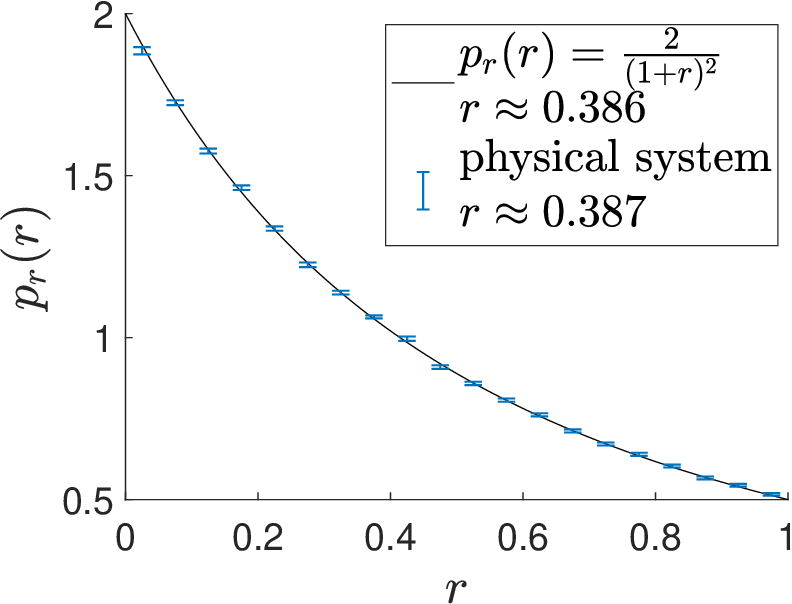}
		\label{pr_free_2d_loc}
	\end{minipage}%
}%
	\caption{(a)~Heat map of the density $\rho_c(x,y)$ of complex eigenvalues
	of the 2D non-Hermitian SU(2) model in
	the 
	weak disorder regime ($W=0.4$, $W^{\prime}=0$, $55 \times 55$ sites). (b,d) Integrated density of complex eigenvalues, $\bar{\rho}_c(y) \equiv \frac{1}{2 x_1}\int_{-x_1}^{x_1} \rho_c(x,y) dx$, as a function of small 
	$y = \mathrm{Im} \left( E \right)$ 
	around the real axis 
	$y=0$  
	(b) in the weak disorder regime ($W=0.4$, $W^{\prime}=0$, $x_1=3$) 
	and (c) in the strong disorder regime ($W=W^{\prime}=40$, $x_1=20$). 
    (d)~Mean value of level-spacing ratios as a function of system size in the metal phase ($|E|<3$ , $W=0.4$, $W^{\prime}=0$). 
    (e)~Level-spacing distribution $p(s)$ of real eigenvalues
	in the metal phase of the non-Hermitian SU(2) model ($|E|<3$, $W=0.4$, $W^{\prime}=0$), and its comparison to $p(s)$ from random matrices  
	in symmetry class AII + $\eta_{+}$ (red line). Inset: Asymptotic 
	behavior of $\int^s_0  p(s^{\prime}) ds^{\prime}$ in the metal phase for $s \ll 1$. 
    (f)~$p(s)$ in the 
	localized phase of the 2D non-Hermitian SU(2) model (all the real energies $E$, $W =80$, $W^{\prime}=0$), and its comparison to the Poisson distribution. 
	(g)~Level-spacing-ratio distribution $p_r(r)$ of real eigenvalues in the metal phase of the 2D non-Hermitian SU(2) model ($\lvert E \rvert < 3$, $W=0.4$, $W^{\prime}=0$) and its comparison to $p_r(r)$ from  
	non-Hermitian random matrices in symmetry class AII + $\eta_{+}$. The difference between the physical model and random matrices cannot be neglected. (h)~$p_r(r)$ of real eigenvalues in the localized phase (all real energy $E$, $W=40$, $W^{\prime}=0$) of the 2D non-Hermitian SU(2) model and its comparison to  $p_r(r)$ of uncorrelated real numbers.
	The mean value $\langle r \rangle = \int_0^1 p_r(r)dr$ of each level-spacing-ratio distribution is also shown in the figures. 
	(d)-(h)~The error ranges are evaluated by the bootstrap method~\cite{press07}.
 }
\label{ciid_2d_ps_DoS}
\end{figure*}

We study a non-Hermitian extension of the disordered SU(2) model~\cite{Asada02} on the two-dimensional (2D) square lattice:
\begin{equation}
 \begin{aligned}
		{\cal H}_{\text{2D}}&=\sum_{i} \left( \varepsilon_{i} c_{i}^{\dagger}\sigma_0 c_{i} +\varepsilon_{i}^{\prime} d_{i}^{\dagger}\sigma_0 d_{i} \right) \\
		&+ \sum_{\langle i, j\rangle}t_1 \left[c_{i}^{\dagger} R(i,j)c_{j} + d_{i}^{\dagger} R^{\prime}(i,j)d_{j} \right] 
		\\
		&+ \sum_{\langle i, j\rangle}t_2 \left[c_{i}^{\dagger} U(i,j)d_{j} + d_{i}^{\dagger} U^{\prime}(i,j)c_{j} \right] \, , \\
\end{aligned}  
\label{H_ciid}
\end{equation}
where $c_i$ and $d_i$ are annihilation operators for  
two different %
orbitals
defined on %
site $i$.  
Both operators have pseudo-spin-1/2 
degree of freedom. 
$\langle i, j\rangle$ denotes the nearest-neighbor square-lattice 
sites. $\sigma_{0}$ stands for the %
two-by-two
identity matrix 
acting on the spin degree of freedom.
$\varepsilon_{i}$ and 
$\varepsilon_{i}^{\prime}$ are on-site random potentials %
that distribute independently %
and 
uniformly in %
the range 
$[-W/2,W/2]$. $t_1$ and $t_2$ are 
real numbers. $R(i,j)$, $R^{\prime}(i,j)$, $U(i,j)$, and 
$U^{\prime}(i,j)$ are SU(2) matrices %
that
distribute uniformly 
with respect to the Harr measure of SU(2)~\cite{Asada02} and satisfy the following symmetry properties, 
\begin{align}
R(i,j) &= R^{\dagger}(j,i), \\
R^{\prime}(i,j) &= R^{\prime \dagger}(j,i), \\
U(i,j) &= -U^{\prime \dagger}(j,i). 
\end{align}
The %
term
with $t_2$ is %
anti-Hermitian %
and the others are Hermitian. %
Let $\tau_z$ be a %
two-by-two
matrix acting on the orbital 
space, satisfying %
$\tau_z(c_i,d_i)^T \equiv (c_i,-d_i)^T$.
The Hamiltonian in Eq.~(\ref{H_ciid}) satisfies %
TRS 
\begin{equation}
{\cal H}_{\text{2D}} = \sigma_y {\cal H}^*_{\text{2D}} \sigma_y
\end{equation}
and pH %
\begin{equation}
{\cal H}_{\text{2D}} = \tau_z {\cal H}^{\dagger}_{\text{2D}} \tau_z,
\end{equation}
where
the TRS operator and the %
pH 
operator commute 
with each other. 
Thus, the Hamiltonian belongs to %
symmetry 
class AII + $\eta_+$. This Hamiltonian is a minimal model 
to study
the 
interplay between the spin-orbit coupling~\cite{geissler2014random} and 
pH.

We investigate the weak (strong) disorder regime %
of Eq.~(\ref{H_ciid}) %
with 
the
parameters $t_1 = 1$, $t_2 = 0.1$,
$W = 0.4$ ($W = 80$) and with 
the
periodic boundary conditions. 
In the weak disorder regime, 
we find that
eigenstates with
real eigenvalues $E$ 
undergo the Anderson transition at a certain mobility 
edge %
$E=E_c$. %
The
normalized localization length 
$\Lambda(E,L) \equiv {\xi_x(E,L)}/{L}$ shows a scale 
invariant behavior at $E=E_c$  
(Fig.~\ref{ciid_2d_len}). 
Here, the localization length 
$\xi_x(E,L)$ is calculated along 
the $x$ direction by the transfer 
matrix method in the quasi-1D geometry 
$L \times L_x$ ($L_x \gg L$)  ~\cite{MacKinnon81,MacKinnon83,Slevin14,Luo21TM}. 
For the weak disorder regime, $\lvert E \rvert < E_c \approx 3.3 $ 
and $\lvert E \rvert > E_c$ correspond to  
the
metal and localized phases, respectively (Fig.~\ref{ciid_2d_len}). 
We diagonalize the Hamiltonians %
in 
Eq.~(\ref{H_ciid}) with 
240 different disorder realizations %
for
each system size (the maximal system size is 
$55 \times 55$
sites),  
where eigenstates near the mobility edge are 
excluded from the %
level statistics
of real eigenvalues. 

In the weak disorder regime, the density $\rho_c(x,y)$ of complex eigenvalues
 shows the soft gap $\rho_c(x, y) \propto |y|$
in the metal phase 
($|x| = |\mathrm{Re}(E)| < E_c$) for $y = \mathrm{Im} (E)$ much smaller than 
the
mean level-spacing (Figs.~\ref{ciid_2d_DoS} and \ref{ciid_2d_DoS_ImE}). 
The soft-gap behavior
is consistent 
with the DoS of non-Hermitian random matrices in class AII + $\eta_{+}$.
The number of real eigenvalues
within the metal phase, 
$\bar{N}_{\text{real}}^{\rm metal} \equiv \int^{E_c}_{-E_c} dx \rho_r(x)$, shows %
the square-root scaling with respect to the dimensions $N$
of the Hamiltonian, $\bar{N}^{\rm metal}_{\rm real} \propto \sqrt{N}$ 
(Fig.~\ref{n_real_ciid_2d}), which is also consistent with 
the random matrix theory. On the other hand, 
the number of real eigenvalues
in the localized phase, 
$\bar{N}_{\text{real}}^{\rm loc} \equiv \int_{|x|>E_c} dx \rho_r(x)$, 
is scaled by $N$, $\bar{N}_{\text{real}}^{\rm loc} %
\propto N$ (Fig.~\ref{n_real_ciid_2d}).

The mean value $\langle r \rangle$ of the level-spacing 
ratios in the metallic phase converges to the 
value $\langle r \rangle \approx 0.41$ for larger 
system size (Fig.~\ref{ciid_2d_meanr_L}). 
This value 
is %
larger than the mean value $\langle r \rangle \approx 0.37$ 
of non-Hermitian random matrices in class AII + $\eta_+$, 
suggesting a possible discrepancy of the level statistics 
between the physical model and the random matrices. 
In fact, the level-spacing distribution $p(s)$ and 
level-spacing-ratio distribution $p_r(r)$ in the metallic phase 
are also different from those of the random matrices (Figs.~\ref{ciid_2d_ps} and \ref{ciid_2d_pr}), 
although the small $s$ behavior of 
$p(s)$ %
is consistent with the random matrix theory 
(inset of Fig.~\ref{ciid_2d_ps}). 
The Kolmogorov-Smirnov distances between $p(s),p_r(r)$ in the metallic phase  
and those of the random matrices are around $0.06$ (Appendix~\ref{sec_ks}) 
while the distances %
are less than $0.01$ in the other four symmetry classes.
Similar discrepancies of the level statistics 
were previously reported in physical models near the mobility edge of the Anderson 
transition~\cite{Evers08} or %
the MBL transition~\cite{sierant19}.
Nonetheless, in our study, an energy window $|E|<3$ for the level 
statistics of the physical model is well within the metallic 
phase $|E|<E_c\approx 3.3$ (Fig.~\ref{ciid_2d_len}). No previous works 
found such discrepancies of $p(s)$ and $p_r(r)$ between physical models in the metallic phases and the random matrices.

A possible reason for the discrepancies 
is unusual level interactions in non-Hermitian random matrices in 
class AII + $\eta_+$. 
As discussed in Sec.~\ref{sec_rm}, 
the small level-spacing ratio $\langle r\rangle < \langle r \rangle_{\rm Poisson}$ and the large spectral 
compressibility $\chi > \chi_{\rm Poisson}$, 
which are also supported by the 
small random matrix analyses, 
suggest that
the attractive interaction between real eigenvalues in the finite $s$ regime dominates the repulsive interaction in the small $s$ regime on average. 
We speculate that this unconventional level attraction makes the level 
statistics of the random matrices in the finite $s$ region %
unstable against details of physical models. Only for level 
spacing $s \ll 1$, the behaviors between the %
physical model
and random matrices are consistent. 
It is unclear whether $p(s)$ and $p_r(r)$ of physical 
models in class AII + $\eta_+$ are universal or not, and 
we leave this issue for future study. 
It is also interesting to investigate whether the non-Hermitian 
generalization of other random matrix ensembles, such as the power-law 
random banded matrix ensemble~\cite{mirlin96}
and the Rosenzweig-Porter random-matrix ensemble~\cite{rosenzweig60,kravtsov15}, 
can describe the physical models in class AII + $\eta_+$.

In the strong disorder regime ($W=80$), the soft gap in 
the DoS around the real axis $y=0$
disappears (not shown here). The 
level-spacings 
of real eigenvalues
show the Poisson 
distribution (Fig.~\ref{ciid_2d_ps_l}), 
and the level-spacing ratios of real eigenvalues show the same distribution as uncorrelated real numbers in Eq.~(\ref{pr_uncorrelated}) (Fig.~\ref{pr_free_2d_loc}).
The average number of all the real eigenvalues becomes 
proportional to $N$ (Fig.~\ref{n_real_ciid_2d}). %
In this phase,
the Hamiltonian is dominated by 
the Hermitian part and almost all the eigenvalues are real, 
resulting in
the %
linear scaling $\bar{N}_{\text{real}} \propto N$. %
To confirm that %
this linear scaling
is a general
property of the Anderson localized phase, 
we generalize the model in Eq.~(\ref{H_ciid}) and study the following model ${\cal H}_{\text{2D}}^{\prime}$:
\begin{align}
& {\cal H}_{\text{2D}}^{\prime} = {\cal H}_{\text{2D}} 
+ \Delta {\cal H}, \nonumber \\
    & \Delta {\cal H} = \sum_i \omega_i c_{i}^{\dagger}\sigma_0 d_{i} - \omega_i  d_{i}^{\dagger}\sigma_0 c_{i}. 
\end{align}
Adding $\Delta {\cal H}$ does not change the symmetry class of 
${\cal H}_2$ (class AII + $\eta_{+}$). 
Here, $\omega_i$ 
distributes 
uniformly
in %
$[-W^{\prime}/2,W^{\prime}/2]$. 
We calculate
the DoS, the level-spacing distribution $p(s)$, 
and the number of real eigenvalues 
with the parameter $W^{\prime} = W = 40$.
Thereby, $\rho_c(x,y)$ has no soft gap 
around $y=0$ (Fig.~\ref{ciid_2d_DoS_ImE_loc}), 
$p(s)$ is consistent with the Poisson distribution, 
and $\bar{N}_{\text{real}} \equiv \int^{+\infty}_{-\infty} dx \rho_r(x)$ 
remains 
linear in $N$ (Fig.~\ref{n_real_ciid_2d}).

\section{Conclusions}
\label{conclusion}

Non-Hermitian Hamiltonians in 
the
seven symmetry classes (classes A + $\eta$, AI, AI + $\eta_{\pm}$, AII, and AII + $\eta_{\pm}$) have %
TRS 
or %
pH.
Real eigenvalues respect these symmetries with complex or Hermitian conjugation. 
We find that
a sub-extensive number of eigenenergies of 
non-Hermitian
random matrices in classes A + $\eta$, AI, AI + $\eta_{\pm}$, and AII + $\eta_{+}$ are real, where the
DoS
on the  
complex plane
has a delta function peak %
on
the real axis. 
We clarify
the universal
level-spacing distributions on the real axis
in these five symmetry classes
for non-Hermitian random matrices, as well as bosonic many-body Hamiltonians in 
the ergodic phases and fermionic non-interacting Hamiltonians in the metal phases. 
In classes A + $\eta$, AI, and AI + $\eta_{\pm}$, 
the level statistics of real eigenvalues %
show 
good agreement
between the ergodic physical models and random matrices, while 
we find
discrepancies
in class AII + $\eta_+$.
The average number of 
real eigenvalues in the ergodic phases universally shows 
the square-root scaling with respect to 
the dimensions of the Hamiltonians.
We explain the
universal asymptotic behaviors of the DoS around the real 
axis and %
the level-spacing distributions 
on the real axis
by small random matrix analyses. 
We also clarify the
level statistics 
of the physical models in the 
Anderson localized %
and %
MBL
phases
by extensive numerical calculations
together with 
the
phenomenological interpretations.

The universal level-spacing and level-spacing-ratio distributions of real eigenvalues and the scaling relationship of the number of real eigenvalues obtained in this paper
provide %
powerful methods of studying level statistics of non-Hermitian systems with TRS and/or pH. 
They are also useful for detecting quantum chaos, many-body localization, and real-complex transitions in %
non-Hermitian systems with the symmetries. 

For example, the level statistics of real eigenvalues help 
us answer fundamental questions on 
non-Hermitian disordered systems.
Analyzing level spacings or level-spacing ratios of 
real eigenvalues in the different energy windows, we 
can determine the presence of mobility edges in non-Hermitian 
many-body systems. 
By the finite-size scaling analysis of 
spacing ratios~\cite{bertrand16}, we can also evaluate the 
critical exponents of the Anderson transitions in 
non-Hermitian systems with TRS and/or pH.
Comparisons of the critical exponents with systems without TRS or pH 
tell us whether TRS and/or pH change the universality classes of MBL transitions in non-Hermitian systems. 
We believe that with the help of the %
results
in this paper, %
a number of 
spectral 
analysis method %
and quantum chaotic study 
used in Hermitian systems
are 
ready to be applied to non-Hermitian systems with TRS and/or pH. 

The real spectrum is related to the stability of  
non-Hermitian systems in their long-time dynamics~\cite{Bender98, hamazaki2019non, Hu-Hughes-11, Kawabata20real}.
The square-root scaling of the number $N_{\rm real}$ of 
real eigenvalues hints at possible dynamical instability in
ergodic
non-Hermitian systems. In contrast, the linear scaling of 
$N_{\rm real}$ in the Anderson localized phase may imply that disorder can %
stabilize the %
dynamics of non-Hermitian systems.
The non-universal scaling of the number of %
real eigenvalues, together with our explanation, also serves as %
evidence for 
the emergent integrability in the %
MBL phase of non-Hermitian systems.

The eigenstates on the real (imaginary) axis can only 
respect TRS (PHS$^{\dagger}$), TRS$^{\dagger}$, and pH (CS) among 
all the symmetries, as long as they are away from zero in the complex 
plane. Thus, for all the 38 symmetry classes, we believe 
that the level-spacing distributions of real (purely imaginary) 
eigenvalues away from zero satisfy one of the five universal level 
spacing distributions found in this paper. For example, for non-Hermitian 
random matrices in symmetry class BDI (i.e., with TRS whose sign is 
$+1$ and PHS whose sign is $+1$), PHS affects the spectral statistics 
only around zero, and hence the real-eigenvalue spacing distribution 
away from zero should be the same as that in class AI. We leave 
testing this for all the 38 symmetry classes for future work.

While we have focused on non-Hermitian Hamiltonians in this paper, our results should also be relevant to Lindbladians, which govern the open quantum dynamics of the Markovian master equation~\cite{breuer02}. 
Similarly to closed quantum systems, it was conjectured that the level-spacing statistics of Lindbladians obey those of non-Hermitian random matrices in the non-integrable phases and the Poisson statistics in the integrable phases~\cite{Hamazaki20,Akemann19,Sa20}.
This conjecture was previously verified for generic complex eigenvalues of Lindbladians away from the real axis.
Notably, Lindbladians always respect TRS, where time reversal is effectively defined by the combination of complex conjugation and the swap operation between the bra and ket spaces.
Thus, we expect that our results
of the level statistics of real eigenvalues for non-Hermitian random matrices 
should coincide with those of non-integrable Lindbladians in the same symmetry classes.
It is also notable that the quadratic Lindbladians can be classified by the $\mathrm{AZ}^{\dag}$ symmetry class~\cite{Lieu20}, which is the same as the ten-fold symmetry class studied in this paper.

Before concluding this paper, we note in passing that a non-Hermitian extension of the disordered Su-Schrieffer-Heeger model was shown to exhibit the level-spacing statistics of the Gaussian orthogonal ensemble~\cite{Mochizuki20}.
This is due to an additional symmetry that enables a similarity transformation between this non-Hermitian model and the Hermitian Su-Schrieffer-Heeger model.
Thus, in this non-Hermitian model, only the Hermitian degrees of freedom are present, and all eigenvalues are real, which are different from generic non-Hermitian random matrices studied in this paper.

\section*{Acknowledgement}
Z.X. thanks Haoran Chen, Lingxian Kong, and Yeyang Zhang for helpful discussions.
We thank Jiachen Li for helpful comments on Appendix~\ref{sec_be}.
Z.X. and R.S. were supported by the National Basic Research Programs of China (No.~2019YFA0308401) and by National Natural Science Foundation of China (No.~11674011 and No.~12074008). X.L. was supported by National Natural Science Foundation of China of Grant No.~12105253. %
K.K. was supported by JSPS Overseas Research Fellowship, and Grant No.~GBMF8685 from the Gordon and Betty Moore Foundation toward the Princeton theory program.
T.O. was supported by JSPS KAKENHI Grants 19H00658 and 22H05114.

\appendix

\section{Analyses of small random matrices}
    \label{B}
    
Small 
non-Hermitian
random matrices in the seven %
symmetry classes are written in the following unified form,
\begin{equation}
{\cal H}^{(s)} =a_0 I + \sum_{i=1}^{m} a_i L_i + \sum_{i=m+1}^{m+n}{\rm i} a_i L_i \, .
\end{equation}
Here, $I$ is the identity matrix, %
$L_1 ,L_2,\cdots,L_{m+n}$ are anti-commuting Hermitian traceless matrices, 
$\{L_i,L_j\} = 2\delta_{i,j}I$, and $a_0,a_1,\cdots,a_{m+n}$ 
are real numbers. Note that 
\begin{equation}
{\rm Tr}({\cal H}^{(s)\dagger}{\cal H}^{(s)}) = \left( \sum_{i=0}^{m+n}a_i^2 \right) {\rm Tr}(I) \,.
\end{equation}
The real numbers $a_0,a_1,\cdots,a_{m+n}$ are independent %
of
each other and obey %
the
identical standard Gaussian distribution.
Note that 
$\sum_{i = 1}^{k} a_i^2$ obeys the $\chi^2$ distribution with 
the
degree $k$~\cite{Hamazaki20}.  
The probability of $\sum^k_{i=1} a^2_{i}=X$ is given by  
\begin{equation}
p(X ; k)= \begin{cases}\frac{X^{\frac{k}{2}-1} e^{-\frac{X}{2}}}{2^{\frac{k}{2}} \Gamma\left(\frac{k}{2}\right)} & X\geq 0, \\ 0, & X<0 \,.\end{cases}
\end{equation}

The square
of the traceless part of ${\cal H}^{(s)}$ is proportional 
to $I$, 
\begin{equation}
\Big(\sum_{i=1}^{m} a_i L_i + \sum_{i=m+1}^{m+n}{\rm i} a_i L_i\Big)^2
 = \left( \sum_{i=1}^{m} a_i^2  - \sum_{i=m+1}^{m+n} a_i^2 \right) 
 I\, .
\end{equation}
Thus, eigenvalues of ${\cal H}^{(s)}$ are given by 
\begin{equation}
\lambda = a_0  \pm  \sqrt{X  - Y}, \ \ 
X \equiv \sum_{i=1}^{m} a_i^2, \ \  Y \equiv \sum_{i=m+1}^{m+n} a_i^2.  
\end{equation}
The probability that the eigenvalues are real is given by %
the
probability of $X  \geq Y$: %
\begin{align}
    p_{\lambda = \lambda^{*}} = 
    \int^{\infty}_{-\infty} dX \int^{+\infty}_{-\infty} dY 
     \!\ \theta(X-Y) \!\ p(X;m) \!\ p(Y;n),
\end{align}
where $\theta(u)$ is the step function satisfying $\theta(u) \equiv 1$ for $u \geq 0$ and $\theta(u)\equiv 0$ for $u < 0$.

In terms of Pauli matrices $\sigma_{\mu}$ and $\tau_{\mu}$ ($\mu=0,x,y,z$) and their Kronecker products $\tau_{\mu}\sigma_{\nu}$, the traceless parts of 
the small random matrices %
in
the seven symmetry 
classes are given as
\begin{equation}
\begin{aligned}
&\tilde{\cal H}^{(s)}_{\text{AI}} = a_1 \sigma_z + a_2 \sigma_x + {\rm i}a_3 \sigma_y   \,  ,\\
&\tilde{\cal H}^{(s)}_{\text{A}+\eta} = a_1 \sigma_z + {\rm i} a_2 \sigma_x + {\rm i}a_3 \sigma_y  \, , \\
&\tilde{\cal H}^{(s)}_{\text{AI}+\eta_+} =  a_1 \sigma_z +  {\rm i}a_2 \sigma_y\,  ,\\
&\tilde{\cal H}^{(s)}_{\text{AII}} =  {\rm i}a_1 \sigma_z + {\rm i} a_2 \sigma_x + {\rm i}a_3 \sigma_y \, ,\\
&\tilde{\cal H}^{(s)}_{\text{AI}+\eta_-} = a_1 \tau_z +  a_2\tau_x  +  a_3 \tau_y \sigma_y + {\rm i}a_4 \tau_y\sigma_x + {\rm i} a_5 \tau_y \sigma_z \, , \\
&\tilde{\cal H}^{(s)}_{\text{AII}+\eta_-} = {\rm i}a_1 \sigma_x + {\rm i} a_2 \sigma_y  \, , \\
&\tilde{\cal H}^{(s)}_{\text{AII}+\eta_+} = a_1 \tau_z + {\rm i}a_2 \tau_y  + {\rm i} a_3\tau_x \sigma_x +  {\rm i} a_4\tau_x \sigma_y + {\rm i} a_5\tau_ x\sigma_z \, .
\end{aligned}
\end{equation}
In fact, the small matrix in each class respects the following symmetries, 
\begin{equation}
\begin{aligned}
&\tilde{\cal H}^{(s)}_{\text{AI}} =\tilde{\cal H}^{(s)*}_{\text{AI}}   \,  ,\\
&\tilde{\cal H}^{(s)}_{\text{A}+\eta} =  \sigma_z \tilde{\cal H}^{(s) \dagger}_{\text{A} + \eta} \sigma_z  \, , \\
&\tilde{\cal H}^{(s)}_{\text{AI}+\eta_+} = \tilde{\cal H}^{(s)*}_{\text{AI}+\eta_+}    , \tilde{\cal H}^{(s)}_{\text{AI}+\eta_+} =  \sigma_z \tilde{\cal H}^{(s)\dagger}_{\text{AI}+\eta_+}  \sigma_z  \, , \\
&\tilde{\cal H}^{(s)}_{\text{AII}} = \sigma_y \tilde{\cal H}^{(s)*}_{\text{AII}} \sigma_y \, ,\\
&\tilde{\cal H}^{(s)}_{\text{AI}+\eta_-} = \tilde{\cal H}^{(s)*}_{\text{AI}+\eta_-} \, ,
\tilde{\cal H}^{(s)}_{\text{AI}+\eta_-} =  \sigma_y\tilde{\cal H}^{(s)\dagger}_{\text{AI}+\eta_-} \sigma_y\, ,\\
&\tilde{\cal H}^{(s)}_{\text{AII}+\eta_-} =   \sigma_y \tilde{\cal H}^{(s)*}_{\text{AII}+\eta_-}  \sigma_y \, ,
\tilde{\cal H}^{(s)}_{\text{AII}+\eta_-} =   \sigma_z \tilde{\cal H}^{(s)\dagger}_{\text{AII}+\eta_-} \sigma_z \, ,\\
&\tilde{\cal H}^{(s)}_{\text{AII}+\eta_+} = \sigma_y \tilde{\cal H}^{(s)*}_{\text{AII}+\eta_+} \sigma_y\, ,
\tilde{\cal H}^{(s)}_{\text{AII}+\eta_+} = \tau_z \tilde{\cal H}^{(s)\dagger}_{\text{AII}+\eta_+} \tau_z \, .
\end{aligned}
\end{equation}
The number 
$m$
of the real degrees of freedom %
and the number 
$n$
of 
the imaginary degrees of freedom %
for
the small random matrices %
in each symmetry class are summarized in Table~\ref{mn}. 
The
probability of real eigenvalues is finite for 
${\cal H}^{(s)}$ in classes AI, A + $\eta$, AI + $\eta_+$, 
AI + $\eta_-$, and AII + $\eta_+$ %
because of
$m \ne 0$; %
on the other hand,
the probability of real eigenvalues is zero for ${\cal H}^{(s)}$ in 
classes AII and AII + $\eta_-$ because 
of $m=0$.

\subsection{Real-eigenvalue spacing} 
With a finite probability, the small random 
matrices %
in classes AI, A + $\eta$, AI + $\eta_{\pm}$, and 
AII + $\eta_+$ have a pair of real eigenvalues. 
The
distance $s$ between the two real eigenvalues is given by %
\begin{align}
    s = 2 \sqrt{\sum^m_{i=1} a^2_i - \sum^{m+n}_{i=m+1} a^2_i}. 
\end{align}
The probability of $(s/2)^2 = X$ under the condition that the eigenvalues 
are real is calculated as follows, 
\begin{align}
& P(X) = \frac{\int_{-\infty}^{\infty} dX^{\prime} 
\int_{-\infty}^{\infty}  dY^{\prime} \delta(X^{\prime}-Y^{\prime}-X ) p(X^{\prime};m)p(Y^{\prime};n)}{\int_{-\infty}^{\infty} dX^{\prime} \int_{-\infty}^{\infty}  dY^{\prime} \theta(X^{\prime}-Y^{\prime}) 
p(X^{\prime};m)p(Y^{\prime};n)}  \nonumber \\
 & \ = \frac{ \int_0^{\infty} dx (x+X)^{\frac{m}{2}-1} e^{-\frac{x+X}{2}} x^{\frac{n}{2}-1} e^{-\frac{x}{2}} }{\int_0^{\infty} dx  \int_0^{\infty} dX (x+X)^{\frac{m}{2}-1} e^{-\frac{x+X}{2}} x^{\frac{n}{2}-1} e^{-\frac{x}{2}} } \nonumber \\
 & \ =   
 \begin{cases} \frac{1}{2} e^{-\frac{X}{2}} & (m,n)= (2,1), \\
 \frac{1}{2(\sqrt{2}-1)} e^{\frac{X}{2}}\text{erfc}(\sqrt{X}) & (m,n) = (1,2), \\
 \frac{1}{\pi} K_{0}(\frac{X}{2})  &(m,n) = (1,1), \\
  \frac{e^{-\frac{X}{2}} \left(\sqrt{\pi } e^X \text{erfc}\left(\sqrt{X}\right)+2 \sqrt{X}\right)}{2 \left(2 \sqrt{2}-1\right) \sqrt{\pi }} & (m,n )= (3,2), \\
  \frac{e^{-\frac{X}{2}} \left(2 \sqrt{X}-\sqrt{\pi } e^X (2 X-1) \text{erfc}\left(\sqrt{X}\right)\right)}{2 \sqrt{\pi } \left(4 \sqrt{2}-5\right)} &  (m,n) = (1,4), \, \nonumber \\ %
 \end{cases}
\end{align}
with $\int^{\infty}_{0} d X P(X)=1$. 
Here, $\text{erfc}(u) \equiv  \frac{2}{\sqrt{\pi}} \int_{u}^{\infty} e^{-t^{2}} d t$ is 
the complementary error function and $K_{\nu}(u)$ is the 
modified Bessel function of the second kind. 
Then, %
the 
probability of the 
distance being $s$ under %
the
condition that the eigenvalues are real is given by 
\begin{align}
P_{s} 
\left( s \right)
= \frac{s}{2} P\left(\left(\frac{s}{2}\right)^2\right),  \label{app-b-2}
\end{align}
with $\int^{\infty}_{0} P_s \left( s \right) ds =1$. 
Note that the real-eigenvalue spacing distribution function $p(s)$ in the main text is defined for the distance normalized by %
the
mean value of the distance: %
$\int^{\infty}_{0} p(s) sds = 1$. 
Thus, we have %
\begin{align}
p(s) \equiv \bar{s} P_{s}(\bar{s} s), \label{app-b-3}
\end{align}
with the mean value 
\begin{align}
    \bar {s} \equiv \int^{\infty}_{0} s P_{s} 
    \left( s \right)
    ds.   
\end{align}
From Eqs.~(\ref{app-b-2}) and (\ref{app-b-3}),  
we obtain the probability distribution functions of the normalized 
distances between the two real eigenvalues %
as follows,
\begin{widetext}
\begin{equation}
p(s) = \begin{cases} \frac{1}{2} \pi  s e^{-\frac{\pi }{4} s^2 } & (m,n )= (2,1), \\
 \frac{1}{4} \left(\sqrt{2}+1\right) \bar{s}_{2}^2 s e^{\frac{1}{8} \bar{s}_{2}^2 s^2 } \text{erfc}\left(\frac{ \bar{s}_{2}s}{2}\right) & (m,n) = (1,2), \\
 \frac{8 s \Gamma \left(\frac{3}{4}\right)^4 K_0\left(\frac{2 s^2 \Gamma \left(\frac{3}{4}\right)^4}{\pi ^2}\right)}{\pi ^3}  &(m,n) = (1,1), \\
  \frac{\bar{s}_{3}^2 s e^{-\frac{\bar{s}_{3}^2 s^2}{2}} \left(\sqrt{\pi } e^{\bar{s}_{3}^2s^2} \text{erfc}(\bar{s}_{3} s)+2 \bar{s}_{3} s\right)}{\left(2 \sqrt{2}-1\right) \sqrt{\pi }} & (m,n )= (3,2), \\
  \frac{\bar{s}_{4}^2s e^{-\frac{1}{8}  \bar{s}_{4}^2 s^2}  \left(\bar{s}_{4}s -\sqrt{\pi } e^{\frac{1}{4} \bar{s}_{4}^2s^2 } \left(\frac{1}{2} \bar{s}_{4}^2s^2 -1\right) \text{erfc}\left(\frac{\bar{s}_{4}s }{2}\right)\right)}{4 \left(4 \sqrt{2}-5\right) \sqrt{\pi }} &   (m,n) = (1,4), \,
 \end{cases}
\end{equation}
\end{widetext}
with
\begin{align}
    \bar{s}_2 &\equiv \frac{2 \left( 1 + \sqrt{2} \right) \left( 2 - \sqrt{2} \sinh ^{-1}(1) \right)}{\sqrt{\pi }} \approx 2.053, \\
    \bar{s}_3 &\equiv \frac{6-\sqrt{2} \sinh ^{-1}(1)}{\left(2 \sqrt{2}-1\right) \sqrt{\pi }} \approx 1.467, \\ \bar{s}_4 &\equiv \frac{20-14 \sqrt{2} \sinh ^{-1}(1)}{\left(4 \sqrt{2}-5\right) \sqrt{\pi }} \approx 2.190.
\end{align}
Note that $p(s)$ takes the following asymptotic forms for $s \ll 1$,   
\begin{equation}
p(s) \sim 
\begin{cases} s & (m,n )= (2,1),(1,2),(3,2),(1,4), \\
 -s\log(s) & (m,n) = (1,1). \,  \\ 
 \end{cases}
\end{equation}

\subsection{Density of states (DoS)}
We calculate the
DoS of the small random matrices. %
For $m\ne 0$, the DoS $\rho(E=x+{\rm i}y)$ is decomposed into
the %
density $\rho_c(x,y)$ of complex eigenvalues and the density $\rho_r(x)$ of real eigenvalues, %
$\rho(E)=\rho_c(x,y)+\delta(y)\rho_r(x)$,
while for $m=0$, %
the density of real eigenvalues always vanishes, %
$\rho(E)=\rho_c(x,y)$. 
The density $\rho_c(x,y)$ of complex eigenvalues is %
the
probability of $E$ being $x+{\rm i}y$. For $m \ne 0$, it is given by 
\begin{equation}
\begin{aligned}
        \rho_c(x,y) &= \frac{2 \lvert y \rvert e^{- \frac{x^2}{2}}}{\sqrt{2 \pi}} 
        \int_{-\infty}^{\infty} dX \int_{-\infty}^{\infty}  dY \!\ \delta(Y-X-y^2 ) \\ 
        & \hspace{1cm} \times p(Y;n) \!\ p(X;m) \\        
& =   \begin{cases}  \sqrt{2} \lvert y \rvert e^{\frac{y^2}{2}-\frac{x^2}{2}}\text{erfc}(\lvert y \rvert) & (m,n)= (2,1), \\
\sqrt{2} \lvert y \rvert e^{-\frac{y^2}{2}-\frac{x^2}{2}} & (m,n) = (1,2), \\
 \sqrt{\frac{2}{\pi}} \lvert y \rvert K_{0}(\frac{y^2}{2}) e^{- \frac{x^2}{2}}  &(m,n) = (1,1), \\
  \sqrt{\frac{1}{2}} \lvert y \rvert e^{-\frac{y^2}{2}-\frac{x^2}{2}} & (m,n)= (3,2), \\
  \sqrt{\frac{1}{2}} (2y^2 + 1 ) \lvert y \rvert e^{-\frac{y^2}{2}-\frac{x^2}{2}} &  (m,n) = (1,4), \,  \\
 \end{cases}
\end{aligned}
\end{equation}
with %
the
normalization $\int^{\infty}_{-\infty}  dx \int^{\infty}_{-\infty} dy \rho(x,y)=2$. 
Here, the constant $2$ is the number of different eigenvalues of small matrices.
Note also that we have $\int^{\infty}_{-\infty} dx \int^{\infty}_{-\infty} dy \rho_{c} (x,y) \neq \int^{\infty}_{-\infty}  dx \int^{\infty}_{-\infty} dy \rho (x,y) = 2$ under this normalization condition.
For $m = 0$, the DoS in the complex plane $\rho(x,y)=\rho_{c}(x,y)$ is given by
\begin{equation}
  \begin{aligned}
        \rho_c(x,y) & = \frac{2|y|e^{-\frac{x^2}{2}}}{\sqrt{2 \pi}}   \int^{+\infty}_{-\infty} dY \!\ 
        \delta(Y-y^2) \!\  p(Y;n) \\
&= \frac{1}{ 2^{\frac{n-1}{2}} \sqrt{\pi} \Gamma(\frac{n}{2})}
\lvert y \rvert ^{n-1} e^{-\frac{y^2}{2}-\frac{x^2}{2}}\\
 & =\begin{cases} \frac{1}{\sqrt{2 \pi}}  \lvert y \rvert e^{-\frac{y^2}{2}-\frac{x^2}{2}} & (m,n )= (0,2), \\
 \frac{1}{\pi}  \lvert y \rvert^2 e^{-\frac{y^2}{2}-\frac{x^2}{2}} & (m,n )= (0,3), \,
 \end{cases}
\end{aligned}  
\end{equation}
with the same normalization $\int^{\infty}_{-\infty}  dx \int^{\infty}_{-\infty} dy \rho(x,y)=2$. 
For $%
|y|
\ll 1$, the density $\rho_c(x,y)$ of complex eigenvalues takes the following asymptotic forms for each symmetry class, %
\begin{equation}
\rho_c(x,y) \sim 
\begin{cases} \lvert y \rvert & (m,n )= (2,1),(1,2),(3,2),\\
  &  \hspace{1.4cm} (1,4),(0,2), \\
  \lvert y \rvert^2 & (m,n )= (0,3), \\ 
 -\lvert y \rvert \log(\lvert y \rvert) & (m,n) = (1,1). \, \\ 
 \end{cases}
\end{equation}

\section{%
Random matrix ensembles}
\label{sec_be}

\subsection{Class AI}
For a random matrix in class AI, let ${\cal U_T}$ %
be the identity matrix $I$. %
In this choice, ${\cal H}$ is a real matrix.
The probability distribution function in the Gaussian ensemble is given by 
\begin{equation}
p({\cal H}) d{\cal H} 
= C_N \exp{ \left( -\beta\sum_{ij} {\cal H}_{ij}^2\right) } \prod_{ij} d{\cal H}_{ij}  \, ,
\end{equation}
where $\beta$ is a constant and $C_N$ is a %
normalization
constant. 
With 
${\cal U_T} = I$, the probability distribution function in the 
Bernoulli ensemble is given by 
\begin{equation}
H_{ij}= \begin{cases}
1 & \text{with the probability $1/2$}, \\ %
-1& \text{with the probability $1/2$}. %
\end{cases}
\end{equation}
Notably, the probability distribution in the Bernoulli ensemble is not 
invariant under unitary transformations.

\subsection{Class AI + $\eta_+$}

For a random matrix in class AI + $\eta_+$, let us 
choose ${\cal U_T} = I$ and ${\cal U}_{\eta} = \sigma_z \otimes I_{\frac{N}{2}\times \frac{N}{2}}$ with %
the
identity matrix $I_{\frac{N}{2}\times \frac{N}{2}}$.
Then, 
${\cal H}_{{\rm AI} + \eta_{+}}$ generally takes
\begin{equation}
{\cal H}_{\text{AI} +\eta_+}   = \left( \begin{matrix}
A & B \\
-B^T & C \\
\end{matrix} \right) \, ,
\end{equation}
where $A,B,C$ are $\frac{N}{2}\times \frac{N}{2}$ real matrices with  
\begin{equation}
A_{ij} = A_{ji} \,  ,C_{ij} = C_{ji} \, .
\label{A_BDId}
\end{equation}
The probability distribution function in the Gaussian ensemble is 
\begin{equation}
\begin{aligned}
p({\cal H}) d{\cal H} 
&= C_N \exp{ \left\{ -\beta\left[ 2\sum_{i,j} B_{ij}^2  +  2\sum_{i>j} \left( A_{ij}^2 +C_{ij}^2\right) \right.\right. }\\
&{\left.\left.+ \sum_{i} \left( A_{ii}^2 +C_{ii}^2\right) \right]\right\}  } \prod_{i,j} dB_{ij}  \prod_{i\geq j} dA_{ij} dC_{ij} \, .
\end{aligned}
\end{equation}
The probability distribution function in the Bernoulli ensemble is 
\begin{equation}
\begin{aligned}
&B_{ij},A_{ij}(i \geq j),C_{ij}(i \geq j) \\
& = \begin{cases}
1 & \text{with the probability 1/2}, \\ %
-1& \text{with the probability 1/2}, %
\end{cases}
\end{aligned}
\end{equation}
with Eq. (\ref{A_BDId}).

\subsection{Class AI + $\eta_-$}
For a random matrix in class AI+ $\eta_-$, let us choose 
${\cal U_T} = I$ and ${\cal U}_{\eta} = \sigma_y \otimes I_{\frac{N}{2}\times \frac{N}{2}}$. 
Then, ${\cal H}_{\text{AI} +\eta_-}$ takes a form of   
\begin{equation}
{\cal H}_{\text{AI} +\eta_-}   = \left( \begin{matrix}
A & B \\
C & A^T \\
\end{matrix} \right) \, ,
\end{equation}
where $A,B,C$ are $\frac{N}{2}\times \frac{N}{2}$ real matrices satisfying, 
\begin{equation}
B_{ij} = -B_{ji} \,  ,C_{ij} = -C_{ji} \, .
\label{A_DIIId}
\end{equation}
The probability distribution function in the Gaussian ensemble is 
\begin{equation}
\begin{aligned}
p({\cal H}) d{\cal H} &= C_N \exp{ \left\{ -2\beta\left[ \sum_{i,j} A_{ij}^2  +  \sum_{i>j} \left( B_{ij}^2 +C_{ij}^2\right) \right]\right\} }\\
& \prod_{i,j} dA_{ij}  \prod_{i> j} dB_{ij} dC_{ij} \, .
\end{aligned}
\end{equation}
The probability distribution function in the Bernoulli ensemble is 
\begin{equation}
\begin{aligned}
& A_{ij},B_{ij}(i > j),C_{ij}(i > j) \\ %
&\quad = \begin{cases}
1 & \text{with the probability 1/2}, \\ %
-1& \text{with the probability 1/2}, %
\end{cases}
\end{aligned}
\end{equation}
with Eq. (\ref{A_DIIId}). 

\subsection{Class A + $\eta$}
For a random matrix in class A with pH, let ${\cal U}_{\eta}$ %
be $\sigma_z \otimes I_{\frac{N}{2}\times \frac{N}{2}}$. %
Then, 
${\cal H}_{{\rm A} + \eta}$ is given by  
\begin{equation}
{\cal H}_{{\rm A}+\eta}   = \left( \begin{matrix}
A & B \\
-B^{\dagger} & C \\
\end{matrix} \right) \, ,
\end{equation} 
where
$A,B,C$ are $\frac{N}{2}\times \frac{N}{2}$ matrices satisfying 
\begin{equation}
A_{ij} = A_{ji}^* \,  ,C_{ij} = C_{ji}^* \, .
\label{A_AIIId}
\end{equation}
The probability distribution function in the Gaussian ensemble is given by 
\begin{equation}
\begin{aligned}
p({\cal H}) d{\cal H} &= C_N \exp{ \left\{ -\beta\left[ \sum_{i} \left( A_{ii}^2+C_{ii}^2 \right)  \right.\right. }\\
&{\left.\left.   +  2 \sum_{i>j} \left( |A_{ij}|^2 +|C_{ij}|^2\right) +   2 \sum_{i,j} |B_{ij}|^2 \right]\right\} }\\
& \prod_{i>j} dA_{ij}dA_{ij}^* dC_{ij}dC_{ij}^* \prod_{i} dA_{ii}dC_{ii} \prod_{i,j} dB_{ij} dB_{ij}^* \, .
\end{aligned}
\end{equation}
The probability distribution function in the Bernoulli ensemble is given by 
\begin{equation}
A_{ii},C_{ii} = \begin{cases}
1 & \text{with the probability 1/2}, \\ %
-1& \text{with the probability 1/2}, %
\end{cases}
\end{equation}
and
\begin{equation}
\begin{aligned}
& A_{ij}(i>j),C_{ij}(i>j),B_{ij}  \\ %
&\quad=\begin{cases}
1 + {\rm i} & \text{with the probability 1/4}, \\ %
-1 + {\rm i }& \text{with the probability 1/4}, \\
1 - {\rm i }& \text{with the probability 1/4}, \\ %
-1 - {\rm i }& \text{with the probability 1/4}, %
\end{cases}
\end{aligned}
\end{equation}
and Eq. (\ref{A_DIIId}). 

\begin{figure*}[tb]
	\centering
	\subfigure[class AI]{
		\begin{minipage}[t]{0.32\linewidth}
			\centering
			\includegraphics[width=1\linewidth]{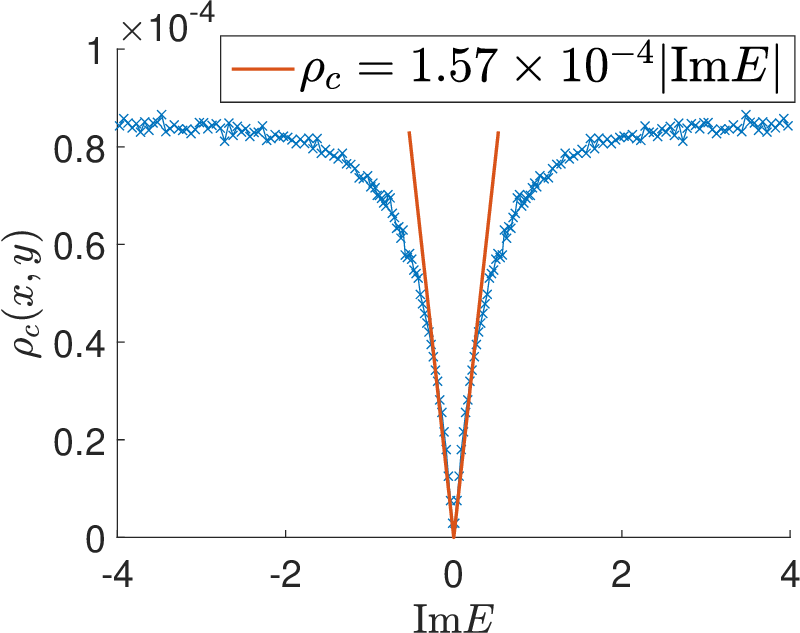}
		\end{minipage}%
	}%
	\subfigure[class AII]{
		\begin{minipage}[t]{0.32\linewidth}
			\centering
			\includegraphics[width=1\linewidth]{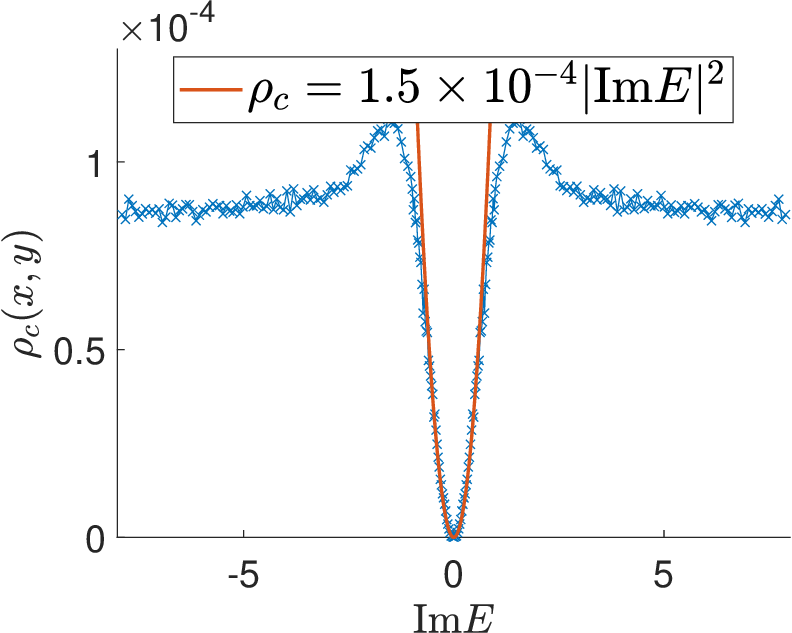}
		\end{minipage}
	}
	\subfigure[class AI + $\eta_+$]{
		\begin{minipage}[t]{0.32\linewidth}
			\centering
			\includegraphics[width=1\linewidth]{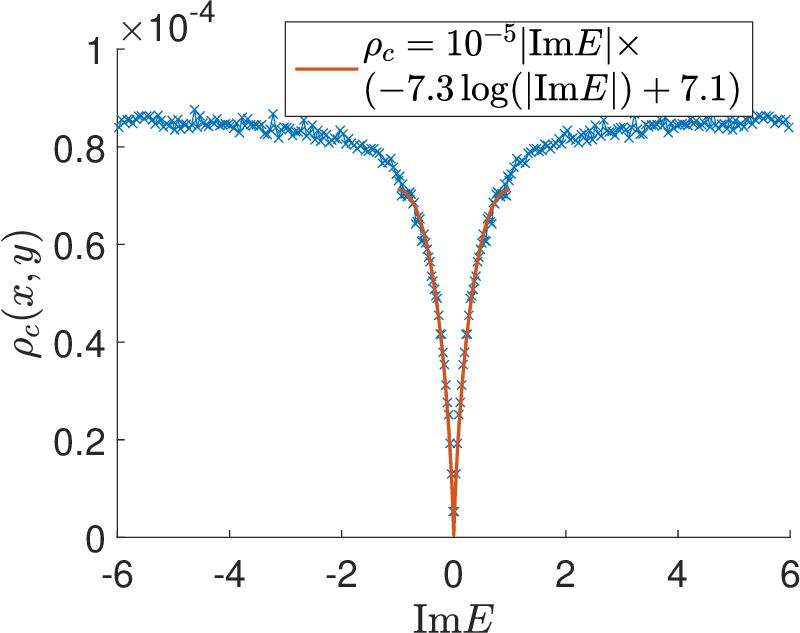}
		\end{minipage}%
	}%
	
	\subfigure[class A + $\eta$]{
		\begin{minipage}[t]{0.25\linewidth}
			\centering
			\includegraphics[width=1\linewidth]{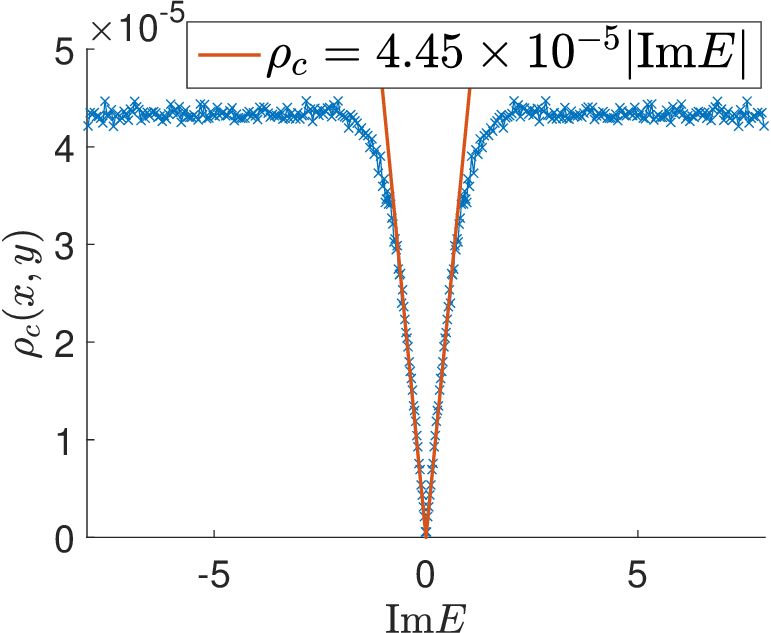}
		\end{minipage}%
	}%
	\subfigure[class AI + $\eta_-$]{
		\begin{minipage}[t]{0.25\linewidth}
			\centering
			\includegraphics[width=1\linewidth]{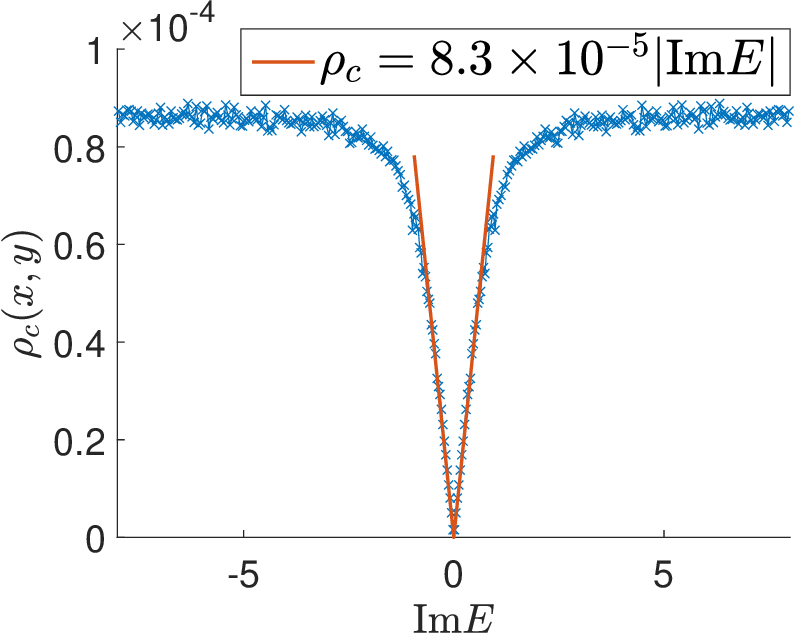}
		\end{minipage}%
	}%
	\subfigure[class AII + $\eta_+$]{
		\begin{minipage}[t]{0.25\linewidth}
			\centering
			\includegraphics[width=1\linewidth]{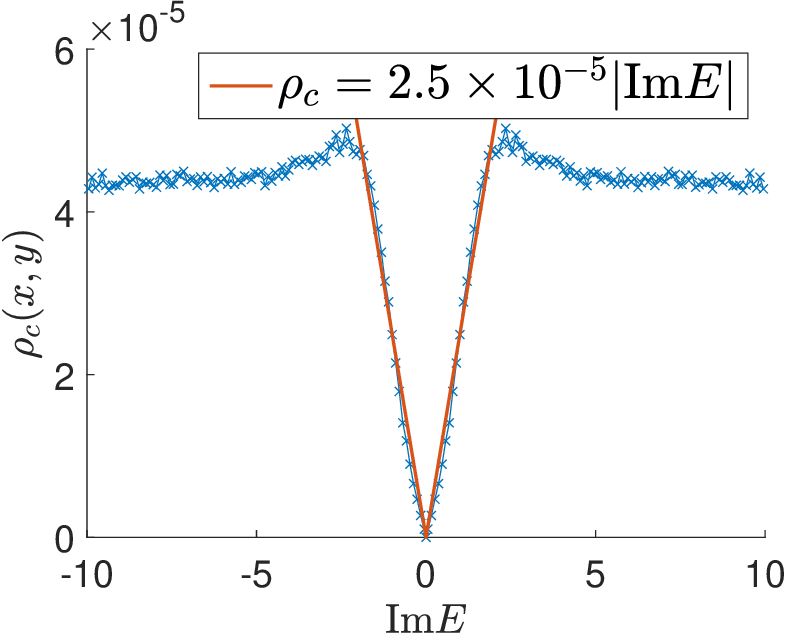}
		\end{minipage}%
	}%
	\subfigure[class AII + $\eta_-$]{
		\begin{minipage}[t]{0.25\linewidth}
			\centering
			\includegraphics[width=1\linewidth]{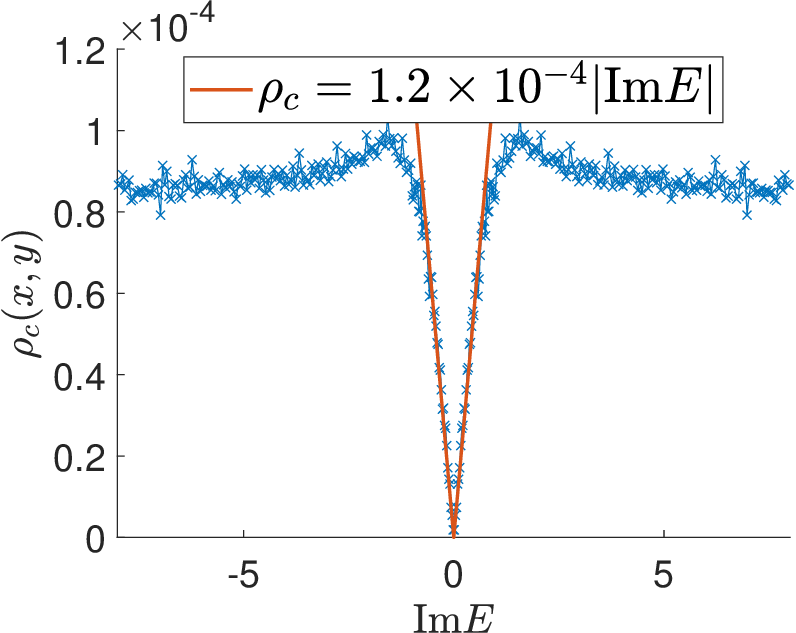}
		\end{minipage}%
	}%
	\caption{Density $\rho_c(x,y)$ of complex eigenvalues
	for non-Hermitian
	random matrices in the Bernoulli ensemble for classes (a) AI, (b) AII, (c) AI + $\eta_+$, (d) A + $\eta$, (e) AI + $\eta_-$, (f) AII + $\eta_+$, and (g) AII + $\eta_-$. Here, $\rho_c(x,y)$ is shown as a function of %
	$y=\mathrm{Im}\left( E \right)$
	for fixed $x = \mathrm{Re} \left(E\right)$ near the real axis $y=0$ of %
	complex energy $E$. %
	The data are
	obtained from diagonalizations of 5000 samples of $4000 \times 4000$ random matrices in each symmetry class. Note that $\rho_c(x,y)$ is almost independent of $x$ %
	as long as
	$E$ is away from 
	the 
	boundary of a circle inside which the complex eigenvalues $E$ distribute.} %
\label{2d_DoS_BE}
\end{figure*}

\subsection{Class AII}
For a random matrix in class AII, let 
${\cal U_T}$ %
be $\sigma_y \otimes I_{\frac{N}{2}\times \frac{N}{2}}$. %
Then,
${\cal H}_{\rm AII}$ is given by 
\begin{equation}
{\cal H}_{\text{AII}}   = \left( \begin{matrix}
A & B \\
-B^* & A^* \\
\end{matrix} \right) \, ,
\end{equation}
with $\frac{N}{2}\times \frac{N}{2}$ matrices $A$ and $B$.
The probability distribution function in the Gaussian ensemble is 
\begin{equation}
\begin{aligned}
p({\cal H}) d{\cal H}& = C_N \exp{ \left[ -2\beta\sum_{ij} \left(|A_{ij}|^2 + |B_{ij}|^2 \right)\right] } \\
&\qquad\qquad\qquad\prod_{ij} dA_{ij} dA_{ij}^* dB_{ij}dB_{ij}^* \,.
\end{aligned}
\end{equation}
The probability distribution function in the Bernoulli ensemble is 
\begin{equation}
A_{ij},B_{ij} = \begin{cases}
1 + {\rm i} & \text{with the probability 1/4}, \\ %
-1 + {\rm i }& \text{with the probability 1/4}, \\
1 - {\rm i }& \text{with the probability 1/4}, \\
-1 - {\rm i }& \text{with the probability 1/4}. %
\end{cases}
\end{equation}

\begin{figure}[bt]
	\centering
	\includegraphics[width=1\linewidth]{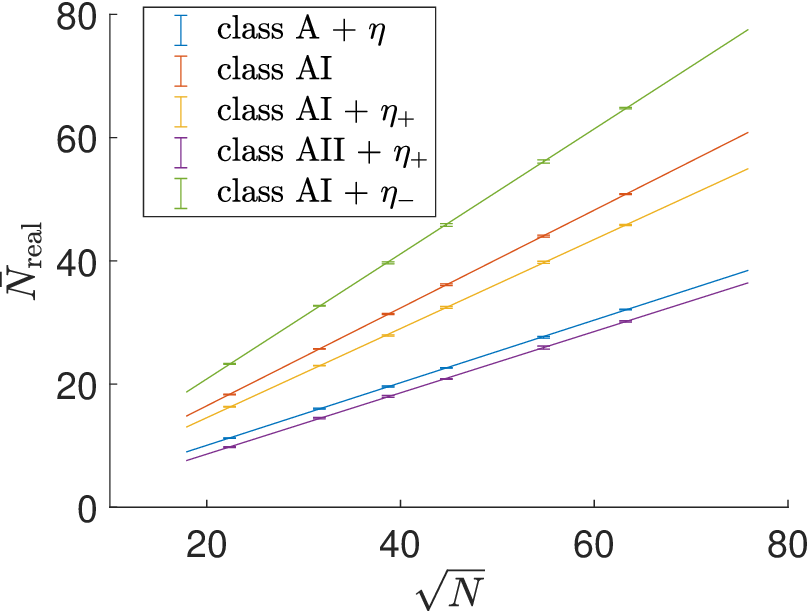}
	\caption{Average number $\bar{N}_{\rm real}$ of real eigenvalues of non-Hermitian random matrices in the Bernoulli ensemble %
	as a function of $\sqrt{N}$ for the five symmetry classes. 
	Here, $N$ is %
	the dimensions
	of the random matrices. %
	The error bars in the plot stand %
	for %
	the
	standard deviations
	of ${\bar N}_\text{real}$. The plots clearly demonstrate %
	the square-root scaling ${\bar N}_{\text{real}} \propto \sqrt{N}$ in all the five symmetry classes. For each symmetry class and matrix size, ${\bar N}_\text{real}$ from the Bernoulli ensemble is almost the same as $\bar{N}_{\rm real}$ from the Gaussian ensemble.}
	\label{N_real_BE}
\end{figure}
\begin{figure*}[bt]
	\centering
	\subfigure[$p(s)$ in class AI]{
		\begin{minipage}[t]{0.28\linewidth}
			\centering
			\includegraphics[width=1\linewidth]{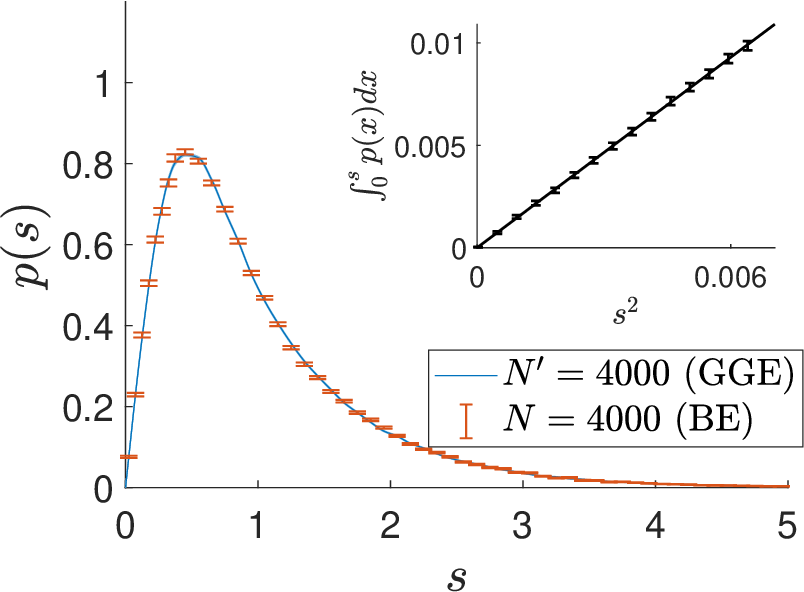}
		\end{minipage}%
	}%
	\subfigure[$p(s)$ in class AI + $\eta_+$]{
		\begin{minipage}[t]{0.28\linewidth}
			\centering
			\includegraphics[width=1\linewidth]{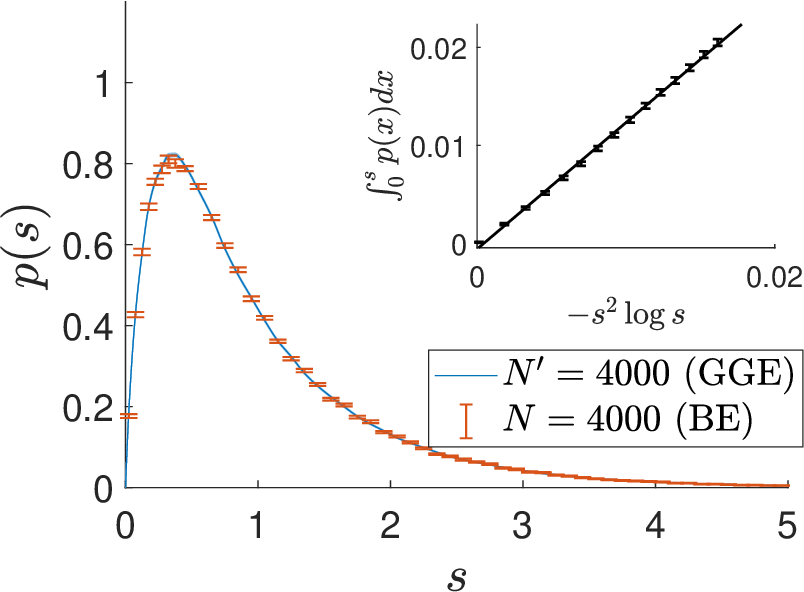}
		\end{minipage}%
	}%
    \subfigure[$p(s)$ in class A + $\eta$]{
		\begin{minipage}[t]{0.28\linewidth}
			\centering
			\includegraphics[width=1\linewidth]{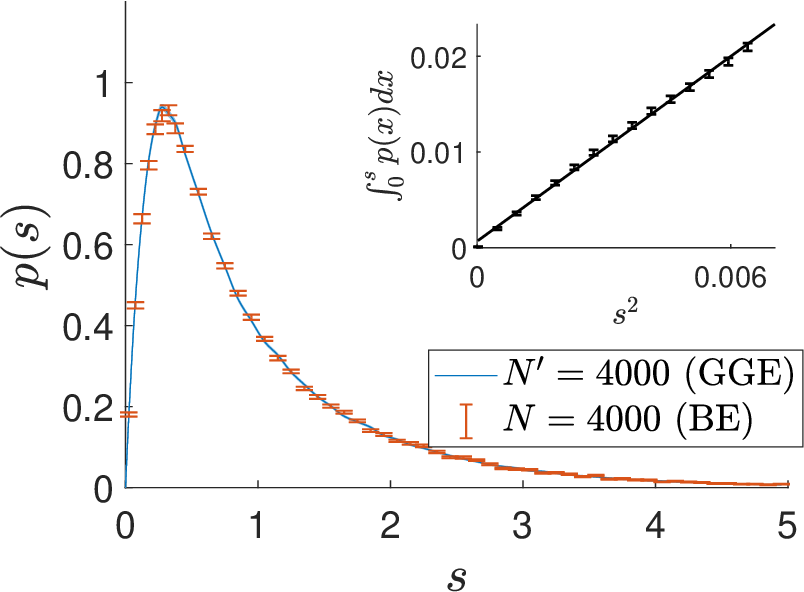}
		\end{minipage}%
	}%
	
	\subfigure[$p(s)$ in class AI + $\eta_-$]{
		\begin{minipage}[t]{0.28\linewidth}
			\centering
			\includegraphics[width=1\linewidth]{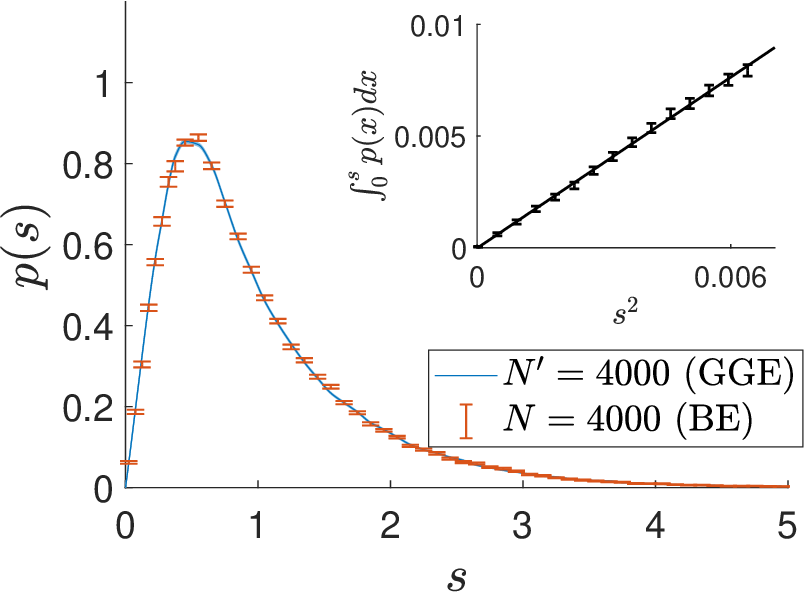}
		\end{minipage}%
	}
	\subfigure[$p(s)$ in class AII + $\eta_+$]{
		\begin{minipage}[t]{0.28\linewidth}
			\centering
			\includegraphics[width=1\linewidth]{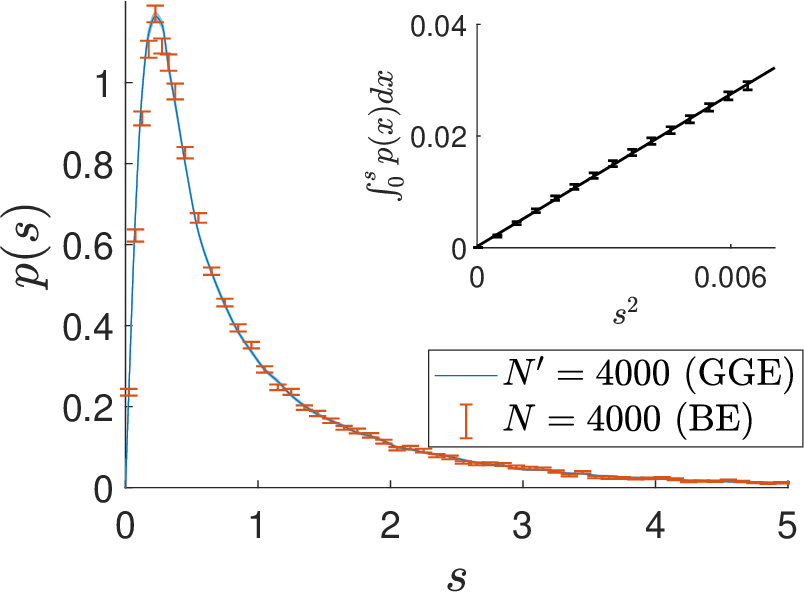}
		\end{minipage}%
	}
	\caption{Real-eigenvalue spacing distribution functions $p(s)$ of $4000 \times 4000$ non-Hermitian random matrices in the Bernoulli ensemble (BE) for the five symmetry classes and their comparisons to $p(s)$ obtained from non-Hermitian random matrices in the generalized Gaussian ensemble (GGE) with $\beta_2/\beta_1 = 16$. Insets: Asymptotic behaviors
	of the distribution functions for $s \ll 1$, where 
	the cumulative distribution function $\int_0^s p(s^{\prime}) ds^{\prime}$ is plotted as a function of either $s^2$ or $-\log(s) s^2$. 
    The error ranges are evaluated by the bootstrap method~\cite{press07}.
   The error ranges of $p(s)$ of the GGE random matrices are much smaller than those of the BE random matrices and not shown here (see also Fig.~\ref{ps_all}).%
 }
	\label{BE_ps}
\end{figure*}
\begin{figure*}[hbt]
	\centering
	\subfigure[$p_r(r)$ in class AI]{
		\begin{minipage}[t]{0.28\linewidth}
			\centering
			\includegraphics[width=1\linewidth]{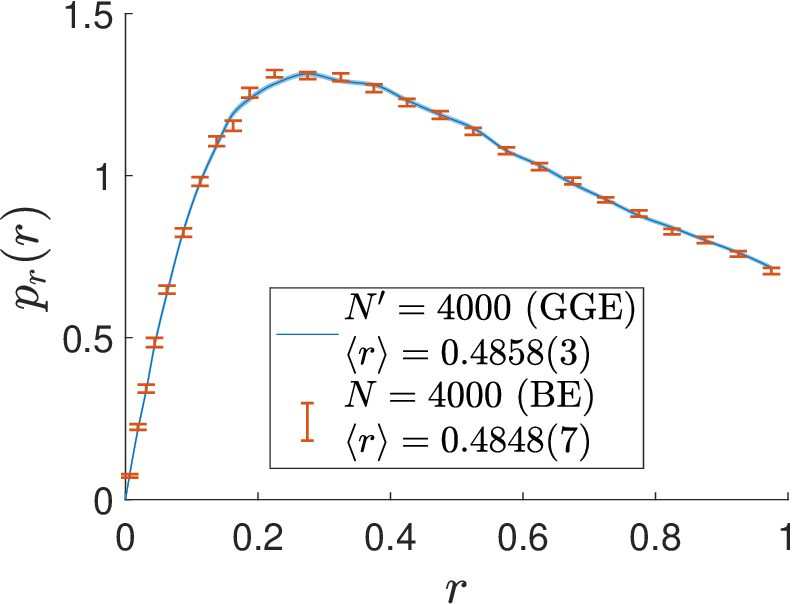}
		\end{minipage}%
	}%
	\subfigure[$p_r(r)$ in class AI + $\eta_+$]{
		\begin{minipage}[t]{0.28\linewidth}
			\centering
			\includegraphics[width=1\linewidth]{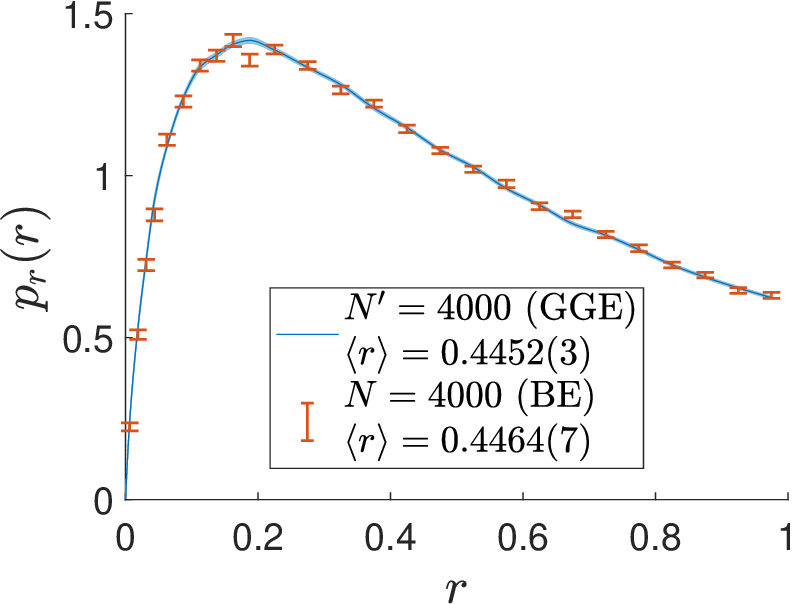}
		\end{minipage}%
	}%
    \subfigure[$p_r(r)$ in class A + $\eta$]{
		\begin{minipage}[t]{0.28\linewidth}
			\centering
			\includegraphics[width=1\linewidth]{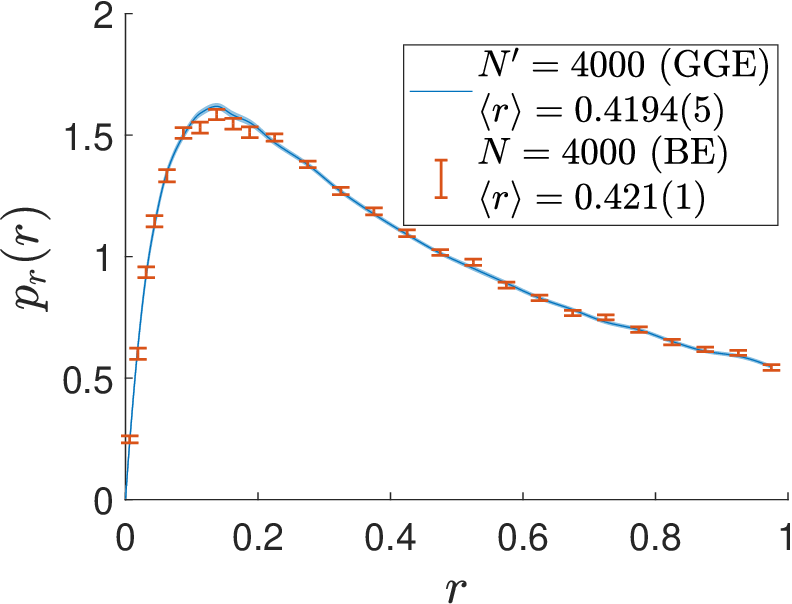}
		\end{minipage}%
	}%
	
	\subfigure[$p_r(r)$ in class AI + $\eta_-$]{
		\begin{minipage}[t]{0.28\linewidth}
			\centering
			\includegraphics[width=1\linewidth]{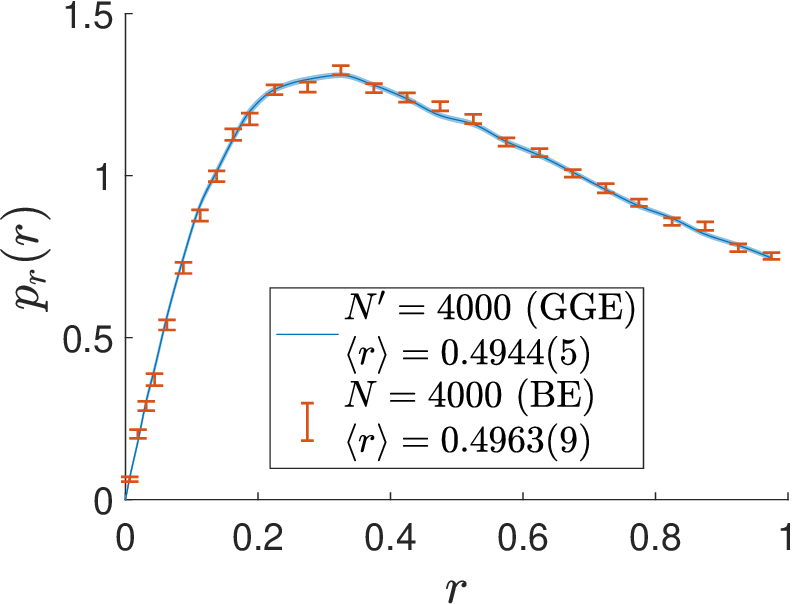}
		\end{minipage}%
	}
	\subfigure[$p_r(r)$ in class AII + $\eta_+$]{
		\begin{minipage}[t]{0.28\linewidth}
			\centering
			\includegraphics[width=1\linewidth]{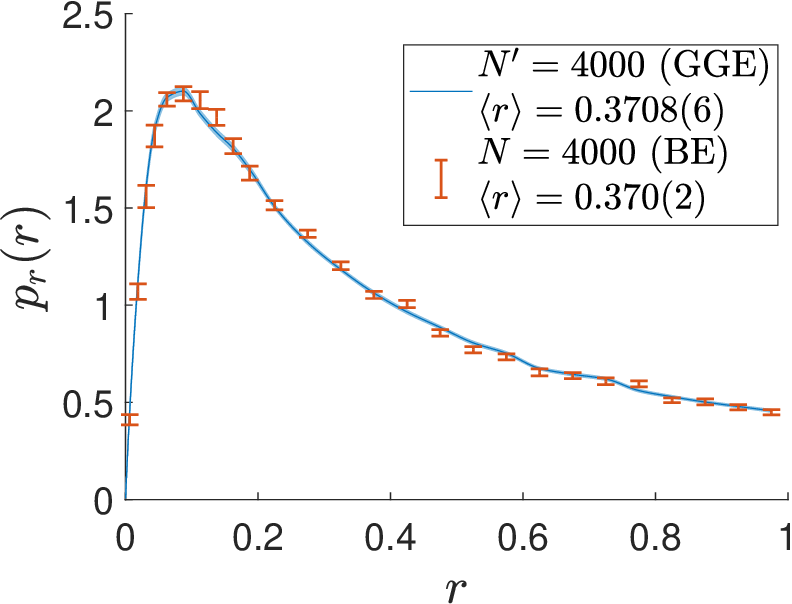}
		\end{minipage}%
	}
	\caption{Level-spacing-ratio distributions $p_r(r)$ of real eigenvalues obtained from $4000 \times 4000$ non-Hermitian random matrices in the Bernoulli ensemble (BE) for %
	the five symmetry classes and their comparisons to $p_r(r)$ obtained from non-Hermitian random matrices in the generalized Gaussian ensemble (GGE) with $\beta_2/\beta_1 = 16$. The mean value $\langle r \rangle = \int_0^1 p_r(r)dr$ of each distribution is also shown in each figure. 
    The error ranges are evaluated by the bootstrap method~\cite{press07}.
    The error ranges of $p_r(r)$ of the GGE random matrices are much smaller than those of the BE random matrices and not shown here (see also Fig.~\ref{pr_all}).}%
	\label{BE_pr}
\end{figure*}

\begin{figure}[bt]
    \centering
    \includegraphics[width=0.9\linewidth]{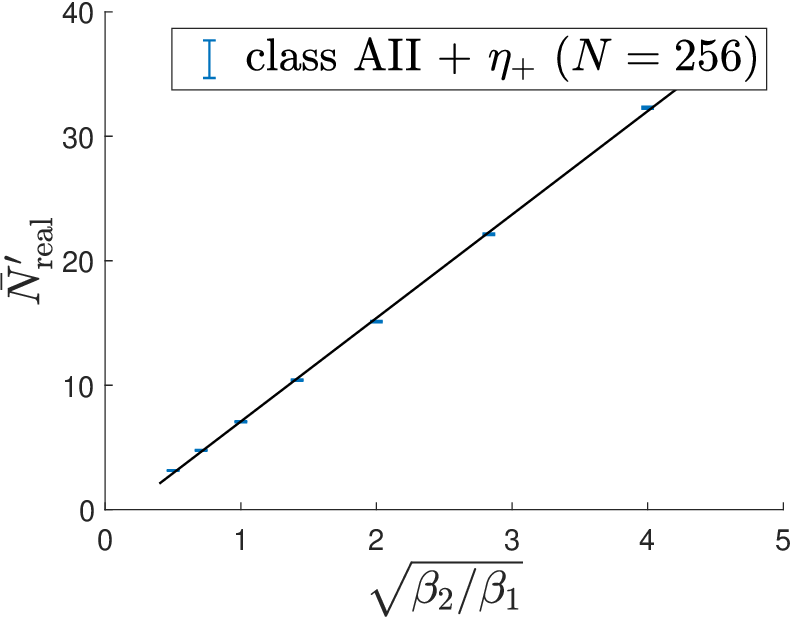}
    \caption{Average number $\bar{N}^{\prime}_{\rm real}$ for real eigenvalues of $256\times256$ non-Hermitian random matrices in the generalized Gaussian ensemble as a function of $\beta_2/\beta_1$ (class AII + $\eta_{+}$). Here, $\beta_1$ and $\beta_2$ are the parameters of the generalized Gaussian ensemble in Eq.~(\ref{p_gge}). The plot clearly demonstrates $\bar{N}^{\prime}_{\rm real} \propto \sqrt{\beta_2/\beta_1}$. For random matrices with different sizes and in the other four symmetry classes, the %
    same scaling
    relation also holds true.}
    \label{n_real_beta}
\end{figure}

\begin{figure*}[tb]
	\centering
	\subfigure[$p(s)$ in class AII + $\eta_+$ \quad \quad \quad ($N^{\prime} =128$ vs. $N= 2000$)]{
		\begin{minipage}[t]{0.3\linewidth}
			\centering
			\includegraphics[width=1\linewidth]{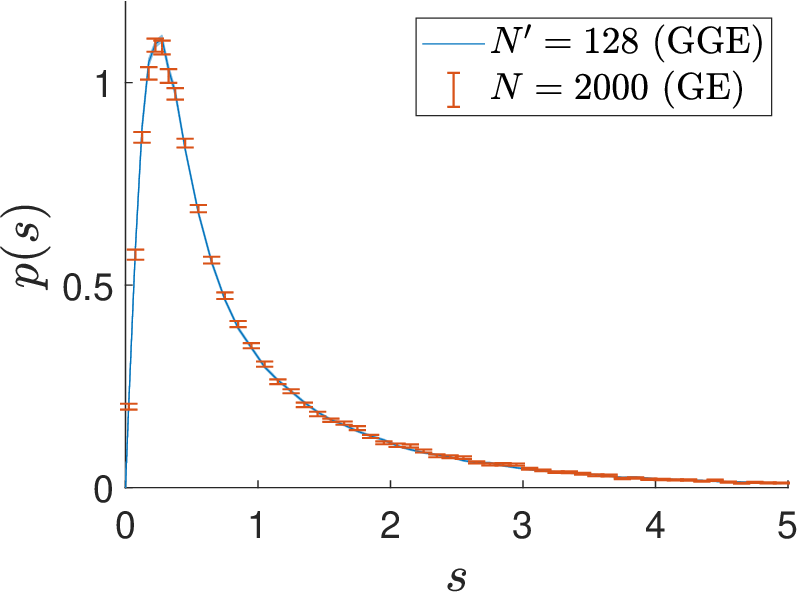}
			\label{mean_r_ge_gge}
		\end{minipage}%
	}
	\subfigure[$p(s)$ in class AII + $\eta_+$ \quad \quad \quad ($N^{\prime} =252$ vs. $N= 4000$)]{
		\begin{minipage}[t]{0.3\linewidth}
			\centering
			\includegraphics[width=1\linewidth]{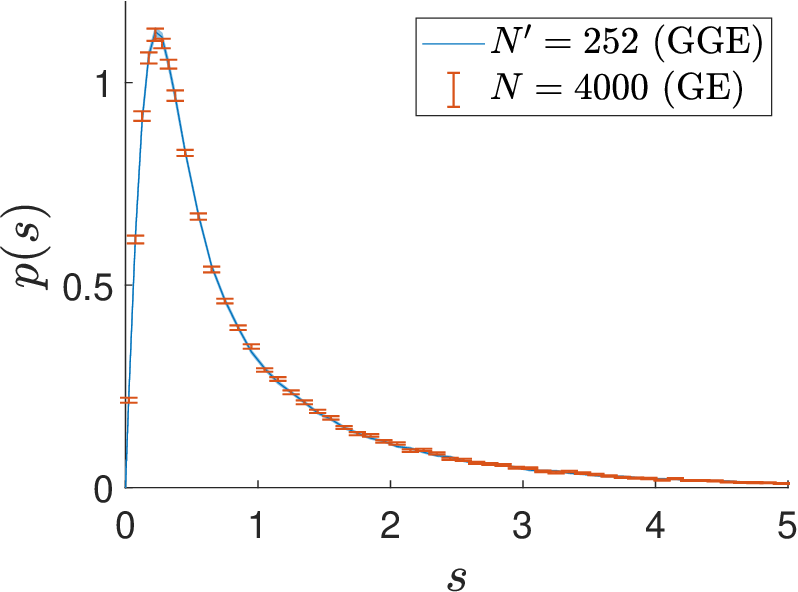}
		\end{minipage}%
	}

	\subfigure[$p_r(r)$ in class AII + $\eta_+$ \quad \quad ($N^{\prime} =128$ vs. $N= 2000$) ]{
		\begin{minipage}[t]{0.3\linewidth}
			\centering
			\includegraphics[width=1\linewidth]{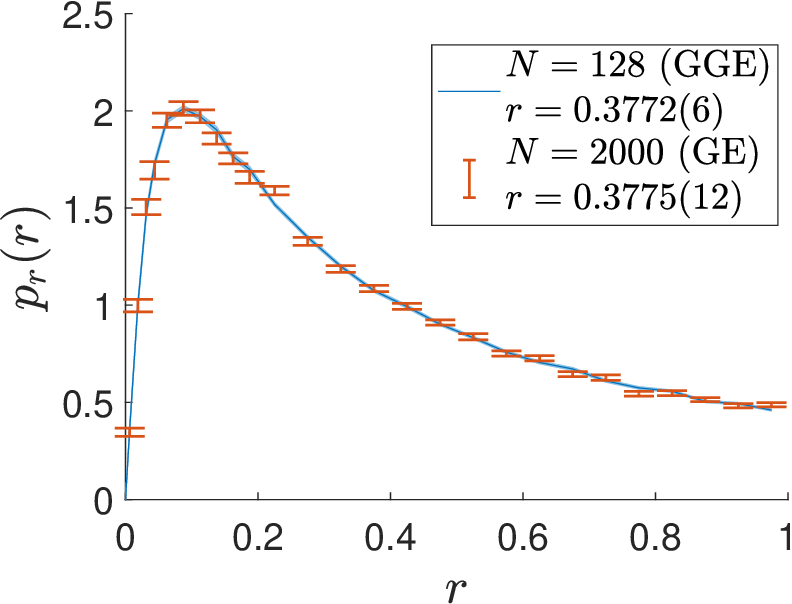}
		\end{minipage}%
	}%
	\subfigure[$p_r(r)$ in class AII + $\eta_+$ \quad \quad ($N^{\prime} =252$ vs. $N= 4000$)]{
		\begin{minipage}[t]{0.3\linewidth}
			\centering
			\includegraphics[width=1\linewidth]{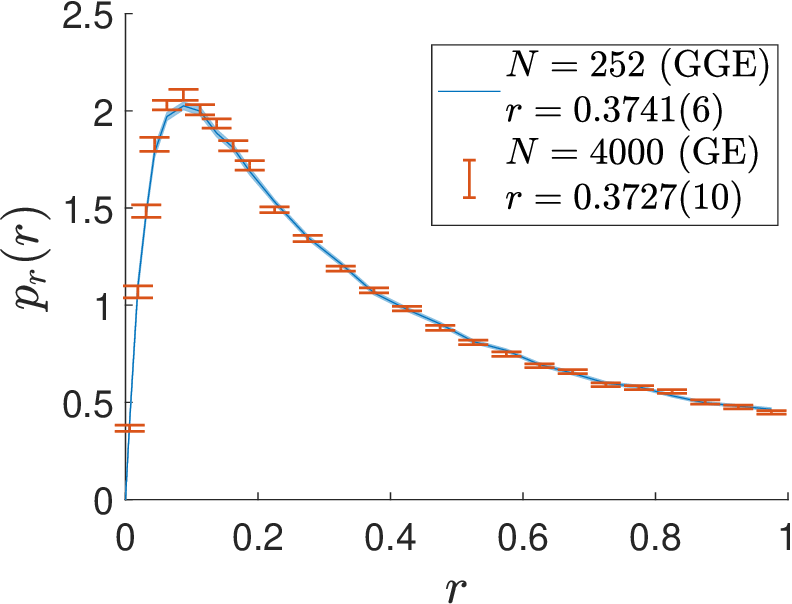}
		\end{minipage}%
	}%
    \subfigure[$\langle r \rangle $ in class AII + $\eta_+$ ]{
		\begin{minipage}[t]{0.3\linewidth}
			\centering
			\includegraphics[width=1\linewidth]{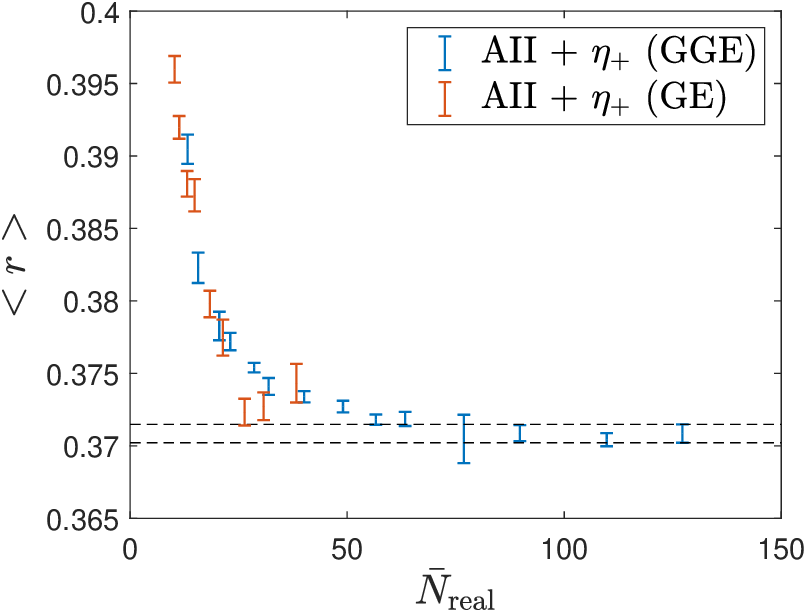}
		\end{minipage}%
		\label{CII_mean_r}
	}%
	\caption{(a)-(d)~Comparison between level-spacing-ratio distributions $p_r(r)$ and level-spacing distributions $p(s)$ of real eigenvalues obtained from $N^{\prime} \times N^{\prime}$ non-Hermitian random matrices in the generalized Gaussian ensemble (GGE) and $N \times N$ non-Hermitian random matrices in the Gaussian ensemble (GE) for class AII + $\eta_+$. 
	(e)~Mean value of the level-spacing ratio $\langle r \rangle = \int_0^1 p_r(r)dr$ as a function of the average number $\bar{N}_{\rm real}$ of real eigenvalues in the GE and GGE for class AII + $\eta_+$. For $\bar{N}_{\rm real} \gtrsim 70$ ($N^{\prime} > 1000$), the error bars of $\langle r \rangle$ for different sizes overlap with one another. 
    The error ranges are evaluated by the bootstrap method~\cite{press07}.
    The error ranges of the distributions of the GGE random matrices are much smaller than those of the GE random matrices and not shown here (see also Figs.~\ref{ps_all} and \ref{pr_all}).%
 }
	\label{compare_elip}
\end{figure*}

\begin{figure*}[pbt]
	\centering
	\subfigure[generalized Gaussian ensemble ($\beta_2/\beta_1 = 16$)]{
		\begin{minipage}[t]{0.3\linewidth}
			\centering
			\includegraphics[width=1\linewidth]{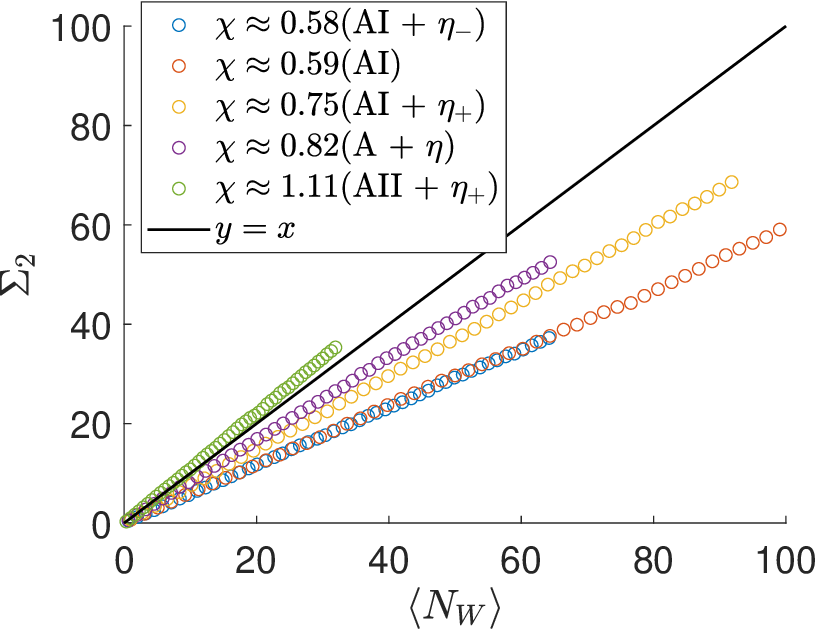}
		\end{minipage}%
	}%
	\subfigure[Gaussian ensemble]{
		\begin{minipage}[t]{0.3\linewidth}
			\centering
			\includegraphics[width=1\linewidth]{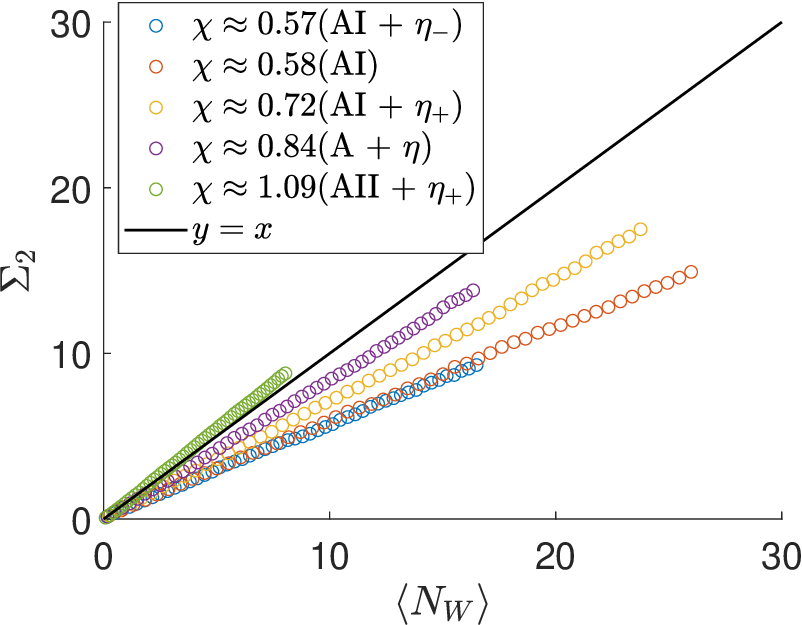}
		\end{minipage}%
	}%
    \subfigure[Bernoulli ensemble]{
		\begin{minipage}[t]{0.3\linewidth}
			\centering
			\includegraphics[width=1\linewidth]{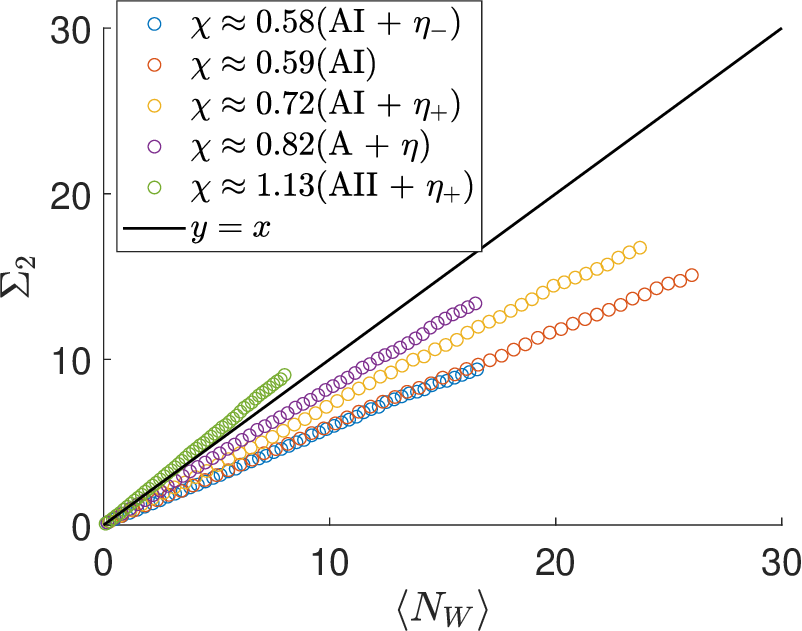}
		\end{minipage}%
	}%
	\caption{Variance $\Sigma_2$ and mean value $\langle N_W \rangle$ of the number of real eigenvalues of non-Hermitian random matrices in the five symmetry classes. $\Sigma_2$ and $\langle N_W\rangle$ within different energy windows are obtained from $4000 \times 4000$ non-Hermitian random matrices in the (a)~generalized Gaussian ensemble with $\beta_2/\beta_1 = 16$, (b)~Gaussian ensemble, and (c)~Bernoulli ensemble.
    The scaling relation $\Sigma_2 \simeq \chi \langle N_W \rangle$ holds for all the five symmetry classes.
	The spectral compressibility $\chi$ is obtained by the linear fitting of the data points for each symmetry class. %
	}
	\label{nv}
\end{figure*}
\subsection{Class AII + $\eta_+$}

For a random matrix in class AII+ $\eta_+$, let us 
choose ${\cal U_T} = \tau_0\sigma_y \otimes I_{\frac{N}{4}\times \frac{N}{4}}$ and ${\cal U}_{\eta} = \tau_z\sigma_0 \otimes I_{\frac{N}{4}\times \frac{N}{4}}$ 
with %
the
identity matrix 
$ I_{\frac{N}{4}\times \frac{N}{4}}$. 
Then, ${\cal H}_{{\rm AII}+\eta_{+}}$ generally takes
\begin{equation}
{\cal H}_{\text{AI} +\eta_+}   = \left( \begin{matrix}
A^1 & A^2  & B^1 & B^2 \\
-A^{2*} & A^{1*} & -B^{2*} & B^{1*} \\
-B^{1\dagger} & B^{2T}  & C^1 & C^2 \\
-B^{2\dagger} & -B^{1T} & -C^{2*} & C^{1*} \\
\end{matrix} \right),
\end{equation}
where $A^{\mu},B^{\mu},C^{\mu} \, (\mu = 1,2)$ are $\frac{N}{4}\times \frac{N}{4}$ matrices satisfying  
\begin{equation}
A^1_{ij} = A^{1*}_{ji} \, ,  C^1_{ij} = C^{1*}_{ji} \, , A^2_{ij} = -A^2_{ji} \,  ,C^{2}_{ij} = -C^{2}_{ji} \, .
\label{A_CIId}
\end{equation}
The probability distribution function in the Gaussian ensemble is 
\begin{equation}
\begin{aligned}
p({\cal H}) d{\cal H} &= C_N \exp{ \left\{ -\beta\left[ 2\sum_{i} \left( (A^1_{ii})^2+(C^1_{ii})^2 \right)  \right.\right. }\\
&{\left.\left.   
+  %
4
\sum_{\mu = 1,2 }\sum_{i>j} \left( |A^{\mu}_{ij}|^2 +|C^{\mu}_{ij}|^2\right)  \right.\right. } \\
&+{\left.\left.    4 \sum_{\mu = 1,2}\sum_{i,j} |B_{ij}^{\mu}|^2 \right]\right\} }\\
& \prod_{i>j} dA^1_{ij}dA_{ij}^{1*} dA^2_{ij}dA_{ij}^{2*}  dC^1_{ij}dC_{ij}^{1*}  dC^2_{ij}dC_{ij}^{2*}  \\
& \prod_{i} dA^1_{ii}dC^1_{ii} \prod_{i,j} dB^{1}_{ij} dB^{1*}_{ij}dB^{2}_{ij} dB^{2*}_{ij}.
\end{aligned}
\end{equation}
The probability distribution function in the Bernoulli ensemble is 
\begin{equation}
A_{ii}^1,C_{ii}^1 = \begin{cases}
1 & \text{with the probability 1/2}, \\ %
-1& \text{with the probability 1/2}, %
\end{cases}
\end{equation}
\begin{equation}
\begin{aligned}
& A^{\mu}_{ij}(i>j),C^{\mu}_{ij}(i>j),B^{\mu}_{ij} \quad  %
\\
&\qquad = \begin{cases}
1 + {\rm i} & \text{with the probability 1/4}, \\ %
-1 + {\rm i }& \text{with the probability 1/4}, \\ %
-1 - {\rm i }& \text{with the probability 1/4}, \\ %
-1 - {\rm i }& \text{with the probability 1/4}, \\ %
\end{cases}
\end{aligned}
\end{equation}
for $\mu = 1, 2$ 
with Eq.~(\ref{A_CIId}).

\subsection{Class AII + $\eta_-$}
For a random matrix in class AII+ $\eta_-$, let us 
choose ${\cal U_T} = \sigma_y \otimes I_{\frac{N}{2}\times \frac{N}{2}}$ 
and ${\cal U}_{\eta} = \sigma_z \otimes I_{\frac{N}{2}\times \frac{N}{2}}$. %
Then, ${\cal H}_{{\rm AII} + \eta_{-}}$ is given by 
\begin{equation}
{\cal H}_{\text{AII}+\eta_-}   = \left( \begin{matrix}
A & B \\
-B^* & A^* \\
\end{matrix} \right) \, ,
\end{equation}
where 
$A,B$ are $\frac{N}{2}\times \frac{N}{2}$ matrices satisfying 
\begin{align}
A_{ij} = A_{ji}^*, B_{ij} = B_{ji}. \label{A_CId}
\end{align}
The probability distribution function 
in the Gaussian ensemble is given by 
\begin{equation}
\begin{aligned}
p({\cal H}) d{\cal H}& = C_N \exp{ \left\{ -\beta \left[\sum_{i} 2\left(A_{ii}^2+|B_{ii}|^2\right) \right.\right.} \\
&+ 
4
{\left.\left. \sum_{i>j}\left(|A_{ij}|^2+|B_{ij}|^2\right)\right]\right\} } \\
&\prod_{i>j} dA_{ij} dA_{ij}^* \prod_{i} dA_{ii} \prod_{i\geq j} dB_{ij}dB_{ij}^* \, .
\end{aligned}
\end{equation}
The probability distribution function in the 
Bernoulli ensemble is given by 
\begin{equation}
A_{ii} = \begin{cases}
1 & \text{with the probability 1/2}, \\ %
-1& \text{with the probability 1/2}, %
\end{cases}
\end{equation}
\begin{equation}
\begin{aligned}
& A_{ij}(i>j),B_{ij}(i\geq j)  %
\\
&\qquad = \begin{cases}
1 + {\rm i} & \text{with the probability 1/4}, \\ %
-1 + {\rm i }& \text{with the probability 1/4}, \\
1 - {\rm i }& \text{with the probability 1/4}, \\
-1 - {\rm i }& \text{with the probability 1/4}, \\
\end{cases}
\end{aligned}
\end{equation}
with Eq.~(\ref{A_CId}).

\subsection{Level statistics in the Bernoulli ensemble}

We diagonalize non-Hermitian random matrices in the Bernoulli ensemble %
for
the seven %
symmetry classes and 
obtain
the universal properties of the DoS as well as 
the level statistics of real eigenvalues %
in the Bernoulli ensemble. 
The DoS %
in each symmetry class shows the same property as the DoS %
in
the Gaussian ensemble. The soft-gap %
behaviors
of the the density $\rho_c(x,y)$ of complex eigenvalues around the real axis ($y=0$) %
are 
consistent with the DoS in the Gaussian ensemble (Fig.~\ref{2d_DoS_BE}). 
For %
non-Hermitian random matrices in classes AII and AII + $\eta_-$, %
no real eigenvalues appear.
By contrast, the average numbers of real eigenvalues  
in classes AI, A + $\eta$, AI + $\eta_{\pm}$, and AII + $\eta_{+}$ are 
proportional to $\sqrt{N}$ in the Bernoulli ensemble (Fig.~\ref{N_real_BE}). 
In these five symmetry classes, both of the %
level-spacing distribution $p(s)$ and level-spacing-ratio distribution $p_r(r)$
of real eigenvalues
in 
the Bernoulli ensemble %
are %
consistent with the distribution functions in the 
Gaussian ensemble (Figs.~\ref{BE_ps} and \ref{BE_pr}). 
These %
results
demonstrate the 
universality of the soft gap of the DoS around the real axis, %
the level-spacing and level-spacing-ratio distributions 
of real eigenvalues, and the scaling relation 
of the average number of real eigenvalues.

\begin{table}[tb]
	\caption{
	Kolmogorov-Smirnov (KS)
	distance among %
	level-spacing distribution functions $p(s)$ 
	of real eigenvalues obtained
	from 
	$4000 \times 4000$ non-Hermitian random matrices in 
	the
	five symmetry classes (classes A+$\eta$, AI, AI + $\eta_{\pm}$, and AII + $\eta_{+}$), and KS distance between $p(s)$ obtained 
	from the physical systems in %
	the ergodic
	phases and $p(s)$ obtained from the random matrices in the five classes.
	The first column specifies the random matrices (RM) %
	and Hamiltonians of 
	the physical models.
	The physical models are %
	the 2D disordered free fermions in class AII + $\eta_{+}$ (${\cal H}_{\rm 2D}$), the 3D disordered free fermions in class AI + $\eta_{+}$ (${\cal H}_{\rm 3D}$), the hard-core boson model with the nonreciprocal hopping (${\cal H}_{\rm HN}$), and the four interacting spin models (${\cal H}_1, {\cal H}_2, {\cal H}_3, {\cal H}_4$).
	The second column denotes  
	the symmetry classes to which the random matrices or the Hamiltonians 
	in the first column belong.  
	The first row labels  
	the %
	RM in the five symmetry classes. 
	The KS distances between $p(s)$ from the first column and $p(s)$ 
	from the first row are shown. 
	For each system in the first column, the shortest distance to 
	random matrices is highlighted with the bold characters.}
 \begin{tabular}{cc|ccccc}
	\hline \hline
	system & \makecell[c]{class}& AI & AI + $\eta_+$ & A + $\eta_+$ & AI + $\eta_-$ & AII + $\eta_+$ \\ \hline 
	RM &AI &\textbf{0}&0.052 &0.089 &0.011 &0.165 \\ 
RM &AI + $\eta_+$ &0.052 &\textbf{0} &0.038 &0.063 &0.114 \\ 
RM &A + $\eta_+$ &0.089 &0.038 &\textbf{0} &0.098 &0.077 \\ 
RM &AI + $\eta_-$ &0.011 &0.063 &0.098 &\textbf{0} &0.174 \\ 
RM &AII + $\eta_+$ &0.165 &0.114 &0.077 &0.174 &\textbf{0} \\ \hline 
${\cal H}_{\text{HN}}$ &AI &\textbf{0.010} &0.043 &0.079 &0.020 &0.155 \\ 
${\cal H}_{1}$ &AI &\textbf{0.009} &0.045 &0.081 &0.018 &0.157 \\ 
${\cal H}_{\text{3D}}$ &AI + $\eta_+$ &0.050 &\textbf{0.003} &0.040 &0.060 &0.115 \\ 
${\cal H}_3$ &AI + $\eta_+$ &0.055 &\textbf{0.004} &0.036 &0.066 &0.111 \\ 
${\cal H}_2$ &A + $\eta_+$ &0.093 &0.043 &\textbf{0.005} &0.102 &0.073 \\ 
${\cal H}_4$ &AI + $\eta_-$ &0.010 &0.062 &0.097 &\textbf{0.003} &0.174 \\ 
${\cal H}_{\text{2D}}$ &AII + $\eta_+$ &0.118 &0.066 &\textbf{0.032} &0.129 &0.052 \\ 
	\hline\hline
	\end{tabular}\label{ks_dis}
\end{table}

\begin{table}[bt]
	\caption{
	Kolmogorov-Smirnov (KS) distance among level-spacing-ratio distribution functions $p_r(r)$ of real eigenvalues obtained
	from $4000 \times 4000$ non-Hermitian random matrices in the five symmetry classes (classes A+$\eta$, AI, AI + $\eta_{\pm}$, and AII + $\eta_{+}$), and KS distance between $p_r(r)$ obtained 
	from the physical systems in the ergodic phases and $p_r(r)$ obtained from the random matrices in the five classes. The notation is the same as Table~\ref{ks_dis}.
	}
	 	 {\begin{tabular}{cc|ccccc}
	\hline \hline
	system & \makecell[c]{class}& AI & AI + $\eta_+$ & A + $\eta_+$ & AI + $\eta_-$ & AII + $\eta_+$ \\ \hline 
	RM &AI &\textbf{0} &0.070 &0.114 &0.016 &0.194 \\ 
	RM &AI + $\eta_+$ &0.070 &\textbf{0} &0.044 &0.085 &0.125 \\ 
	RM &A + $\eta_+$ &0.114 &0.044 &\textbf{0} &0.129 &0.083 \\ 
	RM &AI + $\eta_-$ &0.016 &0.085 &0.129 &\textbf{0} &0.209 \\ 
	RM &AII + $\eta_+$ &0.194 &0.125 &0.083 &0.209 &\textbf{0} \\ \hline
	${\cal H}_{\text{HN}}$ &AI &\textbf{0.007} &0.067 &0.110 &0.021 &0.191 \\ 
${\cal H}_{1}$ &AI &\textbf{0.003} &0.071 &0.114 &0.017 &0.195 \\ 
${\cal H}_{\text{3D}}$ &AI + $\eta_+$ &0.069 &\textbf{0.006} &0.045 &0.084 &0.126 \\ 
${\cal H}_3$ &AI + $\eta_+$ &0.067 &\textbf{0.004} &0.047 &0.083 &0.128 \\ 
${\cal H}_2$ &A + $\eta_+$ &0.119 &0.049 &\textbf{0.006} &0.134 &0.081 \\ 
${\cal H}_4$ &AI + $\eta_-$ &0.015 &0.085 &0.128 &\textbf{0.004} &0.208 \\ 
${\cal H}_{\text{2D}}$ &AII + $\eta_+$ &0.138 &0.069 &\textbf{0.025} &0.154 &0.062 \\ 
	\hline\hline
	\end{tabular}}\label{ks_dis_r}
\end{table}

\section{Generalized Gaussian ensemble} 
    \label{GGE}

\subsection{Probability distribution functions}

Suppose that ${\cal H}$ is a non-Hermitian random matrix in one of the seven symmetry classes (i.e.,  classes A + $\eta$, AI, AI + $\eta_{\pm,}$, AII, and AI + $\eta_{\pm}$). 
Then, 
\begin{equation}
	{\cal{H}}^{\prime}  = \frac{x}{2} ({\cal H} + {\cal H}^{\dagger}) + \frac{1-x}{2}({\cal H} - {\cal H}^{\dagger}) 
	\label{GGE_Trans}
\end{equation}
is a non-Hermitian random matrix in the same symmetry class as ${\cal H}$ 
when $x$ is a real number %
satisfying $x \neq 0, 1$.
Thus, we can generalize the Gaussian ensemble in each symmetry 
class into the generalized Gaussian ensemble. ${\cal H}^{\prime}$ 
in the generalized Gaussian ensemble realizes with the same probability 
as ${\cal H}$ in the Gaussian ensemble,  
\begin{align}
    p^{\prime}({\cal H}^{\prime}) d{\cal H}^{\prime} = p({\cal H}) d{\cal H}. 
\end{align}

The inverse transform of Eq.~(\ref{GGE_Trans}) is given by
\begin{align}
	{\cal{H}}  = \frac{1}{2x} ({\cal H}^{\prime} + {\cal H}^{ \prime \dagger}) + \frac{1}{2(1-x)}({\cal H}^{\prime} - {\cal H}^{\prime \dagger}) \,.
\end{align}
As Eq.~(\ref{GGE_Trans}) is a linear transform, the Jacobian matrix of 
the transform depends only on $x$,
\begin{equation}
	d {\cal H}^{\prime} = C_x d {\cal H } \, ,
\end{equation}
where $C_x$ is an $x$-dependent constant. The probability 
distribution in the generalized Gaussian ensemble is given by 
\begin{equation}
	\begin{aligned}
		& p^{\prime} ({\cal H}^{\prime}) d {\cal H}^{\prime}  = p({\cal H}) d {\cal H} \\
		& =  C_N^{-1} e^{- \beta \text{Tr}( {\cal H}^{\dagger} {\cal H} )} d {\cal H} \\
		& =  C_N^{-1}C_x^{-1} e^{ -{\rm Tr }
		\left[
		\frac{\beta}{4} \left( \cal{H} + \cal{H}^{\dagger} \right)^2 -\frac{\beta}{4} \left( \cal{H} - \cal{H}^{ \dagger} \right)^2 
		\right]} d  {\cal H}^{\prime} \\
		& =  C_N^{-1}C_x^{-1} e^{ -{\rm Tr }
		\left[
		\frac{\beta}{4 x^2} \left( \cal{H}^{\prime} + \cal{H}^{\prime \dagger} \right)^2 -\frac{\beta}{4(1-x)^2} \left( \cal{H}^{\prime} - \cal{H}^{\prime \dagger} \right)^2 
		\right]} d  {\cal H}^{\prime} \,,\\
	\end{aligned}
\end{equation}
which reduces to
Eq.~(\ref{p_gge}) with 
$C^{-1}_{N,(\beta_1,\beta_2)} \equiv C^{-1}_{N} C^{-1}_x$, 
$\beta_1 \equiv \frac{\beta}{4 x^2} $, and $\beta_2 \equiv \frac{\beta}{4(1-x)^2}$.

\subsection{Level statistics}

We show that level-spacing and level-spacing ratio distributions, $p(s)$ and $p_r(r)$, in the generalized Gaussian 
ensemble (GGE) with $\beta_2>\beta_1$ converge faster than those 
in the Gaussian ensemble (GE). 
Figures~\ref{ps_CII_GE} and \ref{pr_CII_GE} 
show that $p(s)$ and $p_r(r)$ in class AII + $\eta_+$ converge 
more slowly than those in the other symmetry classes. Thus, 
we focus on $p(s)$ and $p_r(r)$ in class AII + $\eta_+$ 
and compare their convergence in the GGE and GE. 
For the other four symmetry classes, random matrices 
in the GGE have the same properties.  

We find that for $N\times N$ non-Hermitian random matrices in the %
GGE
with different parameter $\beta_1,\beta_2$, %
the average number of real eigenvalues 
$\bar{N}_{\rm real}$ is approximately 
scaled by %
(see Fig.~\ref{n_real_beta})
\begin{equation}
	\bar{N}_{\rm real} \propto \sqrt{\frac{\beta_2}{\beta_1}} \, ,
\end{equation}
which is compatible with Eq.~(\ref{eq: GGE - ellipse}).
When the average numbers of real eigenvalues of two GGE random matrices with different $\beta_2/\beta_1$ and different $N$ are approximately the same,
we also numerically find that
the level-spacing and level-spacing ratio distributions, $p(s)$ and $p_r(r)$, from the two matrices are %
the same
In Fig.~\ref{compare_elip}, we compare $p(s)$ and $p_r(r)$ from $252\times 252$ ($128 \times 128$) GGE random matrices with $\beta_2/\beta_1=16$ and those from $4000\times 4000$ ($2000 \times 2000$) GE random matrices. %
The error bars for almost all the data points overlap with each other, and the average number of real eigenvalues is approximately the same (e.g., $\sqrt{252 \times 16}=\sqrt{4032} \approx \sqrt{4000}$). 
In the limit of $N\rightarrow \infty$, $p(s)$ and $p_r(r)$ in the GGE with different $\beta_2/\beta_1$ converge to the same universal distribution. For the same matrix size $N$ of the random matrices, $p(s)$ and $p_r(r)$ converge faster in the GGE with larger $\beta_2/\beta_1$, as shown in Fig.~\ref{CII_mean_r}. The square-root scaling of $N_{\rm real}$ with respect to the dimensions $N$ of the matrices and the soft gap of density of complex eigenvalues around the real axis are also universally observed in random matrices of the %
GGE
(not shown).

\section{Kolmogorov-Smirnov distance}
\label{sec_ks}

It is more feasible to use cumulative distribution functions 
than %
probability distribution functions, when comparing ergodic phases of physical 
systems with random matrices. 
Here, we calculate the Kolmogorov-Smirnov 
(KS) distance from the cumulative spacing distribution function 
$\int^{s}_{0} p(s^{\prime}) ds^{\prime}$ and cumulative 
spacing-ratio distribution function $\int^{r}_{0} p_r(r^{\prime}) dr^{\prime}$ among physical systems and random matrices 
in different symmetry classes (Tables~\ref{ks_dis} 
and \ref{ks_dis_r}). In the calculations, we 
first obtain the empirical cumulative distribution function $F_e(x)$ 
from a set of real random variables $\{x_1,x_2,\cdots,x_n\}$ for 
the spacing and spacing ratio, 
\begin{equation}
F_e(x) \equiv \frac{1}{n} \sum_i^n \theta(x_i-x)\, ,
\end{equation}
with the step function $\theta$. This function $F_e(x)$ 
corresponds to the cumulative level-spacing and level-spacing-ratio 
distribution function of real eigenvalues. 
The KS distance between the two empirical cumulative distribution 
functions $F_{e1}(x)$ and $F_{e2}(x)$ is defined by 
the maximum value of the difference between the two functions over all $x$,
\begin{equation}
D_{e1, e2} \equiv \sup _{x}\left|F_{e1}(x)-F_{e2}(x)\right| \, .
\end{equation}

Tables~\ref{ks_dis} (Table~\ref{ks_dis_r}) summarizes the KS distance among 
$p(s)$ ($p_r(r)$) from $4000 \times 4000$ random matrices in the generalized Gaussian ensemble with $\beta_2/\beta_1 = 16$ for the five symmetry classes.
Table~\ref{ks_dis} (Table~\ref{ks_dis_r}) also give 
the KS distance between $p(s)$ ($p_r(r)$) from the physical systems in the ergodic phases with the maximal system size and $p(s)$ ($p_r(r)$) from the random matrices in 
the five symmetry classes. 
Tables~\ref{ks_dis} and \ref{ks_dis_r} show that in classes AI, AI + $\eta_{\pm}$, and A + $\eta$, the probability distribution functions $p(s)$ and $p_r(r)$ from the physical systems in the ergodic phases have the shortest distance to $p(s)$ and $p_r(r)$ from the random matrices in the same symmetry class, and the shortest distances are less than %
$0.01$. 
Tables~\ref{ks_dis} and \ref{ks_dis_r} also show that $p(s)$ and $p_r(r)$ of the physical systems in class AII + $\eta_+$ 
have the shortest KS distance with $p(s)$ and $p_r(r)$ of the random matrices in 
class A + $\eta$ %
but
have the larger distance ($> 0.05$) with $p(s)$ and $p_r(r)$ of the random matrices in class AII + $\eta_+$.

\section{Number variance and spectral compressibility}
    \label{sec_nv}

For an ensemble of non-Hermitian random matrices, we count the number $N_W(E)$ of real eigenvalues in an energy window $[-E,E]$ %
with $E \geq 0$
in each sample. 
We evaluate
the mean value $\langle N_W(E) \rangle$ and the variance $\Sigma_2(E) = \langle N_W(E)^2 \rangle - \langle N_W(E) \rangle^2 $ of the number in the energy window %
for different $E$. 
In our evaluation, only less than $50\%$ of all the real eigenvalues are included in the energy window. Note also that in classes AI + $\eta_{-}$ and AII + $\eta_{+}$, we regard each Kramers pair as one real eigenvalue and characterize the level interaction between neighboring Kramers pairs by $\Sigma_2(E)$ and $\langle N_W(E)\rangle$. 

In all random matrix ensembles studied in this paper (i.e., generalized Gaussian ensemble with $\beta_2/\beta_1 = 16$, Gaussian ensemble, and Bernoulli ensemble), 
we have the scaling relation
\begin{align}
    \Sigma_2(E) \simeq \chi N_W(E)
\end{align}
for all the five symmetry classes (Fig.~\ref{nv}).
The spectral compressibility $\chi $ in each symmetry class takes the same value for the different ensembles, suggesting the universality of $\chi$. 
While $\chi$ is less than $\chi_{\rm Poisson} = 1$ in classes AI, AI + $\eta_{\pm}$, and A + $\eta_+$, 
$\chi$ is larger than $\chi_{\rm Poisson} = 1$ in class AII + $\eta_+$. 
This unusual relation $\chi > \chi_{\rm Poisson} = 1$ in class AII + $\eta_+$ indicates that %
attractive interactions are more dominant than repulsive interactions in this symmetry class, %
which has no analogs in Hermitian random matrices and also non-Hermitian random matrices in the other four symmetry classes.

\clearpage

\bibliography{paper}
\end{document}